\pdfoutput=1
\documentclass[aps,prb,twocolumn,superscriptaddress,amsmath,amssymb,floatfix]{revtex4-2}

\usepackage{graphicx}
\usepackage{placeins}
\usepackage{bm}
\usepackage{xcolor}
\graphicspath{{./}}
\definecolor{linkblue}{rgb}{0.0,0.2,0.6}
\usepackage[colorlinks=true,linkcolor=linkblue,citecolor=linkblue,urlcolor=linkblue,bookmarksnumbered=true]{hyperref}

\newtheorem{proposition}{Proposition}
\begin{document}
\renewcommand{\figurename}{Figure}\renewcommand{\tablename}{Table}
\renewcommand{\thetable}{\arabic{table}}

\title{When the electrons go marching in:\\ Resonances and accuracy of pairwise approximations\\ in multiterminal Josephson junctions}

\author{A. Bar{\i}\c{s} \"Ozg\"uler}
\email{barisozguler@geeqchiqtech.com}
\affiliation{GeeQChIQ Technologies LLC, Berkeley, California, USA}
\affiliation{Haas School of Business, University of California, Berkeley, California 94720, USA}

\date[]{Received: September 30, 2026; Revised: November 14, 2026}

\begin{abstract}
Multiterminal Josephson junctions connect three or more superconductors through a common weak link. We quantify when their equilibrium currents admit a pairwise description. Multiterminal irreducible harmonics are energy components that depend jointly on three or more terminal phases and cannot be reduced to a sum of pair couplings. In microscopic planar-junction models, pairwise terms can capture nearly all energy variation while leaving appreciable current errors: phase differentiation amplifies the omitted harmonics. A terminal-independent Fourier projection quantifies this effect while allowing arbitrary nonsinusoidal pair couplings. Shared normal-region resonances provide a means of controlling the error. Selecting a mode from normal-state data and shifting it with a gate while retaining a frozen background reproduces the normalized error trends in the tested devices; prediction of the complete current waveform also depends on the background contribution. Finite-gap calculations retain the gate dependence at substantially lower absolute currents. An analytic single-level model supplies a sufficient detuning for pairwise accuracy. Together, these results connect normal-resonance control to a quantitative, current-based criterion for pairwise modeling.

\end{abstract}

\keywords{Andreev bound states; chaotic cavities; Fraunhofer interference; quantum transport; supercurrent diode effect}

\maketitle

\section{Introduction}
\label{sec:intro}
\begin{figure*}[!t]
  \includegraphics[width=\textwidth]{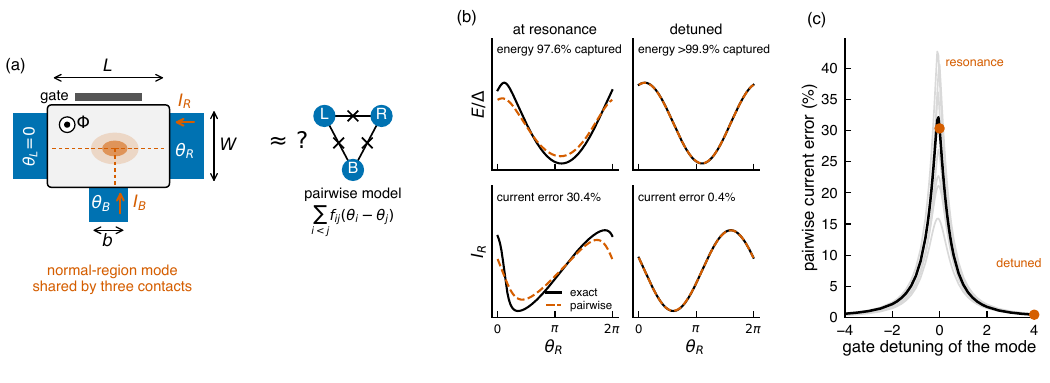}
  \caption{Pairwise accuracy of a multiterminal Josephson junction and its control by a gate. (a) Schematic of a
  planar three-terminal junction (not to scale). A normal InAs region of length $L$ and width $W$ is contacted by
  left, right and bottom superconducting leads, the bottom one of width $b$, with phases $\theta_L=0$, $\theta_R$ and
  $\theta_B$, currents $I_R$ and $I_B$ positive from the terminals into the junction, a perpendicular flux $\Phi$ and a gate. A normal-region mode is shared by the
  three contacts. The pairwise model beside it is built from two-terminal couplings. (b) Exact (black) and best
  pairwise (orange, dashed) energy and current $I_R$ along a phase cut through the point of largest current
  deviation ($\theta_L=0$, $\theta_B=3.34$), for a clean $160\times120$~nm$^2$ three-terminal device with
  $40$-nm contacts at zero flux, computed with an energy-independent scattering matrix. At the Hermitian-mode center
  ($x=0$) the pairwise terms capture 97.6 percent of the energy variation, %
an energy root-mean-square (RMS) error of
  15.5 percent, while the current RMS error averaged over all phases is 30.4 percent. With the mode detuned by the gate, the two curves agree and the current error falls to
  0.4 percent. (c) Pairwise current error versus the gate detuning $x=2\epsilon/\gamma$ of the mode, with $\epsilon$ its
  energy from the Fermi level and $\gamma$ its width, for this device (black) and seven further devices (grey).
  Orange dots mark the two settings shown in (b), the Hermitian-mode center ($x=0$) and a detuned setting ($x=4$).}
  \label{fig:concept}
\end{figure*}

Quantum transport between superconducting reservoirs connected through a weak
link gives rise to the Josephson effect, which relates equilibrium supercurrents
to the phase differences between the reservoirs \cite{annett2004superconductivity}. A multiterminal Josephson
junction (MTJJ) connects three or more reservoirs through one common weak link. For
$n$ terminals its energy depends on $n-1$ independent phase differences, and its
phase derivatives give terminal currents that sum to zero; with three
terminals, two phase differences control two independent currents. Such
junctions have drawn attention for several reasons. Their Andreev bound states
depend on several phases at once and can form topological band structures with
Weyl points
\cite{yokoyama2015singularities,riwar2016multi,xie2017topological,barakov2023abundance}.
Cooper pairs can also be transferred collectively among three or more
terminals, as quartets and other multiplets
\cite{freyn2011quartets,feinberg2015equilibrium,huang2022quartets}; in
equilibrium these processes add energy terms that depend on phase combinations
such as $\theta_1+\theta_2-2\theta_3$ \cite{feinberg2015equilibrium}.
Gate- and flux-tunable nonreciprocal supercurrents have been observed
\cite{gupta2023diode,chiles2023triode,coraiola2024diode}, and MTJJs have been proposed as elements of superconducting quantum circuits
\cite{matutecanadas2024circuits}. Their standard transport observable is the
critical-current contour. It is the boundary of the terminal currents
that admit a stable zero-voltage state and generalizes the two-terminal critical
current \cite{pankratova2020,ozguler2020thesis,lee2022thesis}. Measured
switching contours also depend on which stable branches the bias protocol
reaches \cite{pankratova2020,thieme2026hgte}.

Experiments now probe these junctions in detail. Phase-resolved spectroscopy
maps Andreev spectra as functions of several phases
\cite{coraiola2023phase,coraiola2024spin,nichele2025}, gate-controlled devices
select individual conduction modes
\cite{graziano2020transport,graziano2022selective}, and critical-current
contours have been measured and modeled in three-terminal devices
\cite{pankratova2020,thieme2026hgte}. The platforms include epitaxial Al on InAs
quantum wells \cite{shabani2016two,mayer2019superconducting,fornieri2019evidence}, HgTe
\cite{thieme2026hgte} and monocrystalline gold \cite{polevoy2025gold}. Theory
predicts tunable topological phases
\cite{ohnmacht2025reflectionless,ram2025tunable} and explains nonreciprocal
critical currents \cite{correa2024universal,chirolli2025diode}, which have also
been observed without an applied magnetic field \cite{zhang2024fieldfree}.

Theory also identifies processes that couple more than two terminals. Resonant
Josephson transport through a level near the Fermi energy, and pairwise
cotunneling far from resonance, are established mechanisms
\cite{beenakker1992resonant,matutecanadas2024circuits}. Equilibrium quartets can be extracted from current--phase relations
\cite{freyn2011quartets,feinberg2015equilibrium,ohnmacht2024quartet}. In
quantum-dot models of four-terminal junctions, the amplitudes of sextets, which
transfer three Cooper pairs among four terminals, depend on the dot-level
detuning, on the tunnel coupling between the dots and on interference between
tunneling paths \cite{ebert2025sextets}. At
the same time, networks of ordinary two-terminal Josephson junctions (JJs) reproduce nonlocal
and multiplet responses
\cite{melo2022multiplet,prosko2024flux,arnault2025multipletcircuit} and can
host Weyl points and topological transitions
\cite{fatemi2021weylcircuit,peyruchat2024spectral}. Eliminating their internal
superconducting islands can itself generate couplings among three or more
external phases. Coherent phase dragging occurs even within a pairwise-additive
ballistic model \cite{amin2001nonlocal}. A multiterminal signature therefore
does not by itself reveal whether, or why, the energy goes beyond pairwise
couplings in the external phases.
Reduced network models have also been used to interpret transport in three-terminal devices and arrays of four-terminal junctions~\cite{kolzer2023,behner2025,teller2025}; Sec.~\ref{sec:discussion} distinguishes these circuit descriptions from a microscopic pairwise decomposition.

Multiterminal irreducible harmonics are components of the energy that depend jointly on three or more terminal phases and cannot be written as a sum of functions of phase differences between pairs. The classification concerns phase dependence in the external variables; identifying a microscopic transfer mechanism requires additional information.

A pairwise description writes the junction energy as a sum of couplings
between pairs of terminals, as in a network of two-terminal junctions that
connect the terminals directly. A natural test of such a description is how
much of the energy's phase dependence the pair terms capture. Current accuracy depends on the phase derivatives of the energy, which weight each Fourier harmonic by its phase indices. Harmonics that couple three or more terminals carry larger
indices than the leading pair couplings, so a remainder that is small in the
energy weighs more in the currents. Critical-current contours, nonreciprocal
supercurrents and circuit responses are all measured through currents, so this
difference matters for the experiments and applications above. This raises two questions. How large is the current error of the best
pairwise description in realistic junctions, and which physical features
control it?

We address these questions by comparing the exact equilibrium currents with
those of a pairwise approximation,
$F_{\rm pair}=\sum_{i<j}f_{ij}(\theta_i-\theta_j)$, in which the $f_{ij}$ may be
arbitrary nonsinusoidal functions~\cite{golubov2004current}. Nonsinusoidal current--phase relations are
thus already included, so any approximation error measures departures from
pairwise additivity alone. This approximation keeps only the Fourier harmonics
that couple two terminals. It minimizes the root-mean-square (RMS) error of all
terminal currents, averaged uniformly over the phases, and it does not depend
on which terminal serves as the phase reference.

We compute equilibrium energies and currents from a microscopic scattering model
of planar junctions, where ``planar'' means a two-dimensional normal region with
superconducting contacts along its boundary. The calculations cover ensembles
of disordered three- and four-terminal junctions and designed devices with up
to sixteen terminals. Even where the pairwise terms carry nearly all of the energy variation, the
median current error, averaged over all phases, ranges from 8.0 to
21.2 percent across device groups. That is nearly twice the RMS energy error. We then identify a physical
mechanism that controls this error. The normal region confines electrons like a small quantum dot, with discrete modes; when one mode lies near the Fermi level and couples
to three or more terminals, electrons moving between any two terminals can propagate
coherently through that shared mode, and the energy acquires terms that couple all of
these terminals at once. Moving the mode away from the Fermi level with a gate
suppresses these terms [Fig.~\ref{fig:concept}]. %
In eight three- and four-terminal devices, selecting a mode from normal-state properties and shifting it while retaining a frozen background reproduces the gate dependence of the scalar normalized pairwise error without fitting currents. The construction evaluates phase-resolved superconducting currents; seven of the eight devices also satisfy the separate full-current-waveform criterion. Calculations with a finite
superconducting gap reproduce the predicted gate dependence in two clean (disorder-free) and two disordered devices, where current errors of tens of percent near resonance fall below one
percent far from it. For a single level coupled to any number of terminals, an
analytic model in the infinite-gap limit gives a detuning beyond which pairwise
accuracy is guaranteed within that model.

The surveys characterize the energy-independent scattering model. Selected finite-gap comparisons cover twelve ensemble devices, two eight-terminal devices and four of the eight resonance devices; Sec.~\ref{sec:discussion} explains their implications for normalized errors, absolute currents and physical implementation. For three terminals, we also compare harmonic accuracy with stable current regions, contour shapes, adiabatic current ramps and flux responses. These observables probe distinct aspects of the current field and its accessible branches. Voltage-biased multiplet transport~\cite{huang2022quartets,nowak2019supercurrent} requires a dynamical treatment beyond the equilibrium setting considered here.

The main argument has three parts: the energy-to-current weighting in Sec.~\ref{sec:multiterminal}, the single-level bound and planar resonance test in Sec.~\ref{sec:prediction}, and their experimental scope in Sec.~\ref{sec:discussion}. Section~\ref{sec:methods} defines the models and Sec.~\ref{sec:conclusion} states the conclusions. The Supplemental Material separates supporting work by purpose: Appendices~\ref{sec:diagnostics}--\ref{sec:island-circuit} concern contour observables, switching protocols and circuit comparisons; Appendix~\ref{sec:topology} is a separate topology illustration; Appendices~\ref{app:disorder-ensemble}--\ref{app:planar-controls} report the ensembles, terminal-count studies and predictor controls. Numerical refinement and mode checks are in Appendices~\ref{app:strengthening} and \ref{app:omitted-modes}, the isolated-resonance derivation is in Appendix~\ref{app:isolated-resonance}, and the local-gate test is in Appendix~\ref{app:local-gate}. These analyses provide the observable comparisons, numerical checks and physical conditions supporting the main argument.

\section{Model and methods}
\label{sec:methods}

\subsection{System}
We model a two-dimensional rectangular normal scattering region of width $W$ and length $L$ on a square tight-binding lattice with
spacing $a$, hopping $t=\hbar^2/(2m^\ast a^2)$, and chemical potential $\mu$ (set by the gate), with $m^\ast$ the InAs
effective mass [Fig.~\ref{fig:concept}(a)]. Two superconducting leads (left, right) and, for the three-terminal
device, a bottom lead of contact width $b$ are attached; the superconducting gap is $\Delta=0.18~\mathrm{meV}$. Unless
stated otherwise the bottom contact spans the full length of the normal region, $b=L$. The reference phase
$\theta_L=0$ is grounded, leaving $(\theta_R,\theta_B)$ as the independent phases. The baseline model is spin degenerate and clean. It omits spin--orbit coupling and Zeeman splitting,
and on-site disorder enters only the disorder calculations. The perpendicular magnetic flux, which acts on the orbital motion, is the only
external parameter that breaks time-reversal symmetry. With full-width contacts the clean device approximates a transmitting waveguide,
and the narrow and split contacts of Sec.~\ref{sec:multiterminal} define further scattering geometries.

The superconducting terminals are conventional spin-singlet
reservoirs with local $s$-wave Bardeen--Cooper--Schrieffer (BCS) pairing~\cite{bardeen1957theory}.
Mean-field decoupling in the pairing channel gives a quadratic Hamiltonian;
its Bogoliubov--de Gennes (BdG) eigenproblem determines the electron--hole
quasiparticle amplitudes and energies (Sec.~1.5.1 of
Ref.~\cite{zhu2016bdg}). At each phase configuration, we fix the reservoir order
parameters at $\Delta_j=\Delta e^{i\theta_j}$, uniform in each lead, instead of solving the
gap equation self-consistently. These pairing fields determine the Andreev processes that
carry Josephson current through the shared normal
region~\cite{beenakker1991qpc,beenakker1991Smatrix}.
Although electrons form Cooper pairs, the energy need not be a sum of terms that
each involve two terminals. Coherent propagation through
the weak link can generate terms that depend on all three phases at once, such as a quartet
term $E_Q\cos(\theta_R+\theta_B-2\theta_L)$, within this prescribed-reservoir model.

\subsection{Supercurrent conventions}
\label{sec:conventions}
At $T=0$, terminal $j$ carries the pairing term $\Delta e^{i\theta_j}c^\dagger_{j\uparrow}c^\dagger_{j\downarrow}$
and its Hermitian conjugate. We define $I_j$ as the conventional charge current flowing from terminal $j$ into the
junction, with $e>0$ the magnitude of the electron charge,
\begin{equation}
  I_j = \frac{2e}{\hbar}\frac{\partial E_{\rm gs}}{\partial\theta_j}
      = -\frac{2e}{\hbar}\sum_{E_p>0}\frac{\partial E_p}{\partial\theta_j},
  \label{eq:supercurrent}
\end{equation}
where $\{E_p\}$ are the positive eigenvalues of the spin-reduced Bogoliubov--de Gennes problem in the Nambu basis
$(c_\uparrow,c_\downarrow^\dagger)$. Each such level is spin degenerate, and the ground-state energy of that basis is
$E_{\rm gs}=-\sum_{E_p>0}E_p$ up to a phase-independent constant, with no double counting, so no further spin factor may be applied; we verified
this by exact many-body diagonalization of a small spinful $s$-wave chain, which reproduces $E_{\rm gs}(\theta)$ from
the reduced-block sum to machine precision. Equation~\eqref{eq:supercurrent} is the convention of
Refs.~\cite{riwar2016multi,meyer2017nontrivial}; the current from the junction into the terminal is its negative, and
every contour diagnostic below is invariant under the global sign. In the short-junction approximation of Route~A
(Sec.~\ref{sec:methods}\,C) the continuum above $\Delta$ does not depend on the phases, and the sum reduces to the
Andreev levels $0<E_p<\Delta$; Route~B evaluates the full sum.

The flux is reported in units of the superconducting flux quantum $\Phi_0=h/2e$, while the single-electron magnetic (Peierls) phase entering the tight-binding hoppings is
$\chi=(e/\hbar)\int\bm A\cdot d\bm l$ (Secs.~5.1 and 7.1 of Ref.~\cite{zhu2016bdg}); these two conventions are fixed independently and cross-checked against the
position of the first Fraunhofer minimum (Appendix~\ref{sec:baselines}). With site coordinates in nm and $f=(\Phi/\Phi_0)/(WL)$ in nm$^{-2}$, the horizontal hopping from site $s$ to
site $t$ is
\begin{equation}
 t_{ts}=-t\exp[-i\pi f(x_t-x_s)(y_t+y_s)/2],
 \label{eq:implemented-peierls}
\end{equation}
while vertical hoppings are $-t$ and reverse hoppings are complex conjugates, which fixes a Landau gauge. The semi-infinite normal leads have real hoppings,
as do the bonds coupling their cells to the device; thus the field prescription is terminated at the contacts.

The superconducting reservoirs enter through their Andreev reflection matrices
$r_{A,j}=i e^{-i\theta_j}I_{N_j}$, where $I_{N_j}$ is the identity on the $N_j$ orbital channels of lead $j$ in the spin-reduced formulation. No self-consistent screening field or
additional interface phase is imposed.
Currents are quoted in nA; with
$\Delta=0.18~\mathrm{meV}$ one has $e\Delta/\hbar=43.8~\mathrm{nA}$, the per-channel ceiling of a perfectly
transmitting short junction, $I_c=Ne\Delta/\hbar$ for $N$ spin-degenerate channels \cite{beenakker1991qpc}.

\subsection{Two computational routes}
\emph{Route A} (short junction) obtains the Andreev spectrum from the scattering matrix $s$ of the normal region,
computed with Kwant \cite{groth2014kwant} at energy zero in its basis of incoming and outgoing modes, through the
operator
\begin{equation}
  A=\tfrac12\left(r_A s + s^{\mathsf T} r_A\right),\qquad r_A=i\,\mathrm{diag}\big(e^{-i\theta_j}I_{N_j}\big),
  \label{eq:Aop}
\end{equation}
with $A^\dagger A\,\Psi=(E/\Delta)^2\Psi$ \cite{beenakker1991Smatrix,irfan2018geometric,van2014single}; $N_j$ is the
number of modes of lead $j$. In this basis the plus sign reproduces the single-channel spectrum
$E=\Delta\sqrt{1-\tau\sin^2(\phi/2)}$. Ref.~\cite{irfan2018geometric} writes the operator with a minus sign, in a basis in which the outgoing
modes are the time-reversed partners of the incoming ones. We verified that the
$A^\dagger A$ spectrum coincides, to $10^{-14}$ at zero and finite flux, with the spectrum of Beenakker's secular
equation $\det[1-\alpha(E)^2R^\dagger s^*Rs]=0$, $\alpha=e^{-i\arccos(E/\Delta)}$, $R=\mathrm{diag}(e^{-i\theta_j})$~\cite{beenakker1991Smatrix}.
Each physical subgap level appears twice in the spectrum of
$A^\dagger A$ (a structural electron--hole degeneracy of the operator built on the full scattering space), and the
sums in Eq.~\eqref{eq:supercurrent} are halved accordingly; a single perfectly transmitting channel then gives
$I_c=e\Delta/\hbar=43.8~\mathrm{nA}$ \cite{beenakker1991qpc}.

\emph{Route B} diagonalizes the finite-system Bogoliubov--de Gennes Hamiltonian \cite{zhu2016bdg} directly, with superconducting leads
of length $2\xi$ ($\xi=\hbar v_F/\Delta$), and evaluates the phase derivative of the full ground-state energy, that is, of the sum over all negative-energy levels of the finite matrix. The subgap levels alone do not give the full current. In a two-channel
benchmark junction, the levels above the gap, out to about $20\Delta$, carry a
phase-dependent energy \cite{furusaki1991dc,bagwell1992suppression} that changes the current of the finite system by
9--61\% of its subgap value, independently of the lead length (leads of 2, 5, 10 and 20$\xi$;
Appendix~\ref{sec:baselines}). In the ideal equal-gap,
zero-field short-junction limit the continuum makes no phase-dependent contribution to the equilibrium current
\cite{beenakker1991qpc}. The finite-system benchmark measures the subgap-versus-full-current discrepancy of one coarse junction; its scope and relation to Route A are detailed in Appendix~\ref{sec:baselines}. Both routes were first set up for MTJJs
in the author's thesis \cite{ozguler2020thesis}, which summed only the subgap levels; the conventions,
normalization and error control of this
section supersede those used there.

Route A retains a finite superconducting gap; its short-junction approximation
treats the normal scattering matrix as energy independent on that scale.
The infinite-gap single-level model in Sec.~\ref{sec:prediction} is a
separate limit. The selected finite-gap tests in
Sec.~\ref{sec:planar-resonance} retain dispersive superconducting leads and
phase-dependent continuum contributions with the same fixed-gap BCS reservoirs. These tests cover four of the eight selected resonance devices, two clean and two disordered. The device matrices and disorder ensembles, referred to below as surveys, use Route A. Their dwell times are not small compared with $\hbar/\Delta$ (Appendix~\ref{app:disorder-ensemble}), so these surveys characterize the energy-independent scattering model. Finite-gap checks of twelve selected ensemble devices and two eight-terminal devices give a largest observed change of about 4.3 percentage points in the normalized pairwise error.

Normalized Fourier content and absolute current amplitude test different aspects of the approximation. The selected finite-gap comparisons quantify both for those devices; extending the survey results to material devices requires corresponding checks of their dispersive scattering and contact spectra.

\section{Pairwise accuracy from three to sixteen terminals}
\label{sec:multiterminal}

\subsection{A terminal-independent definition}
For $n$ terminals, write the dimensionless equilibrium energy as
\begin{equation}
 \mathcal E(\bm\theta)=c_0+\sum_{\substack{\bm q\ne0\\\sum_iq_i=0}}
 c_{\bm q}e^{i\bm q\cdot\bm\theta},\qquad
 \bm g=\nabla_{\bm\theta}\mathcal E,
 \label{eq:nt-energy}
\end{equation}
where $\mathcal E=E_{\rm gs}/\Delta$, $\bm g=\hbar\bm I/(2e\Delta)$, $q_i$ are integers,
and $c_{-\bm q}=c_{\bm q}^*$. A harmonic is pairwise precisely when its charge
vector has support on two terminals, $\bm q=k(\bm e_i-\bm e_j)$, with $\bm e_i$ a coordinate unit vector.
All other nonconstant charges have support on at least three terminals; these are the multiterminal irreducible harmonics. This is a classification of energy in the retained external phases, with microscopic multipair transfer providing one possible physical origin. This classification is independent of
which terminal is grounded. It allows arbitrary higher harmonics within each
pair coupling and therefore does not assume sinusoidal current--phase relations.

Let $P_2\mathcal E$ denote the orthogonal projection onto these pair modes, and let
$\mathcal R$ be the set of omitted charges. With normalized uniform phase
averaging on the $(n-1)$-torus, Parseval's identity gives
\begin{align}
 f_E&=\frac{\langle|\mathcal E-P_2\mathcal E|^2\rangle}{\langle|\mathcal E-c_0|^2\rangle}
 =\frac{\sum_{\mathcal R}|c_{\bm q}|^2}{\sum_{\bm q\ne0}|c_{\bm q}|^2},
 \label{eq:nt-energy-error}\\
 \epsilon_I^2&=\frac{\langle\|\bm g-\nabla P_2\mathcal E\|^2\rangle}
 {\langle\|\bm g\|^2\rangle}
 =\frac{\sum_{\mathcal R}\|\bm q\|^2|c_{\bm q}|^2}
 {\sum_{\bm q\ne0}\|\bm q\|^2|c_{\bm q}|^2}.
 \label{eq:nt-current-error}
\end{align}
The constant term is retained in $P_2\mathcal E$. We call $f_E$ the energy-power fraction and $\epsilon_E\equiv\sqrt{f_E}$ the energy RMS error, plotted in Fig.~\ref{fig:ensemble-headline} and in Fig.~\ref{fig:multiterminal} of Appendix~\ref{app:terminal-tables}. The current norm includes all $n$
terminals and is invariant under terminal permutation; it is not the two-component
$(I_R,I_B)$ norm of Eq.~\eqref{eq:disorder-current-error}, used in Appendix~\ref{sec:pairwise-accuracy}.
Differentiation preserves each Fourier sector, so the same projection also gives the smallest
phase-averaged current error of any pairwise model. Equation~\eqref{eq:nt-current-error} holds whenever the
currents are square integrable; it does not require a gap at every phase.

In the main text, ``pairwise current error'' means the all-terminal quantity $\epsilon_I$ in Eq.~\eqref{eq:nt-current-error}. The two-current error in the three-terminal appendices omits the grounded-terminal component and is not numerically interchangeable with it. The predictor waveform error instead compares the entire predicted current field with the full microscopic field, including the retained pairwise sector; matching $\epsilon_I$ therefore need not imply a small waveform error.

\begin{table*}[t]
\caption{Error measures used in this paper. Phase averages use the normalized uniform torus measure. Energy power,
current-waveform error and current-region discrepancy are different observables; a small value of one does not certify the others.}
\label{tab:error-notation}
\begin{ruledtabular}
\begin{tabular}{p{0.15\textwidth}p{0.43\textwidth}p{0.34\textwidth}}
Quantity & Definition or reference & Interpretation\\
\hline
$f_E$ & Eq.~\eqref{eq:nt-energy-error} & Omitted energy power divided by nonconstant energy power\\
$\epsilon_E$ & $\sqrt{f_E}$ & Relative RMS energy error\\
$\epsilon_I$ & Eq.~\eqref{eq:nt-current-error}, all terminal currents & Relative RMS error of the best pairwise model\\
Two-current error & Eq.~\eqref{eq:disorder-current-error}, only $(I_R,I_B)$ & Different norm used in Appendix~\ref{sec:pairwise-accuracy}\\
Predictor waveform error & $\langle\|\bm I_{\rm pred}-\bm I_{\rm full}\|^2\rangle^{1/2}/\langle\|\bm I_{\rm full}\|^2\rangle^{1/2}$ & Error of the mode predictor, including pairwise currents\\
$d_H$, $d_A$ & Eq.~\eqref{eq:region-distance} & Current-region boundary and area discrepancies\\
\end{tabular}
\end{ruledtabular}
\end{table*}

Energy dominance does not alone bound current accuracy. For example, the smooth
three-terminal energies
\begin{equation}
 \mathcal E_K=-\cos\theta_R-\cos\theta_B
       +K^{-1}\cos[K(2\theta_R-\theta_B)]
 \label{eq:nt-counterexample}
\end{equation}
have $f_E=(2K^2+1)^{-1}\to0$, but $\epsilon_I^2=3/5$ for every positive integer
$K$. This is a mathematical counterexample to an energy-only inference. More generally, Eqs.~\eqref{eq:nt-energy-error} and \eqref{eq:nt-current-error} give
\begin{equation}
 \frac{\epsilon_I^2}{1-\epsilon_I^2}=\frac{f_E}{1-f_E}\,
 \frac{\langle\|\bm q\|^2\rangle_{\mathcal R}}{\langle\|\bm q\|^2\rangle_{P}},
 \label{eq:weight-ratio}
\end{equation}
where $\langle\|\bm q\|^2\rangle_{\mathcal R}$ and $\langle\|\bm q\|^2\rangle_{P}$ are the energy-weighted means of
$\|\bm q\|^2$ over the omitted and retained harmonics. For small errors the current error is therefore
$\epsilon_E$ times the square root of that ratio. If the omitted energy sits in quartet harmonics
($\|\bm q\|^2=6$) and the retained energy in fundamental pair harmonics ($\|\bm q\|^2=2$), the factor is
$\sqrt3\approx1.7$, close to the ratios of 1.8--1.9 in Fig.~\ref{fig:ensemble-headline}. For a square-integrable
Hessian, the corresponding Frobenius error instead weights $\|\bm q\|^4$;
that additional regularity need not hold at a zero-temperature cusp.
Even small average errors do not supply a pointwise stability guarantee. Positivity at a specified phase is guaranteed if the reduced pair Hessian's
smallest eigenvalue exceeds the operator norm of its difference from the full
Hessian. The vanishing margin at a fold explains why current-region boundaries
and switching protocols require separate tests.

\subsection{Contact and channel controls at two device sizes}
We apply this definition to 36 zero-field lattice cases with
$n=3,4,5,6$, keeping $W=L=120$~nm, $a=5$~nm and $\mu=20$~meV.
In family A, each contact is $40$~nm wide and carries two channels.
The three-terminal device has left, right and bottom contacts; a top contact
is added for $n=4$. For $n=5$ the bottom contact is replaced by two contacts
centered at $\pm35$~nm; for $n=6$ the top contact is also split [Fig.~\ref{fig:multiterminal-controls}(a) in Appendix~\ref{app:terminal-tables}]. Family B keeps twelve channels in total, with contact widths of $100$~nm for $n=3$, $80$~nm for $n=4$
and, for $n=5$, $80$~nm at left and right with three $40$~nm contacts. At $n=6$ it coincides with
family A [Fig.~\ref{fig:multiterminal-controls}(b)]. Channel counts are checked from the normal scattering matrix.
These controls compare fixed channels per lead with fixed total channels while varying contact placement and width together with terminal count.

For $n=3$--$6$, family A includes clean devices, one matched disorder pattern at
$U_0/\mu=0.5$ and $2$, and an $80$~meV on-site contact-barrier control for
each $n$. Family B includes the clean and moderate-disorder devices.
Four further illustrative disorder realizations are scaled through $U_0/\mu=0.25,0.5,1,2$ at $n=4$. The on-site disorder is uniform in $[-U_0/2,U_0/2]$; the ratios $U_0/\mu$ are dimensionless numerical controls of the lattice model and are not calibrated to a particular material. All calculations use the equilibrium energy and currents of Sec.~\ref{sec:methods}.

We add 20 physical cases at $n=8,12,16$. The eight-terminal extension
of family A has two $40$ nm contacts on each side of the same $120$ nm square,
centered at $\pm35$ nm, in clean and matched $U_0/\mu=0.5$ conditions.
Families C and D instead keep the square size fixed at $240$ nm
[Fig.~\ref{fig:multiterminal-controls}(c,d)]. C has two channels per terminal, with contact centers along each side at $\pm60$ nm for $n=8$,
$-80,0,80$ nm for $n=12$ and $-90,-30,30,90$ nm for $n=16$. Each has clean, $U_0/\mu=0.5$,
$U_0/\mu=2$ and $80$ meV contact-barrier controls. D keeps 32 channels in total. For $n=8$, each $100$ nm contact has four channels; for $n=12$, contact widths
of $70$, $40$ and $70$ nm give $3$, $2$ and $3$ channels per side without shared corner interfaces.
Two additional D controls retain $80,40,80$ nm widths, for which neighboring
contacts share four normal-region corner sites; these are reported separately.
D includes the clean and
moderate-disorder cases and coincides with C at $n=16$. Both sizes retain
$a=5$ nm and $\mu=20$ meV.

The phase integration, its convergence checks and the implementation tests are described in
Appendix~\ref{app:terminal-tables}. The resolved charge-two decomposition remains limited to
$n\le6$; the larger devices test the full pairwise remainder.

\subsection{Energy dominance and current corrections}
Pairwise terms retain at least 96.9 percent of the energy-variation
power across the combined 56-case matrix, yet all-terminal current errors span
0.7--33.8 percent. The energy RMS error $\epsilon_E$ reaches 17.6 percent; Fig.~\ref{fig:multiterminal}
(Appendix~\ref{app:terminal-tables}) compares both errors on an RMS basis. Thus predominantly pairwise energy
coexists with appreciable current corrections in explicitly connected junctions
through sixteen terminals. Within the $n=3$--$6$ matrix, the largest error occurs in the four-terminal
family-A device at $U_0/\mu=2$ for the third matched pattern. There an energy-power remainder
of 3.08 percent accompanies a 33.8-percent current error,
and adding the charge-two modes still leaves 23.40 percent.
Contact geometry matters even at fixed
$n$. For the clean three-terminal controls, the narrow-contact and twelve-channel
families have current errors of 18.01 and 4.45 percent,
respectively (Table~\ref{tab:multiterminal} in Appendix~\ref{app:terminal-tables}). Neither family supports a universal
monotonic dependence of the error on terminal count.

The larger-terminal controls reinforce the distinction
(Table~\ref{tab:large-n} in Appendix~\ref{app:terminal-tables}). In the 240 nm family C, clean current errors are
29.53, 23.98 and 30.32 percent at $n=8,12,16$;
the corresponding energy-power remainders are 2.50, 1.88
and 2.92 percent. At fixed $n=8$ and the same device size, widening
the contacts to obtain 32 total channels lowers the clean current error from
29.53 to 19.84 percent. The clean twelve-terminal comparison
is 23.98 versus 19.25 percent. The C/D controls therefore
compare terminal partition and contact width at fixed device size, while A/C
at $n=8$ also changes the size and contact positions; these are distinct tests.
The corner-sharing twelve-terminal D controls have current errors
of 20.42 percent (clean) and 21.75 percent (disordered),
compared with 19.25 and 20.27 percent for separated interfaces
at the same total channel count. Both geometries are retained in the case matrix.
The comparisons identify contact geometry and channel allocation, alongside terminal count, as relevant variables for current accuracy.

The ratio $R=\epsilon_I/\epsilon_E=\epsilon_I/\sqrt{f_E}$ ranges from $1.750$ to $2.637$ across the 26 distinct devices plotted in Fig.~\ref{fig:multiterminal}(e,f). For clean family C it is $1.867$, $1.750$ and $1.774$ at $n=8,12,16$, whereas the corresponding fixed-total-channel family D decreases from $2.009$ through $1.916$ to $1.774$. The variation around two reflects the derivative weighting in Eq.~\eqref{eq:weight-ratio} and the different spectra of the specified contact geometries. With a finite pair-harmonic cutoff, the omitted sector can contain both unretained high-order pair harmonics and multiterminal irreducible harmonics.

The lowest irreducible transferred-pair order has charges that are permutations
of $(2,-1,-1,0,\ldots)$ or $(1,1,-1,-1,0,\ldots)$, and their negatives.
Their current weights $\|\bm q\|^2$ are $6$ and $4$, compared with $2$ for a
fundamental pair mode. Four-terminal support first becomes possible at $n=4$;
its existence does not imply numerical dominance. Counting conjugate pairs once,
there are $3\binom{n}{3}$ three-terminal and $3\binom{n}{4}$ four-terminal
charge-two modes. In the moderate-disorder family-A cases with $n=4,5,6$,
the four-terminal subset carries 0.9--12.6 percent of the
omitted current power. Retaining all charge-two multiterminal irreducible harmonics lowers the
moderate-disorder errors from 16.41, 23.60 and 24.98 percent
to 6.40, 13.36 and 13.85 percent, respectively.
This identifies a useful leading correction, without exhausting the remainder.
Support on five or more terminals requires higher transferred-pair order and therefore remains in that unresolved remainder.

The contact-barrier controls show a strong dependence on geometry. Their current errors are 0.71, 2.22, 9.99 and 2.55 percent for $n=3,4,5,6$. The validity of a tunneling expansion therefore depends on the transmission eigenvalues as well as the imposed boundary potential. For example, the five-terminal
barrier device has a largest lead-to-rest transmission eigenvalue of 0.70,
computed from $\mathbb{1}-r_i^\dagger r_i$ for each lead reflection block $r_i$;
not all its channels are weakly transmitting.

\begin{figure}[t]
 \includegraphics[width=\columnwidth]{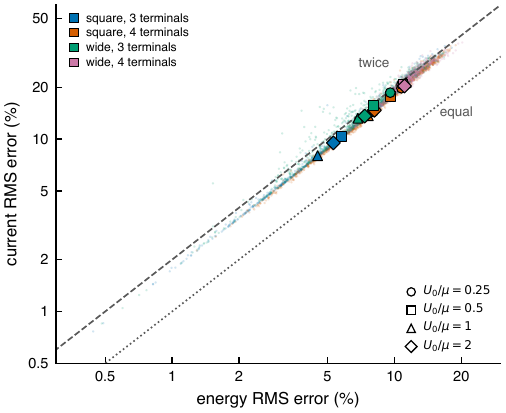}
 \caption{Current error versus energy error in the matched disorder ensembles. Small points show all 4096
 realizations, 256 for each of four device classes at four disorder strengths; large symbols are the sixteen group
medians. The dotted line marks equal errors. The dashed line labelled ``twice'' is $\epsilon_I=2\epsilon_E$.}
 \label{fig:ensemble-headline}
\end{figure}

A fixed ensemble of 256 matched realizations for each of four device
classes, square and rectangular junctions with three and four terminals,
tests disorder dependence at four strengths.
The 4096 disordered calculations quantify both sample variability and the
actual current penalty of a pairwise description
[Fig.~\ref{fig:multiterminal-controls}(e), Appendix~\ref{app:disorder-ensemble}].
Across the sixteen device/strength groups, median relative RMS energy errors
are 4.5--11.3 percent and median all-terminal
current RMS errors are 8.0--21.2 percent [Fig.~\ref{fig:ensemble-headline}].
The recorded per-realization phase-sampling sensitivities reach 0.2 percentage points.
The RMS currents that normalize these errors have group medians of 27--47~nA, and every group's 10th
percentile is at least 16~nA, for $\Delta=0.18$~meV ($2e\Delta/\hbar=87.6$~nA); small currents
therefore do not inflate the relative errors. The nonpairwise RMS currents have group medians of 2.1--8.7~nA.
The corresponding median energy-power remainders are
0.2--1.3 percent.
Restoring all charge-two multiterminal irreducible harmonics reduces those medians to
2.5--12.7 percent. Charge-two corrections thus recover part of the omitted current, with the remaining error spanning both sides of the 5-percent benchmark. This numerical benchmark was fixed before the calculations; experimental tolerances require a specified measurement or circuit.
Table~\ref{tab:threshold} in Appendix~\ref{app:disorder-ensemble} shows how the fractions of realizations
above and below it change between 2 and 10 percent. The distributions depend
on geometry, terminal count and disorder strength. Two square four-terminal amplitude medians do not pass the prespecified sample-size
stability diagnostic, and a separate 256-realization replication tests that device
(Appendix~\ref{app:disorder-ensemble}).

In the larger family-C barrier controls, current errors are
3.87, 19.62 and 7.51 percent for $n=8,12,16$.
These values characterize the particular contact potentials studied.

The contour and ramp analyses of Appendices~\ref{sec:diagnostics}--\ref{sec:beyond} are three-terminal. Stable current regions and switching for $n\ge4$, and the limit of large $n$, remain open.

\section{Controlled limits and predictive tests}
\label{sec:prediction}

The error in Eq.~\eqref{eq:nt-current-error} measures the resolved phase-dependent currents. To connect this diagnostic to physical control parameters, we derive a sufficient bound in a solvable model, survey barrier, gate and temperature controls, and test a normal-resonance mechanism by gate intervention in new devices.

\subsection{A sufficient bound for a single induced level}
Consider a noninteracting spin-degenerate level of detuning $\epsilon_d$ in
the infinite-superconducting-gap limit, with fixed nonnegative induced pairing
amplitudes $\Gamma_i$ and grand-canonical equilibrium occupations. Set
$z=\sum_i\Gamma_i e^{i\theta_i}$, $B=|z|^2$, and
$E=(\epsilon_d^2+B)^{1/2}$. Its phase-dependent free energy is
\begin{equation}
 F=-2k_BT\log[2\cosh(E/2k_BT)],\qquad F_{T=0}=-E.
 \label{eq:dot-free}
\end{equation}
This is a separate model from the planar junction, with energy unit $\Gamma_\Sigma=\sum_i\Gamma_i$. Interactions, parity constraints,
additional levels and self-consistent gaps are excluded. The expansion
$F_{T=0}=-|\epsilon_d|-B/(2|\epsilon_d|)+B^2/(8|\epsilon_d|^3)+\cdots$
has a pairwise leading term; higher powers generate multiterminal irreducible harmonics.
This realizes established cotunneling and resonance mechanisms
\cite{matutecanadas2024circuits,feinberg2015equilibrium,ohnmacht2024quartet}.

Let $\epsilon_I$ use the uniform phase-torus and all-terminal norm of
Eq.~\eqref{eq:nt-current-error}, retaining all pair harmonics, and define
\begin{align}
 E_-&=\sqrt{\epsilon_d^2+\max(2\Gamma_{\max}-\Gamma_\Sigma,0)^2},\nonumber\\
 E_+&=\sqrt{\epsilon_d^2+\Gamma_\Sigma^2},\nonumber\\
 h_T(E)&=\tanh(E/2k_BT)/E.
 \label{eq:dot-endpoints}
\end{align}
The continuous limit is $h_T(0)=1/(2k_BT)$ for $T>0$, while $h_0(E)=1/E$.
\begin{proposition}\label{prop:dot}
Assume a single noninteracting, spin-degenerate level with detuning $\epsilon_d$ in the infinite-gap limit, fixed
couplings $\Gamma_i\ge0$ to $n\ge2$ terminals, grand-canonical equilibrium at temperature $T\ge0$, and a nonzero
phase-averaged current norm. Then the exact pair projection satisfies
\begin{equation}
 \epsilon_I\le \rho_T
 =\frac{h_T(E_-)-h_T(E_+)}{h_T(E_-)+h_T(E_+)}.
 \label{eq:dot-bound}
\end{equation}
Consequently, for every $T\ge0$ and $0<\delta<1$,
\begin{equation}
 \frac{|\epsilon_d|}{\Gamma_\Sigma}\ge
 \frac{1-\delta}{2\sqrt\delta}
 \quad\Longrightarrow\quad \epsilon_I\le\delta.
 \label{eq:dot-sufficient}
\end{equation}
\end{proposition}
At $T=0$ and $E_-=0$, the limiting statement is the trivial bound
$\rho_0=1$. A vanishing full current has undefined relative error.

To prove Eq.~\eqref{eq:dot-bound}, write the current, apart from its common
$2e/\hbar$ factor, as $\bm j=h_T(E)\bm v$ with
$\bm v=-\nabla B/2$. The vector $\bm v$ is purely pairwise because
$B=\sum_i\Gamma_i^2+2\sum_{i<j}\Gamma_i\Gamma_j\cos(\theta_i-\theta_j)$.
Triangle and polygon inequalities give $E_-\le E\le E_+$.
The function $h_T$ is positive and nonincreasing, because
$\tanh y-y\,\mathrm{sech}^2y\ge0$ for $y\ge0$, which follows by differentiating from zero.
Write $a=h_T(E_+)$ and $b=h_T(E_-)$ and choose
$\bm j_p=c\bm v$ with $c=2ab/(a+b)$. This is the current of the pairwise
energy $-cB/2$, and $|1-c/h_T(E)|\le(b-a)/(b+a)$ pointwise.
Integration gives the same relative $L^2$ bound. Orthogonality of the
Fourier sectors ensures that the best pair projection cannot do worse than
this comparator. Measure-zero zero-temperature crossings do not affect the
$L^2$ inequality; the singular endpoint follows by orthogonal contraction.

For Eq.~\eqref{eq:dot-sufficient}, note that $\rho_T$ increases with the ratio $h_T(E_-)/h_T(E_+)$, which is at
most $E_+/E_-$ at every $T\ge0$ because $\tanh$ is increasing, and that $E_-\ge|\epsilon_d|$. Hence
$\epsilon_I\le(E_+-|\epsilon_d|)/(E_++|\epsilon_d|)$, which is at most $\delta$ under the stated detuning. The condition
does not depend on the contact ratios. For $\delta=0.05$, the threshold is 2.1243 when rounded
upward. Section~\ref{sec:planar-resonance} maps an ideal isolated rank-one resonance, with background modes omitted, to the same normalized phase dependence through Eq.~\eqref{eq:resonance-energy}. Proposition~\ref{prop:dot} therefore gives a rigorous sufficient condition for the single-level model. In multilevel junctions, spectator currents and interference require separate control: the threshold guides the resonant-to-cotunneling crossover when an isolated mode dominates, but a conservative bound including those contributions has not been established.

\begin{figure*}[t]
 \centering
 \includegraphics[width=\textwidth]{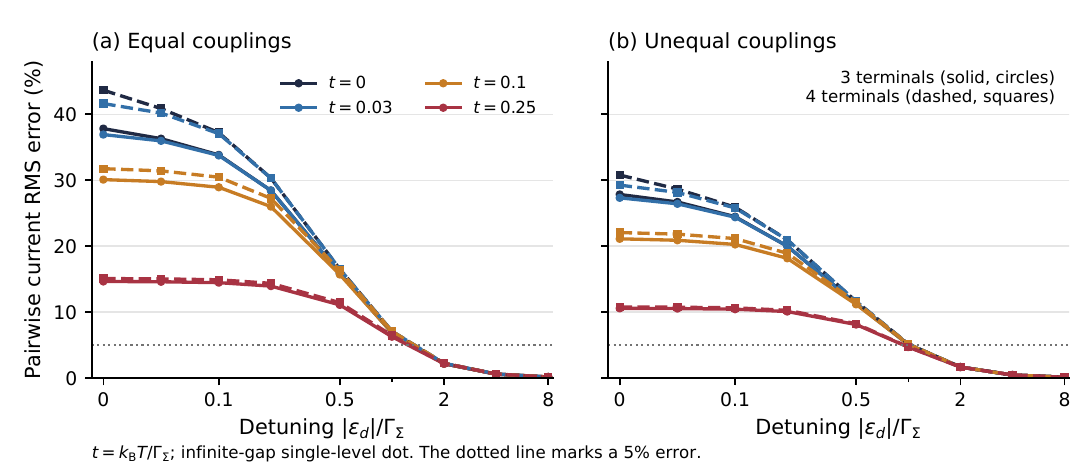}
 \caption{Controlled single-level crossover. The pairwise current error is
 evaluated for three and four terminals with equal couplings (a), or ratios
 $(1,0.6,0.3)$ and $(1,0.6,0.3,0.15)$ normalized to unit sum (b).
 Temperature is in $\Gamma_\Sigma$ units. Detuning suppresses the error;
 thermal smoothing reduces the resonant error but does not bring it below 5 percent
 in the displayed temperature range. The dotted line marks the prespecified 5 percent numerical benchmark, which is not an experimental acceptance threshold.}
 \label{fig:dot-control}
\end{figure*}

The numerical benchmark contains 36 systems at four temperatures,
$k_BT/\Gamma_\Sigma=0,0.03,0.1,0.25$, with detunings
$|\epsilon_d|/\Gamma_\Sigma=0,0.05,0.1,0.2,0.5,1,2,4,8$ and the two
contact families in Fig.~\ref{fig:dot-control}. All 144 computed errors pass independent phase-shift, grid and off-grid checks and respect Eq.~\eqref{eq:dot-bound}; none exceeds it, and the check allowed a numerical margin of $2\times10^{-4}$ in the relative error.
At resonance the equal-contact three- and four-terminal errors are
37.8 and 43.6 percent at $T=0$, reduced to
14.6 and 15.1 percent at the largest temperature.
Independent Fock-space thermodynamics, derivatives and asymptotics validate
the model implementation. These numerical checks test the implementation independently of the analytic proof.

\subsection{Planar barrier, gate and temperature controls}
A controlled matrix of 108 planar systems in the energy-independent scattering model varies
the contact barrier, a normal-region gate or the temperature, one at a time, in square and wide junctions with
three and four terminals, clean or disordered (Appendix~\ref{app:planar-controls}). The pairwise terms keep at
least 96.49 percent of the energy variation, while the current error reaches
32.5 percent. The error varies nonmonotonically with the barrier. In the clean square four-terminal
device it takes the values 15.53, 1.17, 2.22, 32.47 and 0.00017 percent at five increasing barrier heights, and a dense
gate scan resolves narrow features that reach 38.8 percent. Linear predictors fixed before scoring, from a
constant baseline to normal-transport and sampled-spectrum features, did not transfer to held-out geometries
(Appendix~\ref{app:planar-controls}).

\subsection{A planar resonance mechanism}
\label{sec:planar-resonance}
The nonmonotonic barrier dependence and the narrow gate features are compatible
with resonant scattering and motivate a test based on the normal spectrum. Resonant Josephson transport and the recovery of pairwise
couplings by far-detuned cotunneling are established mechanisms
\cite{beenakker1992resonant,matutecanadas2024circuits}; the question here is
whether parameters extracted from a planar lattice predict the error of its
best arbitrary-harmonic pairwise current model.

At zero field, write $H_0=H_N+\operatorname{Re}\Sigma_N(0)$ and
$WW^T=-2\operatorname{Im}\Sigma_N(0)$, using a real lead-local channel basis.
In the eigenbasis of $H_0$, the reaction matrix and unitary scattering matrix are
\begin{equation}
 K(E)=\tfrac12 W^T(E-H_0)^{-1}W,\qquad
 S=(1-iK)(1+iK)^{-1}.
 \label{eq:resonance-reaction}
\end{equation}
This representation reproduces the normal scattering calculation at $E=0$.
The finite-gap calculations below retain the energy dependence of the leads and normal region.
For one isolated mode, let $\gamma_i$ be its projected normal partial width,
$\gamma=\sum_i\gamma_i$, $p_i=\gamma_i/\gamma$, and $x=2\epsilon/\gamma$.
We call the mode resonant when $|x|\lesssim1$, that is, when its energy lies within its half-width of the
Fermi level, and shared when at least three terminal weights $p_i$ are substantial (in the selection below, $p_i\ge0.1$).
Throughout, the Hermitian-mode center means $x=0$, the zero of the selected eigenvalue of $H_0$. The open-system pole is the complex pole of $S$, and the sampled error maximum is the largest superconducting pairwise error found in a gate scan. These three references need not coincide.
At zero temperature, the short-junction phase energy, up to a constant, is
\begin{equation}
 \frac{F}{\Delta}=-\frac{\sqrt{x^2+|\sum_i p_i e^{i\theta_i}|^2}}
 {\sqrt{1+x^2}}.
 \label{eq:resonance-energy}
\end{equation}
Appendix~\ref{app:isolated-resonance} derives this energy from the rank-one scattering matrix. At zero temperature,
the infinite-gap level of Proposition~\ref{prop:dot} with $\Gamma_i=\gamma_i/2$ has the same phase dependence, so the
normalized current error of an isolated resonance is the same in the short-junction and infinite-gap limits, and
Eq.~\eqref{eq:dot-sufficient} bounds it with $|\epsilon_d|/\Gamma_\Sigma$ replaced by $|x|$.
Far detuning makes the square root approximately linear in the pairwise
quantity $|\sum_i p_i e^{i\theta_i}|^2$. Near resonance its nonlinear phase
dependence generates irreducible currents when several terminal weights are
appreciable. This mechanism is controlled by the normal spectrum.

We fixed eight further devices, reserved for this test, before evaluating their
currents. They are $140\times100$ and $160\times120~\mathrm{nm}^2$ rectangles with $n=3,4$, clean or
with one $U_0=10~\mathrm{meV}$ disorder pattern per geometry,
$a=5$~nm, $40$-nm contacts, $\mu=20$~meV and a $200$-meV boundary barrier.
Normal data alone select the nearest nondegenerate mode with
$|\epsilon|\le3$~meV, $\gamma\ge10^{-5}$~meV,
width/nearest-level spacing $\le0.15$, and at least
three $p_i\ge0.1$. All eight qualify. A gate potential applied uniformly to the whole normal region takes the selected mode through
$x=-4,-1,0,1,4$; the full calculation shifts every normal mode and recomputes $S$ at each gate. The prediction moves
only the selected mode, a rank-one change of $K$, %
and holds the background reaction matrix fixed at the Hermitian-mode center $x=0$, without fitting a current.
Agreement at the center holds by construction; nonzero detunings test the prediction.
Table~\ref{tab:evidence-status} distinguishes the original intervention from later analyses on the reused devices.
Three criteria were fixed before any current was evaluated. At every gate, the predictor's full-current
waveform error must be at most 10 percent and its pairwise error within 3 percentage points of the computed one,
and the largest pairwise error at $|x|\le1$ must exceed that at $|x|=4$ by at least 5 percentage points
(Table~\ref{tab:mechanism-criteria}).

\begin{figure*}[t]
 \centering
 \includegraphics[width=\textwidth]{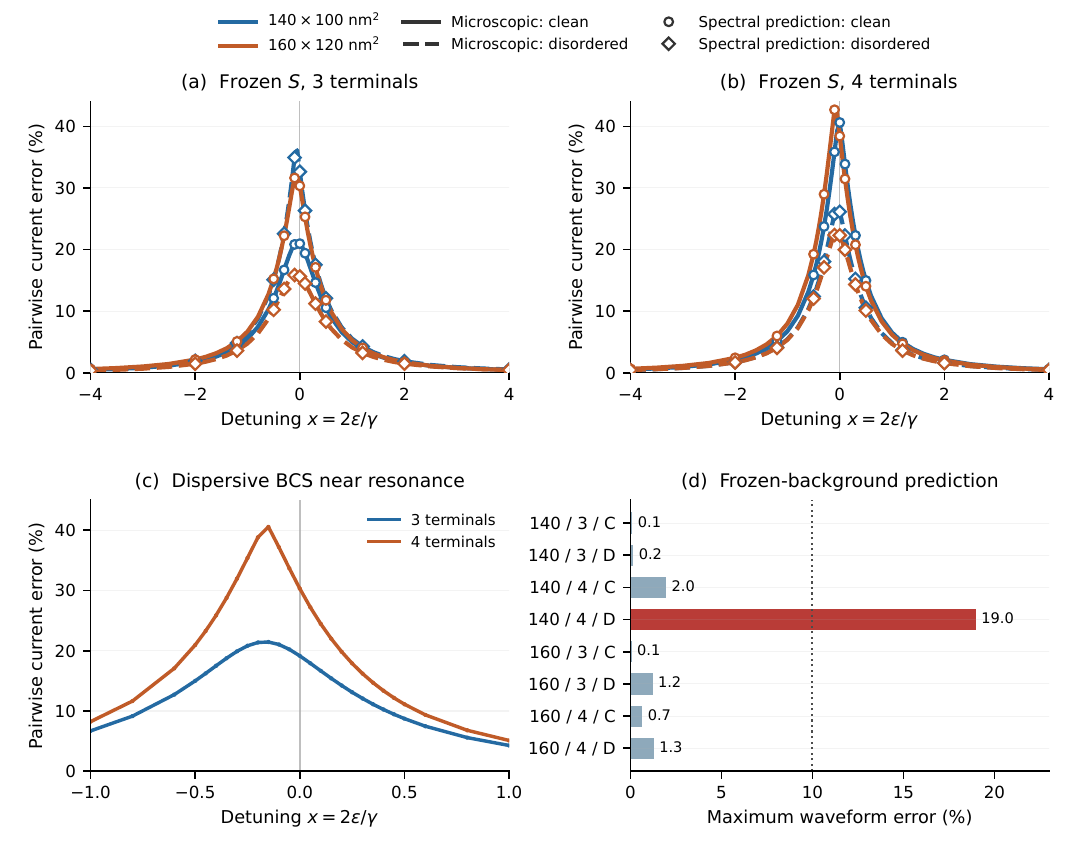}
 \caption{Resonance control of the pairwise error at $T=0$.
(a,b) All-terminal pairwise-current errors $\epsilon_I$ on 45 calculated gates for each of eight planar devices in the frozen-$S$ model. Thick solid and dashed lines show full microscopic calculations for clean and disordered devices, respectively. Open circles and diamonds show the corresponding frozen-background spectral predictions at a displayed subset of gates; matching colors identify the same geometry. The original predictive tests used $x=-4,-1,0,1,4$; the 45-gate scans refine those same devices retrospectively.
 (c) Near-resonance view of the two clean devices with dispersive BCS leads;
 all points use the microscopic finite-gap current calculation.
 The full scan extends to $x=\pm4$; the near-center spacing is $0.05$.
 The vertical line marks the zero of the selected Hermitian normal mode. Connecting segments add no data.
 (d) Maximum full-current-waveform error of the frozen-background predictor
 over the 45 gates in each frozen-$S$ device. Labels give rectangle length,
 terminal count and clean (C) or disordered (D) status. Red marks the device
 outside the prespecified 10-percent tolerance (dotted line).
 Numerical projection and phase-sampling sensitivities meet a
 0.2-percentage-point resolution target; maxima are sampled values.
 For experimental comparison, the approximately $12~\mu$eV resonance width requires temperatures well below
 $70$~mK in the zero-temperature regime; a one-percent residual on the $0.10$--$0.17$~nA detuned currents is
 $1$--$1.7$~pA. Experimental resolution must be assessed at these temperature and current scales.}
 \label{fig:planar-resonance}
\end{figure*}

On the prespecified five-gate grid in the frozen-$S$ model, resonant errors span 15.6--40.8 percent,
whereas both $|x|=4$ endpoints lie below 0.7 percent in every device
[Fig.~\ref{fig:planar-resonance}(a,b)]. Predicted pairwise errors agree within
0.02 percentage points, smaller than the maximum numerical sensitivity
of 0.19 percentage points. Seven devices also meet the predeclared
10-percent full-current-waveform tolerance; the disordered $140\times100$-nm
four-terminal device reaches 19.0 percent. The scalar-error and waveform criteria thus resolve distinct aspects of the prediction.
A retrospective 45-gate extension resolves the narrow peak between the five prespecified points. The near-center spacing is $\delta x=0.05$.
In the frozen-$S$ calculations the sampled maxima range from 15.9 to
42.7 percent, with peak locations between $x=-0.10$ and $0$.
The frozen-background predictor reproduces the resolved scalar errors within
0.012 percentage points and leaves the same single device outside the waveform tolerance on the denser grid.
A gate confined to a central patch of the normal region (27 percent of its sites) also changes the mode
shape; with gates chosen from first-order perturbation theory, the same predictor meets all three criteria in
7 of the eight devices, and the disordered four-terminal device again misses only the waveform criterion
(Appendix~\ref{app:local-gate}).
Both clean finite-gap local-gate calculations also meet the three criteria. The independent frequency and
channel follow-up meets the unchanged numerical criteria for both models at all five gates
(Table~\ref{tab:finitegap-refinement} and Appendix~\ref{sec:finitegap-refinement}). In a follow-up comparison,
the relative operator-norm change of the background reaction matrix, defined in Appendix~\ref{app:local-gate},
reaches 20 to 27 percent in that disordered device and at most 6 percent in the others across the uniform and local
gate sweeps. Its waveform failure is consistent with this variation of the background.

The difference between the Hermitian-mode center and the open-system pole can be made explicit.
For $K(E)=K_b+vv^T/[2(E-\epsilon)]$, define
$a-ib=v^T(1+iK_b)^{-1}v$, where $K_b$ is the frozen real background.
The poles of $S$ are the zeros of $\det[1+iK(E)]$. Since $1+iK_b$ is invertible for real symmetric $K_b$,
the matrix determinant lemma, ${\det(A+uw^T)=(1+w^TA^{-1}u)\det A}$, gives
\begin{equation*}
 \det[1+iK(E)]=\det(1+iK_b)\Bigl[1+\frac{i(a-ib)}{2(E-\epsilon)}\Bigr],
\end{equation*}
which vanishes at
\begin{equation}
 E_{\rm pole}=\epsilon-b/2-ia/2.
 \label{eq:background-dressed-pole}
\end{equation}
Thus the real part of the open-system pole crosses the Fermi level at $x=b/\gamma$, and its width is $a$. In these eight devices $b/\gamma$ ranges approximately between
$-0.090$ and $-0.025$. %
The spectator background thus shifts the normal pole relative to the Hermitian-mode center $x=0$. The superconducting pairwise-error maximum also depends on phase-dependent interference and, for finite-gap leads, energy dispersion.

The broader, overlapping resonance of the $160$-meV-barrier square device in the control
matrix (Appendix~\ref{app:planar-controls}) falls outside this isolated-mode selection.

For these narrow modes, the absolute current scale requires a finite-gap calculation. We repeat the $140\times100$-nm three- and four-terminal gate
interventions, clean and disordered, with $\Delta=0.18$~meV, retaining the full normal lattice and
phase-dependent continuum through an imaginary-frequency determinant.
The ideal superconducting leads have the normal tight-binding dispersion and
uniform BCS pairing. Appendix~\ref{app:omitted-modes} gives the channel truncation and checks for the clean
uniform-gate calculations.
Freezing the spectator modes while moving only the selected normal mode constructs a gate-dependent current prediction without a Josephson fit. Its scalar pairwise-error prediction and its full-current-waveform prediction are tested separately.
This predictor retains all normal modes and evaluates phase-resolved currents. It replaces the
recomputation of the normal region at each gate by the motion of one mode, so its agreement tests whether that
mode controls the gate response; Appendix~\ref{app:omitted-modes} compares it with references that omit or freeze
the other modes.
At the Hermitian-mode center $x=0$, the pairwise errors are 19.1 and 30.3 percent
for $n=3,4$ [Fig.~\ref{fig:planar-resonance}(c)]. Both $|x|=4$ endpoints
in the full scan lie below 0.6 percent. Both devices meet the prespecified waveform,
pairwise-error and gate-contrast criteria. The maximum current-waveform
prediction error on that five-gate grid is 2.4 percent. Replacing the dispersive leads by
wide-band reservoirs instead changes selected current waveforms by up to 14.0 percent.
The two disordered devices were added later, with a plan and unchanged criteria fixed before their finite-gap
currents were computed. A later independent follow-up checks their 64-node frequency quadrature and four
anomalous channels per terminal against 128/256 nodes and all nine channels, for both models at every gate.
All cases meet the unchanged numerical criteria (Table~\ref{tab:finitegap-refinement}). Their $x=0$ errors are 27.2 and 20.9 percent for $n=3,4$, compared with 32.6 and 26.3 percent in
the frozen-$S$ model, and their $|x|=4$ endpoints do not exceed 0.56 percent. The three-terminal device meets
all three criteria. The four-terminal device meets the pairwise-error and gate-contrast criteria and, as in the
frozen-$S$ model, misses the waveform criterion, with a 16.2-percent error at $x=-4$
(Table~\ref{tab:mechanism-criteria}). At $x=0$, keeping four instead of all anomalous channels changes their currents
by less than 0.02 percent on the original 64 sampled phases. The new all-gate checks use a separate
1,024-point Sobol set plus additional conditioning probes and separately quantify frequency and channel sensitivity.
In absolute terms ($2e\Delta/\hbar=87.6$~nA), the finite-gap RMS currents of the four
devices are 0.72--1.20~nA at $x=0$ and 0.10--0.17~nA at $|x|=4$, whereas the frozen-$S$ model gives
22--31~nA at $x=0$. The mode width, about $12~\mu$eV, limits the current, %
while the normalized errors of the two models remain comparable. For an ideal isolated mode at zero temperature, a common current prefactor cancels in that ratio in the short-junction and infinite-gap limits (Appendix~\ref{app:isolated-resonance}). The comparison separates normalized phase dependence from absolute current scale. Extending it to finite-gap multilevel junctions requires treating mode-dependent dispersion and background interference.
These finite planar calculations link the gate dependence of the pairwise error to the selected normal resonance. A spatial-refinement follow-up on the clean
three-terminal example closes its second lead channel and removes the selected mode's bottom-terminal coupling,
eliminating the large nonpairwise response on the five sampled gates (Appendix~\ref{sec:resonance-lattice}). The shared-mode condition must
therefore be checked again when the spatial model or contact spectrum changes.

Across the eight selected devices, the scalar normalized pairwise-error prediction meets its criterion in every case, and seven meet the 10-percent waveform criterion in the energy-independent model. The same disordered four-terminal device exceeds the waveform tolerance with dispersive leads. The selection thresholds and error tolerances define numerical comparison criteria; experimental tolerances depend on the intended measurement or circuit.

A phase-independent part of the finite-gap shift is directly calculable.
Let $u$ be the selected eigenmode of $H_0$ and let
$\Sigma_{{\rm BCS},ee}(0)$ denote the electron block of the superconducting
lead self-energy at zero imaginary frequency. Its real part differs from
that of the normal lead. The first-order level shift is
\begin{equation}
 \delta\epsilon=u^T[\operatorname{Re}\Sigma_{{\rm BCS},ee}(0)
 -\operatorname{Re}\Sigma_N(0)]u.
 \label{eq:dispersive-level-shift}
\end{equation}
For $n=3,4$ this gives $0.747$ and $0.836~\mu$eV, respectively,
or recentering shifts $\delta x=-2\delta\epsilon/\gamma=-0.125,-0.149$.
Direct diagonalization of this shifted normal block changes either estimate
by less than $0.0006$ in $x$; the correction tends to zero as $\Delta\to0$.
A sub-$\mu$eV correction matters here because the normal widths are only
$11$--$12~\mu$eV. This normal-block diagnostic identifies a lead-dispersion contribution
on the scale of the observed displacement. The anomalous block contributes separately to the full superconducting response.

The resolved dispersive-lead scan locates sampled maxima of
21.5 and 40.6 percent at
$x=-0.15$ for both $n=3,4$
[Fig.~\ref{fig:planar-resonance}(c)]. These exceed the errors at the
Hermitian-mode center quoted above, and are close to the corresponding frozen-$S$
peaks (21.1 and 40.6 percent). The gate dependence therefore matters even within the narrow resonance region.
Resolving it requires the denser scan. This extension is a retrospective refinement
on the same devices, with maxima defined over the sampled gates.

\begin{table*}[t]
\caption{Fixed finite-gap numerical refinements for the disordered uniform-gate and clean local-gate tests.
Each row covers full and predictor models at all five gates, ten logical cases.
$\delta_\omega$ compares 64 with 256 frequency nodes, taking the larger change for four or nine anomalous
channels; $\delta_{\rm ch}$ compares four with all nine channels at 256 nodes. RMS values use a separate
1,024-point Sobol set; maxima use the additional conditioning probes. Percent columns are current-waveform changes,
normalized by the corresponding full-target RMS current. $\Delta I$ is the combined RMS change from 64/four
to 256/nine in pA. Entries are observed maxima over the row, which can occur at different gates.
They are numerical sensitivities, without a statistical confidence or phase-uniform error-bound interpretation.
Appendix~\ref{sec:finitegap-refinement} states the unchanged acceptance criteria and independent checks.}
\label{tab:finitegap-refinement}
\begin{ruledtabular}
\begin{tabular}{lccccc}
Device/gate & $\delta_\omega^{\rm RMS}$ (\%) & $\delta_\omega^{\max}$ (\%) & $\delta_{\rm ch}^{\rm RMS}$ (\%) & $\delta_{\rm ch}^{\max}$ (\%) & $\Delta I$ (pA)\\
Disordered, uniform, $n=3$ & 0.0004 & 0.0006 & 0.0247 & 0.0291 & 0.1396 \\
Disordered, uniform, $n=4$ & 0.0021 & 0.0026 & 0.0318 & 0.0513 & 0.1533 \\
Clean, local, $n=3$ & 0.0030 & 0.0010 & 0.0341 & 0.0469 & 0.1085 \\
Clean, local, $n=4$ & 0.0037 & 0.0021 & 0.0370 & 0.0541 & 0.2038 \\
\end{tabular}
\end{ruledtabular}
\end{table*}

The distinction between frozen and omitted spectators can be quantified
by expanding their contact-resolvent contribution to the equilibrium current
(Appendix~\ref{app:omitted-modes}). In the two clean finite-gap devices, freezing the spectator energies gives
full-current waveform errors of at most 2.4 percent over the five gates, whereas keeping only the mode nearest
zero energy gives up to 28 percent (Table~\ref{tab:omitted-mode-response}). A second-order correction that uses no
Josephson data reduces these errors to below 0.001 and about 1 percent, respectively, while the first-order
correction can increase them. The expansion thus identifies, device by device, which approximation the resonance picture requires and how to correct it.

\section{Relation to prior work and experimental outlook}
\label{sec:discussion}

The resonance intervention connects established multipair physics to a quantitative test of pairwise approximation error. A mode selected from the normal spectrum reproduces the gate dependence of that error in eight planar devices, including the two clean and two disordered devices checked with dispersive leads. The distinction between scalar error and full-current prediction makes the role of the background explicit: seven devices satisfy the waveform tolerance, and the same disordered four-terminal exception appears in both models. %
The analytic single-level bound and the planar intervention serve complementary roles. The bound supplies a sufficient detuning in its stated model; the intervention retains the device background and evaluates phase-resolved currents to test the mechanism in a multilevel junction. The contour and circuit comparisons below connect these results to experimental observables and alternative reduced descriptions.

The contour diagnostics developed here are complementary to the main existing approaches. Andreev-band tomography
\cite{coraiola2023phase,coraiola2024spin,nichele2025} targets the resolved bound-state spectrum and needs a tunnel probe
and phase loops. The equilibrium critical-current contour describes the existence of stable zero-voltage states; measured switching
boundaries additionally depend on circuit dynamics, branch accessibility and heating. %
Thermal escape depends on barrier heights and can precede a deterministic stability boundary; Appendix~\ref{sec:ramp} relates these noiseless limits to fluctuation-sensitive switching. Bias sweeps in three- and
four-terminal experiments probe this related, protocol-dependent observable
\cite{pankratova2020,graziano2020transport,graziano2022selective,lee2022thesis,thieme2026hgte}. The Josephson diode
effect \cite{gupta2023diode,chirolli2025diode} compares opposite critical currents under specified controls.
It is not identical to whole-region inversion asymmetry. At zero field, time reversal implies
$E(\bm\theta)=E(-\bm\theta)$ and $\mathbf I(\bm\theta)=-\mathbf I(-\bm\theta)$, with stable states paired.
The complete equilibrium current region is therefore centrosymmetric without requiring a spatial mirror.
For a fixed nonzero transverse bias the relation is instead
$I_R^{\max}(I_B)=-I_R^{\min}(-I_B)$; it does not forbid diode response at the same fixed $I_B$.
The analogous qualification applies to a fixed third phase.
These distinctions also separate our equilibrium construction from switching and self-heating
\cite{kedves2024switching}. The measurements of Thieme et al.\ on HgTe T- and X-junctions \cite{thieme2026hgte}, which decompose the contour into the
contributions of the individual junctions of the network, resolve a field-driven crossover between SQUID-like and
Fraunhofer-like interference and reach diode efficiencies of $0.8$, are the closest experimental counterpart of the
present model. Their network description and our pairwise projection address complementary questions: the former fits circuit transport, while the latter measures the microscopic energy's departure from pairwise additivity. Their resistively-shunted-junction simulations additionally describe switching dynamics beyond the equilibrium stability boundary.

K\"olzer et al.\ describe Bi$_4$Te$_3$/Nb three-terminal transport using two Josephson branches coupled through a resistor~\cite{kolzer2023}. Behner et al.\ interpret current maps and a geometry-induced diode effect in a related topological-insulator device using a triangular network of three two-terminal JJs, each described by a resistively and capacitively shunted junction (RCSJ) model~\cite{behner2025}. Teller et al.\ describe an array of Nb--Pt--Nb four-terminal junctions using four two-terminal JJs per cell, with alternating plaquette frustrations~\cite{teller2025}. These circuit descriptions provide experimental precedents for reduced MTJJ networks; resolving their microscopic irreducible Fourier content requires phase-resolved energy or current information.

Studies of topological invariants
\cite{riwar2016multi,eriksson2017topological,meyer2017nontrivial,xie2017topological,ohnmacht2025reflectionless,
ram2025tunable,barakov2023abundance} compute Chern numbers and Weyl-point densities for gapped, generic junctions,
and superconducting circuits now show spectral signatures of such topology \cite{peyruchat2024spectral}.
For four-terminal junctions, Xie et al.\ explicitly connect Weyl nodes and Chern numbers to quantized nonlocal conductance and nonanalytic current--phase relations~\cite{xievavilov2018}.
The
clean device studied here has no established positive lower gap bound on its torus, whereas Appendix~\ref{sec:disorder}
gives a disordered example with one. A continuum-inclusive \cite{repin2019topological} or full
Bogoliubov--de Gennes treatment of such a device is where those invariants and the contour could be
compared in one model. In the language of phase
transitions, the gate- and geometry-driven evolution of the contour is a crossover with no order parameter; phase windows of nonzero Chern number, Weyl points and parity boundaries
(Appendix~\ref{sec:topology}) concern changes in the Andreev band structure.
Jung et al.\ observed phase-dependent gapped and gapless Andreev spectra in a graphene three-terminal
junction, interpreted in terms of regions with different winding numbers \cite{jung2025tomography};
relating these regions to Chern numbers requires the gap and the Andreev states over the full phase torus. Parity switches have been observed in Al/InAs
\cite{coraiola2024spin}; interactions are predicted to
reshape the gapless region near special phase configurations \cite{erdmanis2022special}.

Several established microscopic mechanisms generate departures from pairwise additivity. In the resonant double-dot bijunction, weak interdot coupling lifts
Andreev degeneracies and creates narrow, large inverse-inductance responses,
which are thermally smoothed when temperature exceeds the splitting
\cite{feinberg2015equilibrium}. Those local derivative features are distinct
from our phase-averaged current norm. Andreev-molecule theory connects spatial
overlap to nonlocal current--phase relations \cite{pillet2019molecules};
the corresponding experimental phase shift decreases with junction separation
and vanishes well beyond the superconducting coherence length
\cite{haxell2023nonlocal}. Transferring this separation criterion to a common normal region requires accounting for its distinct propagation geometry. Quartet
tomography \cite{ohnmacht2024quartet} and four-terminal sextet calculations
\cite{ebert2025sextets} also show why individual spectral branches, coupling
strength or transparency cannot be interpreted as a universal monotonic
measure of the total nonpairwise response. Our Fourier error is an
a posteriori diagnostic; a predictive criterion must be tested independently
of the phase-resolved data used to measure its error.

Pairwise additivity also differs from the absence of frustration. Competing pair couplings can prefer incompatible phase differences and produce nontrivial minima or coupled phase dynamics while the energy remains pairwise in all retained phases. Frustration-driven dynamics~\cite{r2prb105134503} and phase locking in an ac-driven multiterminal Josephson-junction analog~\cite{r2prb106174509} illustrate this circuit setting at the level of phase dynamics, distinct from the Fourier classification of the static external-phase energy. Conversely, eliminating an internal superconducting phase can generate such harmonics even from pairwise circuit elements, as Eq.~\eqref{eq:classical-island} shows. Pairwise additivity, frustrated phase configurations and a microscopic multipair process are consequently distinct statements.

Higher mixed phase derivatives have also been proposed to isolate
contributions involving several terminals \cite{melin2021manyleads}.
A nonzero appropriate derivative is a useful witness, whereas a zero at one
phase point does not establish the absence of all multiterminal irreducible harmonics.
Matrix circuit theory of Andreev reflection~\cite{nazarov1999circuit} retains coherent proximity effects through matrix variables, extending beyond a scalar pairwise-energy description. Diffusive multiterminal proximity-gap calculations~\cite{amundsen2017diffusive} provide a complementary controlled regime whose gap boundaries and pairwise-current errors are separate observables.

The model suggests a three-terminal experimental protocol in four steps. First, a circuit with two independent flux controls,
with calibrated reference-junction response and loop inductances, maps the imposed fluxes onto the junction phases
\cite{ohnmacht2024quartet,prosko2024flux}. Second, both independent currents measured over the phase torus give the
Fourier sectors and the current error of Eq.~\eqref{eq:nt-current-error} directly. Because differentiation weights
energy harmonics by their phase indices, the required sensitivity follows from the measured current spectrum; an
energy-power share does not set it. %
Third, gate-dependent normal-state linear-response transport can identify a candidate shared resonance. With
the other reservoir voltages held fixed, the magnitude of the off-diagonal conductance-matrix element near an isolated
resonance is approximately $(2e^2/h)\,4p_ip_j/(1+x^2)$ for $i\ne j$
(Appendix~\ref{app:isolated-resonance}). %
A correlated peak in conductance between several terminal pairs, its gate width and a calibrated gate-to-energy conversion can estimate partial weights and linewidth within this isolated-resonance form. The quantitative current predictor additionally requires channel-resolved amplitudes, the background reaction matrix and superconducting self-energies, including an assessment of background interference and overlapping resonances. %
Fourth, a low-temperature gate sweep would test the calculated detuning trend: in the selected devices the pairwise current error reaches tens of percent within about one half-width of the Hermitian-mode zero and falls below about one percent at $|x|\approx4$. The sampled current-error maximum is displaced from the Hermitian-mode zero at $x=0$; both are defined separately from the open-system pole, so conductance and current-error peaks should be compared separately (Sec.~\ref{sec:planar-resonance}). In the finite-gap devices studied here, whose selected modes are about
$12~\mu$eV wide, comparison with the zero-temperature predictions requires $k_BT\ll\gamma/2$, corresponding to
temperatures well below about $70$~mK. The near-resonant RMS currents are about one nanoampere; at the
$0.10$--$0.17$~nA detuned currents, a one-percent residual is only $1$--$1.7$~pA. %
The omitted-current signal has all-terminal RMS equal to $\epsilon_I$ times the full-current RMS in the same norm. Resolving a one-percent residual at the quoted detuned current scale therefore requires sensitivity to $1$--$1.7$~pA. Assessing detectability and acceptable error also requires the noise level, phase-space coverage, branch accessibility and measurement protocol. Comparing square and wide--short devices with controlled disorder would test the geometry dependence by measuring harmonic content alongside contour shape. The nonzero-flux calculations begin at $0.10\Phi_0$ or $0.16\Phi_0$, depending on the family; resolving a lower-field experimental onset requires additional sampling.

The surveys retain a spin-degenerate lattice model with energy-independent scattering on the gap scale, although
their dwell times reach a sizable fraction of $\hbar/\Delta$ (median $\Delta\tau_{\max}/\hbar$ of 0.77,
Appendix~\ref{app:disorder-ensemble}). Finite-gap calculations for twelve selected ensemble devices and two
eight-terminal devices give a largest observed change of about 4.3 percentage points in the normalized pairwise errors and reduce the
currents by 20--39 percent in the ensemble samples.
For the narrow resonances of Sec.~\ref{sec:planar-resonance}, the short-junction model overestimates the current
scale by more than an order of magnitude while the normalized errors remain comparable. %
For an ideal single isolated mode at zero temperature, the short-junction and infinite-gap energies share the same phase-dependent square root and differ by a phase-independent prefactor (Appendix~\ref{app:isolated-resonance}). Cancellation of this prefactor in the normalized error explains how similar relative errors accompany markedly different absolute currents. The selected finite-gap devices approximately follow this pattern. In an overlapping-resonance regime, mode-dependent dispersion and background interference can instead reweight the Fourier sectors, requiring a device-specific finite-gap calculation. The numerical surveys therefore describe the specified lattice geometries; connecting their disorder trends and contour shapes to a material device requires matching its contact spectrum and spatially converged current response. Testing whether shared resonances also organize the disorder-enhanced ensemble errors is a separate extension.

A physical realization of the shared-mode mechanism requires appreciable overlap with propagating modes in at least three contacts. Contact widths and carrier density must keep the relevant transverse channels open away from their thresholds, with mode parity compatible with the confined state. Spatial refinement of the selected device in Appendix~\ref{sec:resonance-lattice} closes the odd bottom-lead channel; the surviving even fundamental has zero overlap with the odd tracked state, and the large nonpairwise response disappears on the five sampled gates. This result identifies the contact-spectrum condition that a spatially robust design must satisfy. Contact placement or symmetry breaking offers routes to investigate, with coupling strength and persistence to be verified in a spatially converged device calculation.

Spin--orbit coupling can modify the interference, level separations and contact amplitudes of the spin-degenerate model used here. Spin-dependent scattering treatments of charge-to-spin conversion under confinement~\cite{r2prresearch7043100} and of ballistic Rashba quantum dots~\cite{r2scipost20144} provide starting points for a spin-resolved extension. Such a calculation would combine the normal scattering problem with consistent singlet Andreev reflection and reevaluate the terminal weights, isolated-mode approximation and current errors. These quantities can change even when time-reversal symmetry is preserved.

For an array of superconducting islands, the Fourier split supplies arbitrary-harmonic pairwise and irreducible
multi-island couplings. Gate control of a shared resonance changes these couplings. The validated current-error
measure also supplies a test problem for automated device design \cite{kaplan2025deep,Adler2019CorrelatedDesign};
geometry and gate searches must resolve currents as well as energies.
A possible extension would use calibrated multiterminal cell free energies to define a classical superconducting-island array and compare its equilibrium observables with a specified neural-field ensemble, including explicitly represented topological sectors appropriate to the boundary conditions~\cite{Ferko2026TopologicalNNFT}. The extension requires validation of cell additivity, the physical equilibrium regime and ensemble accuracy at the array level.

\section{Conclusion}
\label{sec:conclusion}

Shared normal-region resonances provide a control parameter for pairwise current accuracy in planar junctions. In eight prospectively selected devices, detuning a mode shifts the error of the best arbitrary-harmonic pairwise model from tens of percent to below one percent. A frozen-background construction predicts this change without fitting Josephson currents, for a local as well as a uniform gate. Finite-gap calculations reproduce the gate dependence in two clean and two disordered devices. %
The scalar error trend is reproduced in all eight devices, and seven also meet the full-current-waveform criterion. The same disordered four-terminal device exceeds that tolerance with dispersive leads, identifying a case where the background requires a more complete treatment.

The energy-to-current distinction follows directly from Fourier differentiation. Across the sixteen matched disorder groups, median energy RMS errors of 4.5--11.3 percent accompany median all-terminal current errors of 8.0--21.2 percent. Equation~\eqref{eq:weight-ratio} relates this amplification to the larger phase indices of the omitted harmonics, and designed devices extend the comparison through sixteen terminals. Contour shapes, stability boundaries and ramp-selected branches provide complementary tests of the current field (Appendices~\ref{sec:diagnostics}--\ref{sec:island-circuit}). %
For an ideal isolated mode at zero temperature, the short-junction and infinite-gap limits give the same normalized error, while their current amplitudes differ. Proposition~\ref{prop:dot} supplies a terminal-count-independent sufficient detuning for the single-level infinite-gap model; applying this guide to multilevel devices requires control of spectator currents and interference.

Physical implementation requires a shared mode that remains coupled to at least three propagating contacts under spatial refinement. The channel closure found in the refinement test makes this contact-spectrum requirement concrete; its satisfaction in a material device remains to be verified. Phase-resolved gate sweeps following Sec.~\ref{sec:discussion} would test the resonance mechanism and quantify the omitted-current signal. %
At the computed detuned current scale, a one-percent residual is $1$--$1.7$~pA, setting a sensitivity requirement to assess against the noise and measurement protocol. The 5-percent pairwise-error benchmark and 10-percent waveform tolerance are numerical comparison criteria; experimental performance specifications must be set for the intended circuit. We developed a way to test which circuit description a multiterminal junction needs, and a mechanism that could guide its gate tuning; applying it to fabricated devices is the next validation step.

\FloatBarrier
\begin{acknowledgments}
This research was supported in part by the National Science Foundation under Grant No. PHY2309135 to the Kavli Institute for Theoretical Physics (KITP).
The author thanks participants of the AI for Quantum Matter 2026 program for discussions at the Kavli Institute for
Theoretical Physics in Santa Barbara, California, United States.
This work originated in the author's doctoral research at the University of
Wisconsin--Madison, during which the author benefited from discussions with \mbox{J.-X.}~Zhu at Los Alamos National
Laboratory. The author thanks M.~G.~Vavilov for guidance on the theory, and
V.~E.~Manucharyan, N.~Pankratova, and H.~Lee for discussions of the multiterminal transport experiments. %
The author used the AI tools Get Physics Done (GPD) by PSI, OpenAI Codex and Claude Code to help write and debug analysis and figure scripts, examine numerical outputs, formulate numerical checks and edit the text. The author directed this work, checked the results and takes full responsibility for the content.
\end{acknowledgments}

\clearpage
\onecolumngrid
\begin{center}
\textbf{\large Contents of the Supplemental Material}
\end{center}
\vspace{2ex}
\noindent\begin{tabular}{@{}l p{0.72\textwidth} r@{}}
Appendix~\ref{sec:diagnostics} & The critical-current contour and its diagnostics & \pageref{sec:diagnostics}\\
Appendix~\ref{sec:pairwise-accuracy} & Current and region accuracy in three-terminal junctions & \pageref{sec:pairwise-accuracy}\\
Appendix~\ref{sec:baselines} & Zero-flux baselines and validation & \pageref{sec:baselines}\\
Appendix~\ref{sec:morphology} & critical-current contour morphology & \pageref{sec:morphology}\\
Appendix~\ref{sec:thirdlead} & Third-lead flux response & \pageref{sec:thirdlead}\\
Appendix~\ref{sec:inversion} & Inversion-symmetry breaking under flux & \pageref{sec:inversion}\\
Appendix~\ref{sec:beyond} & Beyond the clean, static contour & \pageref{sec:beyond}\\
Appendix~\ref{sec:island-circuit} & Direct pairwise couplings versus an internal island & \pageref{sec:island-circuit}\\
Appendix~\ref{sec:topology} & Separate topology diagnostics and lattice illustration & \pageref{sec:topology}\\
Appendix~\ref{app:disorder-ensemble} & Disorder studies & \pageref{app:disorder-ensemble}\\
Appendix~\ref{app:terminal-tables} & Terminal-count comparisons & \pageref{app:terminal-tables}\\
Appendix~\ref{app:planar-controls} & Planar controls and held-out predictor tests & \pageref{app:planar-controls}\\
Appendix~\ref{app:strengthening} & Additional convergence and mode controls & \pageref{app:strengthening}\\
Appendix~\ref{app:omitted-modes} & Current response to omitted normal modes & \pageref{app:omitted-modes}\\
Appendix~\ref{app:isolated-resonance} & Energy of an isolated short-junction resonance & \pageref{app:isolated-resonance}\\
Appendix~\ref{app:local-gate} & Local-gate test & \pageref{app:local-gate}\\
\end{tabular}
\clearpage
\begin{center}
\textbf{\large Supplemental Material}
\end{center}
\vspace{1ex}
\twocolumngrid

\appendix
\makeatletter\@addtoreset{figure}{section}\@addtoreset{table}{section}\makeatother
\renewcommand{\thefigure}{\Alph{section}.\arabic{figure}}
\renewcommand{\thetable}{\Alph{section}.\arabic{table}}

\section{The critical-current contour and its diagnostics}
\label{sec:diagnostics}
\begin{figure}[!t]
  \includegraphics[width=\columnwidth]{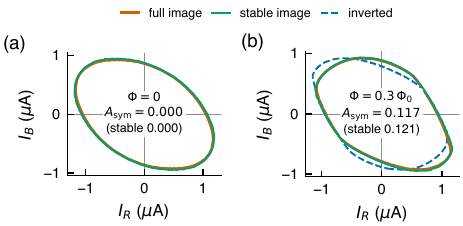}
  \caption{Critical-current contours of a square junction ($W=L=300$~nm, $\mu=90$~meV). (a) Zero flux. The boundaries
  of the image of the stable phase domain (green) and of the full phase-reachable image (orange) coincide with the
  origin-inverted contour (dashed), so $A_{\rm sym}=0$. (b) Flux $\Phi=0.3\,\Phi_0$. The contour and its inverted
  image separate. The labels give the inversion asymmetry $A_{\rm sym}$ of the full image and, in parentheses, of
  the stable image (Appendix~\ref{sec:diagnostics}).}
  \label{fig:schematic}
\end{figure}
\begin{figure}[!t]
  \includegraphics[width=\columnwidth]{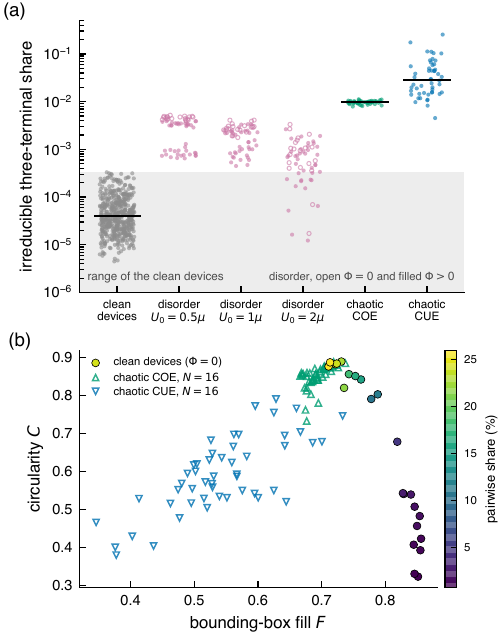}
  \caption{Three-terminal survey. (a) Irreducible three-terminal share of the Josephson energy (the Parseval share of the
  harmonics that depend on $\theta_R$ and $\theta_B$ other than through $\theta_R$, $\theta_B$ or $\theta_R-\theta_B$
  alone) for all 508 device--flux points of the clean rectangular model devices (33 geometries; the bar marks the median), for the square device
  with on-site disorder $U_0/\mu=0.5$, $1$ and $2$ (20 realizations each at $\Phi=0$, open, and at $0.32$ and
  $0.8\,\Phi_0$, filled; Appendix~\ref{sec:disorder}), and for chaotic cavities with $16$ channels per terminal (50 draws
  each; Appendix~\ref{sec:chaos}). The shaded band is the range of the clean devices. (b) Circularity versus
  bounding-box fill of the zero-flux contours of the clean devices, colored by the share of the direct $R$--$B$
  coupling, and of the chaotic cavities. The orthogonal-ensemble cavities fall on the clean square devices.}
  \label{fig:headline}
\end{figure}
Under current bias $\bm I=(I_R,I_B)$ a zero-voltage state is a local minimum of the tilted energy
$G(\bm\theta)=E_{\rm gs}(\bm\theta)-(\hbar/2e)\,\bm I\cdot\bm\theta$, that is, a phase point $\bm\theta$ with
$\bm I(\bm\theta)=\bm I$ and a positive-definite Hessian of $E_{\rm gs}$ \cite{pankratova2020,lee2022thesis}. The set of biases for
which such a state exists is the image of the stable phase domain $\{\bm\theta:\ {\rm Hess}\,E_{\rm gs}\succ0\}$ under
the current map, and we define the critical-current contour as its boundary [Fig.~\ref{fig:schematic}]. The full phase-reachable image (every $\bm\theta$ mapped to
$\bm I(\bm\theta)$), which was the definition used in the author's thesis \cite{ozguler2020thesis},
can be larger. At a regular fold of the map $\bm\theta\mapsto\bm I$ the Hessian is singular, its null vector corresponding
to the normal of the image boundary, and the sign of the other eigenvalue decides whether that point is reached by a
stable branch.

We compute both images on periodic phase grids of $41$ and $81$ points per axis (the Hessian from central differences of the currents,
with every ambiguous point re-evaluated with a local step of $10^{-3}$~rad). The full-image boundary is the alpha shape
(concave hull) of the image cloud; the stable and full images are compared through the maximal current in each of
$360$ bias directions, a sampled angular-neighborhood extent (with a $\pm2^\circ$ running maximum),
computed from both clouds in the same way, and through their support functions; both are extent measures. The stable
image, including its holes and separate pieces, is approximated by the union of the images of the stable grid cells.

Contour boundaries, polygon unions, areas and related geometric operations are evaluated with Shapely~\cite{Shapely_Python_Package}. Kwant~\cite{groth2014kwant} supplies the normal-state scattering calculation described in Sec.~\ref{sec:methods}.

This reconstruction covers a median 97 percent of the envelope area up to $0.32\,\Phi_0$ and
94 percent at $0.8\,\Phi_0$, where it has up to 5 pieces. At and above $1.3\,\Phi_0$ it
fragments into up to 19 pieces covering 62 percent of it, so the envelope then
overstates the stable region. The stable domain covers a median
fraction 0.71 of the torus at fluxes up to $0.32\,\Phi_0$ and 0.53 at a flux quantum and
above, and it is not a single component. The component containing the zero-bias minimum covers 0.62 and
0.35; which part of it a current ramp reaches is answered by the adiabatic continuation of Appendix~\ref{sec:ramp}. Up to $0.32\,\Phi_0$ the stable image reaches the
full image to within 2.0 percent in every direction (50 device--flux points), and the two inversion
asymmetries differ by at most 0.006. At $0.8\,\Phi_0$ the stable image falls short by a median
3.9 percent and up to 18 percent, the component containing the minimum by up to 22 percent. For
the square device at and above a flux quantum the stable image is smaller by up to 21 percent and the
component containing the minimum by up to 61 percent. The low-flux agreement reflects current--phase relations close to
the sawtooth of a ballistic channel, whose energy is convex almost everywhere; the high-flux shrinkage is where the
stability definition matters, and it is reported alongside the full-image curves below. For the square device at
$0.3\,\Phi_0$ the extent deficit is 0.8 percent at $81$ and 0.3 percent at
$161$ points per axis. Switching by thermal or quantum activation is outside the model; the adiabatic, noiseless
ramp is treated in Appendix~\ref{sec:ramp}.

From the contour we extract three dimensionless scalars. (i) The circularity $C=4\pi\,\mathrm{Area}/\mathrm{Perimeter}^2$
($C=1$ for a circle). (ii) The bounding-box fill fraction $F=\mathrm{Area}/[(\max I_R-\min I_R)(\max I_B-\min I_B)]$,
which equals 1 for a product set (two currents that can be set independently) and $\pi/4$ for an axis-aligned
ellipse, and is lower when the contour is tilted because $I_R$ and $I_B$ are correlated. (iii) The current-inversion asymmetry
$A_{\rm sym}=|P\,\triangle\,(-P)|/|P\cup(-P)|\in[0,1]$, the area of the symmetric difference between a specified current region $P$
and its origin-inverted image normalized by their union [Fig.~\ref{fig:schematic}(b)], together with the normalized centroid offset
$d_c=|\mathrm{centroid}(P)|/\sqrt{\mathrm{Area}/\pi}$, which responds to a rigid shift and not to a centrosymmetric
distortion. The resolution diagnostic for each metric is its own grid residual, the change between the $41$- and $81$-point grids,
reported in Appendix~\ref{sec:baselines}. A two-grid residual is not a certified discretization-error bound. The stable-versus-full comparison above uses 86 grids,
50 of them at low flux, and is descriptive. Its unshifted grids can meet nondifferentiable gapless
lines, and neither their finite Hessian assignments nor their angular neighborhoods establish uniform accuracy.
The quantitative pairwise current and region comparisons instead use the independently refined data
of Appendix~\ref{sec:pairwise-accuracy}.

To quantify the coupling between the two currents, we decompose the Josephson energy on the periodic grid into
harmonics, $E_{\rm gs}(\theta_R,\theta_B)=\sum_{m,n}e_{mn}\cos(m\theta_R+n\theta_B+\varphi_{mn})$, and measure the share of the energy variation in each of three groups of harmonics, from the
squared Fourier amplitudes (Parseval's theorem). The separable harmonics ($m=0$ or $n=0$) describe two
independent current paths. The pairwise harmonics ($m+n=0$, $m\neq0$) form a direct Josephson coupling
$\cos k(\theta_R-\theta_B)$ between the right and bottom terminals. The remaining harmonics couple all three
terminals, and we call them multiterminal irreducible harmonics; Fig.~\ref{fig:headline}(a) shows their share across the survey. The image of a separable energy is a product set,
while tilt and fill provide geometric signatures to compare with the shares of the harmonics that depend on both phases [Fig.~\ref{fig:headline}(b)].
With $\theta_L=0$, the lowest quartet phase combinations include $\theta_R+\theta_B$,
$2\theta_R-\theta_B$ and $2\theta_B-\theta_R$ \cite{ohnmacht2024quartet}. All belong to the multiterminal irreducible harmonic sector, which also includes higher harmonics and is therefore broader than a single quartet amplitude.

\section{Current and region accuracy in three-terminal junctions}
\label{sec:pairwise-accuracy}
Small energy-power residuals do not bound derivative observables. We make quantitative clean-current accuracy
claims only for two zero-field controls with independent phase-grid refinements. The square device
($W=L=300$~nm, $\mu=90$~meV) has $\epsilon_I=1.88\%$, and the wide--short device
($W=600$~nm, $L=100$~nm, $\mu=60$~meV) has $\epsilon_I=0.88\%$, using the centered
two-current RMS norm
\begin{equation}
 \epsilon_I=\frac{\langle\lVert\mathbf I-\mathbf I_{\rm pw}\rVert^2\rangle_\theta^{1/2}}
 {\langle\lVert\mathbf I-\langle\mathbf I\rangle_\theta\rVert^2\rangle_\theta^{1/2}},
 \qquad \mathbf I=(I_R,I_B).
 \label{eq:disorder-current-error}
\end{equation}
 We differentiate the singular-value-decomposition (SVD) energy and evaluate
$81^2$, $161^2$ and $321^2$ samples at
$\theta_j=-\pi+2\pi(k+q_j)/n$, with fixed unequal offsets
$q_R=(\sqrt5-1)/2$ and $q_B=\sqrt2-1$. No sampled normalized gap is below $10^{-8}$.
Each successive RMS change must be at most $\max(0.002,0.05\epsilon_I)$; both controls pass.
An independent Fourier projection reproduces the real-space marginal projection, and independent finite differences
of the energy at 48 random phases per device check the current derivatives and current conservation.
These tests establish finite-grid robustness.

For a direct region comparison, each model has its own sampled positive-Hessian domain.
The union of mapped phase-cell triangles retains holes and disconnected components. We measure
\begin{equation}
 \begin{aligned}
 d_H&=\frac{d_{\rm Haus}(\partial P,\partial P_{\rm pw})}{I_*},\qquad
 d_A=\frac{|P\triangle P_{\rm pw}|}{|P|},\\
 I_*&=\max_{\rm grid}\lVert\mathbf I\rVert .
 \end{aligned}
 \label{eq:region-distance}
\end{equation}
Here $P$ is the full or stable current region. The pairwise current uses
own-phase marginals and an antisymmetric phase-difference marginal, with its constant current removed
to respect periodic energy. Table~\ref{tab:pairwise-regions} gives the two clean controls and four
disordered realizations, selected by their energy remainder as stringent cases, at zero flux. The disorder examples are individual selected cases.
For the two finest clean grids, enforcing this joint constraint changes the component-wise projected current
by at most 0.15 percent of the full-current RMS, below the 0.2-percent quadrature tolerance.
For each individual image, both successive phase-grid comparisons must satisfy $d_H\le0.02$ and $d_A\le0.05$;
changing the cell diagonal and the second- versus fourth-order Hessian must meet the same thresholds.
All six pass these numerical-refinement checks, although the physical full--pairwise differences
can exceed those thresholds. To assess resolution of that difference, we also remove the pairwise zero mode
on every grid and report the achieved changes in Appendix~\ref{app:strengthening}.
The largest final individual boundary and area changes are 0.237 and
0.174 percent; each is below one quarter of the corresponding model discrepancy
in its case. The full--pairwise differences themselves change by at most
0.148 and 0.046 percentage points
on the last refinement. These comparisons resolve the approximate size of the effects; the last printed
decimal is not a certified continuum precision. Polygon precision is independently varied from $10^{-10}I_*$ to $10^{-9}I_*$,
requiring a relative induced area change below $10^{-8}$.
The piecewise-linear construction neither resolves a jump inside a cell nor certifies its interior stability.
Thus the table supports selected finite-resolution region comparisons.

\begin{table}[t]
\caption{Selected zero-flux full--pairwise comparisons. Entries are percentages for the stable region;
$s$ is the disorder seed index. Clean grids end at $321^2$, disordered grids at $161^2$.
Model differences are distinct from the refinement tolerances described in the text.}
\label{tab:pairwise-regions}
\begin{ruledtabular}
\begin{tabular}{lcrr}
Device & $U_0/\mu$ ($s$) & $100d_H$ & $100d_A$\\
Square & $0$ & 2.0 & 2.6\\
Wide--short & $0$ & 1.2 & 2.1\\
Square & $0.5$ (9) & 3.8 & 8.2\\
Square & $2$ (15) & 3.9 & 8.9\\
Wide--short & $0.5$ (4) & 1.0 & 2.7\\
Wide--short & $2$ (0) & 2.6 & 6.0\\
\end{tabular}
\end{ruledtabular}
\end{table}

For comparison, a maximum radius in an angular bin estimates only a sampled extent and discards holes.
Varying angular resolution without increasing phase sampling can leave bins empty or undersampled.
The stable-versus-full extent diagnostic of Appendix~\ref{sec:diagnostics} uses 360 directions with a fixed
$\pm2^\circ$ neighborhood; it is a different observable from Eq.~\eqref{eq:region-distance}.

Disorder raises this error. In the square device $\epsilon_I$ is 1.88 percent without disorder and
reaches a median of 12.93 percent over twenty realizations at $U_0/\mu=0.5$, then falls at larger
strengths. The lowest multiterminal harmonics, the quartet modes with phase dependence $\theta_R+\theta_B$,
$2\theta_R-\theta_B$ and $\theta_R-2\theta_B$, carry 68.1 percent of the omitted current power,
and restoring only them lowers the median error to 7.22 percent
(Appendix~\ref{app:disorder-ensemble}).

\section{Zero-flux baselines and validation}
\label{sec:baselines}
\begin{figure}[!t]
  \includegraphics[width=\columnwidth]{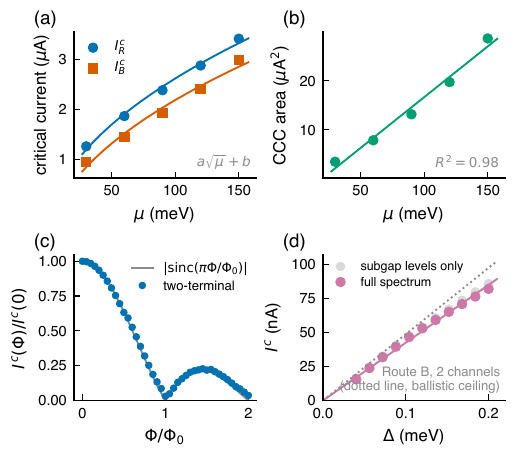}
  \caption{Zero-flux baselines. (a) Critical currents from the right and bottom leads into the junction versus chemical potential
  (Route A, $W=L=600$~nm); lines are $a\sqrt{\mu}+b$ fits, the offset reflecting the integer channel count. (b) critical-current contour
  area versus $\mu$. (c) Two-terminal Fraunhofer pattern with the $|\mathrm{sinc}|$ envelope; the first minimum sits at
  $\Phi/\Phi_0=1.00$. (d) Gap scaling of the Route-B critical current ($W=L=60$~nm, $a=20$~nm, $\mu=10$~meV, two
  channels, leads $2\xi$) with the full spectrum and with the subgap levels only; the dotted line is the ballistic
  ceiling $2e\Delta/\hbar$ of the two channels.}
  \label{fig:validation}
\end{figure}
\begin{figure}[!t]
  \includegraphics[width=\columnwidth]{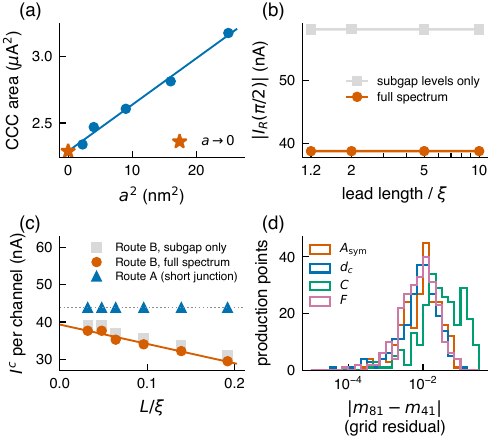}
  \caption{Convergence checks. (a) Route-A critical-current contour area versus $a^2$ with the linear extrapolation to $a\to0$
  (star). (b) Route-B current at $\theta_R=\pi/2$ versus lead length at the benchmark point $L/\xi=0.139$, from the subgap
  sum and from the full spectrum. (c) Route-B critical current per channel versus $L/\xi$ (subgap sum and full
  spectrum), the linear $L\to0$ extrapolation of the full-spectrum values, and the Route-A short-junction value for
  the same normal region; the dotted line marks the ballistic ceiling $e\Delta/\hbar$. (d) Grid residuals of the four contour
  metrics over the device set, the change between the $41$- and $81$-point phase grids.}
  \label{fig:convergence}
\end{figure}

We first verify the expected zero-flux phenomenology (Fig.~\ref{fig:validation}). The critical current scales with the
number of propagating channels $M$, which grows as $\sqrt{\mu}$ in the continuum; at finite width the integer
channel-count offset biases a naive $I_c\propto\mu^p$ fit above $p=\tfrac12$, so we report $I_c\propto M$ as the
primary statement [Fig.~\ref{fig:validation}(a)]. The critical-current contour area grows linearly with $\mu$ [Fig.~\ref{fig:validation}(b)].
For the two-terminal ($b=0$) limit the critical current versus perpendicular flux follows the Fraunhofer envelope
$|\mathrm{sinc}(\pi\Phi/\Phi_0)|$ with its first minimum at $\Phi/\Phi_0=1.00$ [Fig.~\ref{fig:validation}(c)],
confirming the $h/2e$ flux period and the single-electron magnetic phase; the side lobes sit above the envelope
because the current--phase relation of a transparent short junction is nonsinusoidal and the trajectory distribution of a wide junction is nonuniform; the normalized Route-A curve is independent of the coherence length. The critical current of the finite Route-B junction grows linearly with $\Delta$ over $\Delta=0.04$--$0.20$~meV,
with $R^2=0.993$ for a fit through the origin, at 0.87 of the ballistic slope $Ne/\hbar$ with the
full spectrum and 0.90 with the subgap levels alone [Fig.~\ref{fig:validation}(d)]; this tests gap scaling in the finite system.

Three quantitative validations underpin the later analysis (Fig.~\ref{fig:convergence}). A discretization study fixes
the lattice constant. The Route-A critical-current contour area follows the expected $a^2$ law and sits $2.2\%$ from its $a\to0$
extrapolation at $a=1.5$~nm [Fig.~\ref{fig:convergence}(a)]; the contours studied below use $a=5$~nm, where the
area sits 39 percent above the $a\to0$ extrapolation (18 percent in the current scale), a
discretization effect on the absolute currents and areas. The shape metrics were tested under refinement in $a$ over the whole $(\mu,W,L)$
grid at $\Phi\in\{0,0.32\}\,\Phi_0$ with $a=5$, $3.5$ and $2.5$~nm (46 device--flux points). Between $5$ and
$2.5$~nm the circularity changes by a median $|\Delta C|=0.018$ (90th percentile 0.061, maximum
0.12), the fill by 0.008 (0.018; 0.036), the asymmetry by 0.000
(0.021; 0.043) and the pairwise fraction by 0.5 percentage points (maximum
4.1), while the channel numbers change and the relative area changes range from
$\mbox{$-$}28.3$ to $+2.2$ percent; the rank correlation of $C$ with $W/L$ is \mbox{$-$}0.97, \mbox{$-$}0.95 and \mbox{$-$}0.94 at the three
rungs. The strong aspect-ratio association persists, but these changes do not establish pointwise continuum
convergence of the shapes or coupling fractions. Absolute currents and areas retain discretization dependence,
and this refinement covers $\Phi\le0.32\,\Phi_0$ only. For the irreducible share, between the
$41$- and $81$-point grids its median over the 508 device--flux points changes from 0.003995 to
0.003996 percent, but individual values change by 0.32 percent (median) and 1.26 percent (90th
percentile); between $a=5$ and $2.5$~nm the median changes from 0.0035 to 0.0040 percent, with
per-device ratios between 0.8 and 1.3 (10th to 90th percentile) and a maximum of
0.03 percent. The chaotic-cavity shares change by less than
$3\times10^{-3}$ in relative terms for 300 draws recomputed from $41$ to $81$ points per axis.
The clean--chaotic contrast persists under these tested refinements. The observed changes are numerical sensitivity estimates; the enhancement factors in
Appendix~\ref{sec:chaos} retain the discretization and sample-selection limitations of both ensembles. A lead-length
series shows that the Route-B current is converged in lead length from its shortest lead ($1.2\xi$) on, for the
subgap sum and for the full spectrum alike, and that the two differ by 33 percent
at the benchmark point [Fig.~\ref{fig:convergence}(b)]. The continuum contribution is therefore a property of the junction itself. With leads of $25\xi$ the two-channel gap-scaling junction of Fig.~\ref{fig:validation}(d)
gives continuum corrections of 33, 19 and 9 percent at $\theta=\pi/4$, $\pi/2$ and $3\pi/4$, so the contribution
persists in the long-lead limit. A direct comparison of the two routes on the same normal region [Fig.~\ref{fig:convergence}(c)]
places the Route-B per-channel critical current at 0.90 of the Route-A value in the $L\to0$
extrapolation with the full spectrum (0.93 with the subgap sum); the full-minus-subgap part of the difference grows with $L/\xi$, but the subgap current itself extrapolates to
0.93 of the Route-A value at $L\to0$, a 7 percent residual of this coarse
benchmark ($a=20$~nm, two channels) that is not a continuum effect and that we have not resolved. The
9--61\% correction is therefore an observed discrepancy of the benchmark, measured relative to the subgap current. We interpret Route A as the specified
energy-independent-scattering model and compare its shape diagnostics within that model. The finite-device
correction has not been bounded for these geometries, and the shape diagnostics, although ratios, are not
immune to phase-dependent corrections. Finally, Fig.~\ref{fig:convergence}(d) shows the metric-specific grid residuals over the
device set. The change of $A_{\rm sym}$, $d_c$, $C$ and $F$ between the $41$- and $81$-point phase grids has a
median of 0.0097, 0.0082, 0.0307 and 0.0083 and a 90th percentile of
0.033, 0.029, 0.167 and 0.028, respectively. These two-grid differences serve as convergence diagnostics and accompany each number quoted below. The circularity, computed from the perimeter of an alpha shape, is the least converged of the four.

\section{Critical-current contour morphology}
\label{sec:morphology}
\begin{figure*}[!t]
  \includegraphics[width=\textwidth]{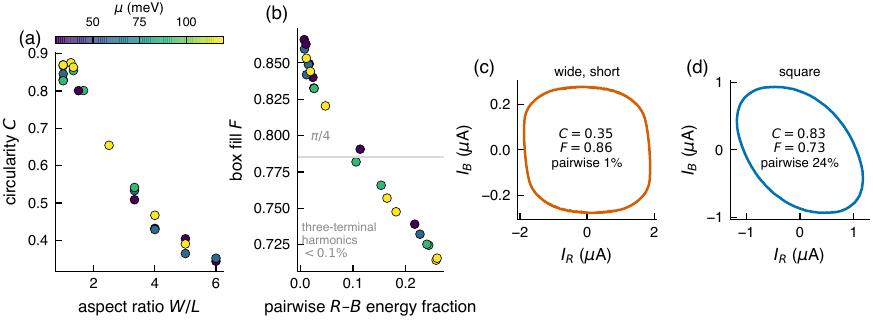}
  \caption{Contour morphology and coupling at zero flux. (a) Circularity versus aspect ratio $W/L$ over the
  $(\mu,W,L)$ grid, colored by $\mu$. (b) Bounding-box fill $F$ versus the pairwise $R$--$B$ energy fraction of the
  Josephson energy; the line marks $\pi/4$, the fill of an axis-aligned ellipse. (c) A wide, short junction
  ($W/L=6$, $\mu=60$~meV; axes scaled independently) and (d) a square junction ($\mu=90$~meV), with their
  circularity, fill and pairwise energy fraction on the 81-point grid. The disorder-panel clean shape
references use the coarser 41-point grid; their small baseline differences are sampling effects.}
  \label{fig:morphology}
\end{figure*}

We map the contour over a 23-point $(\mu,W,L)$ grid with $\mu\in\{30,60,90,120\}~\mathrm{meV}$,
$W\in[300,600]~\mathrm{nm}$ and $L\in[100,400]~\mathrm{nm}$ [Fig.~\ref{fig:morphology}(a)]. The circularity spans
$C=0.34$--$0.88$, and its strongest observed geometric association is with the aspect ratio. $C$ decreases with $W/L$
(Spearman rank correlation \mbox{$-$}0.94), while the rank correlation with $\mu$ is +0.49 and that with the
right-lead channel number +0.14. In the fixed-width sweep ($W=300$~nm, $L=200$--$450$~nm) the zero-flux circularity stays within
$0.79$--$0.87$.

What the circularity measures becomes clear from the two representative contours in Figs.~\ref{fig:morphology}(c,d)
and from the harmonic decomposition [Fig.~\ref{fig:morphology}(b)]. The wide, short junction has $44$ channels in
each of the left and right leads but only $8$ in the bottom contact, so its critical currents are strongly unequal;
drawn to scale the contour is a thin sliver, and $C$ is dominated by this anisotropy. Drawn with independently scaled
axes it is an axis-aligned rounded rectangle with fill $F=0.86$, close to the product-set value, and its
Josephson energy is 99 percent separable. $I_R$ and $I_B$ can then be set almost independently, the
signature of two effectively two-terminal current paths sharing a lead. The square junction has nearly equal
critical currents, a tilted contour with $F=0.73$, below $\pi/4$, and an energy of which 24
percent sits in the pairwise harmonics $\cos k(\theta_R-\theta_B)$. Reaching the maximum of one current then forces the
other away from its own maximum, because a Josephson coupling acts directly between the right and bottom terminals
through the shared normal region. Over the whole grid the fill fraction falls as the pairwise fraction grows (rank correlation
\mbox{$-$}0.99), the pairwise fraction ranges from 1 to 26 percent, and the
genuinely three-terminal harmonics never exceed 0.1 percent of the energy variation. In these
clean junctions a pairwise-additive projection reproduces the tested currents and contour observables to the
accuracy quantified in Appendix~\ref{sec:pairwise-accuracy}. The nonzero energy remainder alone does not establish that accuracy. This does not exclude
Andreev states with support on all three leads, and a network of two-terminal junctions with an internal
superconducting node would itself produce multiterminal irreducible harmonics. A tilted contour signals coupling between phases,
and the Fourier decomposition identifies its predominantly pairwise origin in these devices, in line with the observation of Prosko \emph{et al.} that the
nonlocal current--phase relations of a four-terminal junction can be reproduced by an array of two-terminal junctions
\cite{prosko2024flux}. This replaces the ``rhombus versus circle'' dichotomy of earlier
work \cite{pankratova2020,ozguler2020thesis} by two measurable quantities, $F$ and the pairwise fraction, and it
replaces the reading of the tilt as ``genuinely multiterminal'' transport, used in Ref.~\cite{ozguler2020thesis}, by
the pairwise one.

\section{Third-lead flux response}
\label{sec:thirdlead}
\begin{figure*}[!t]
  \includegraphics[width=\textwidth]{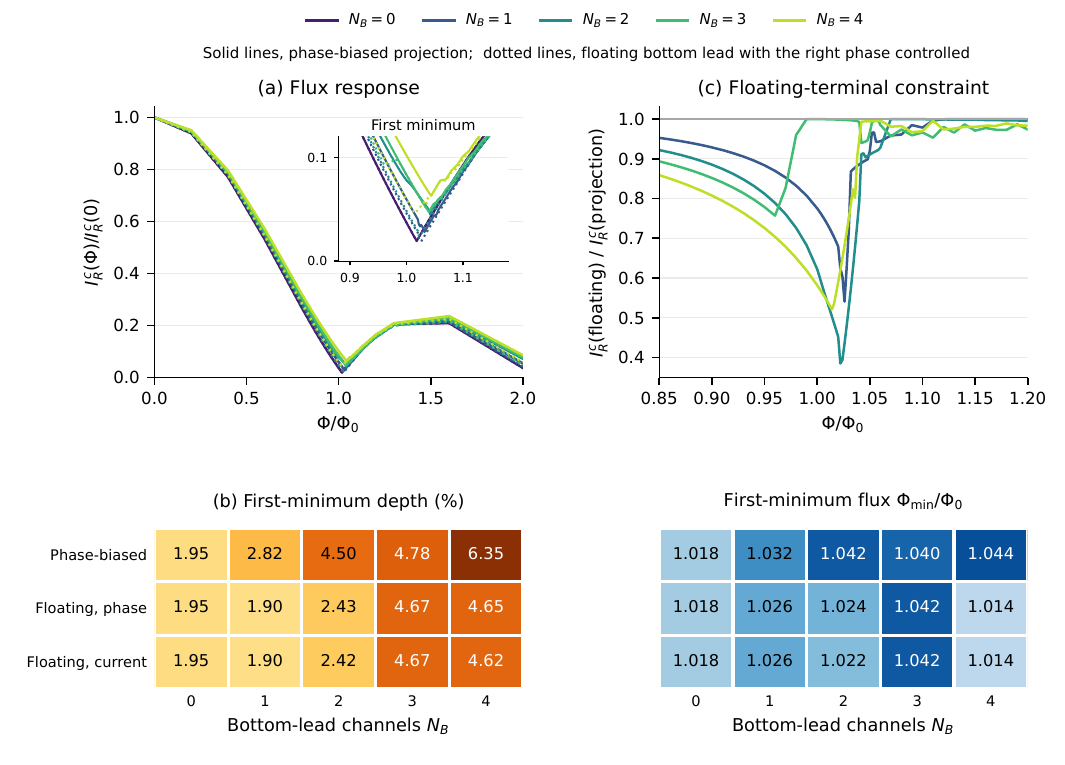}
  \caption{Third-lead Fraunhofer response at fixed $W=600$~nm, $L=60$~nm, $\mu=120$~meV ($a=2$~nm; $60$ channels in the
  left and right leads, $0$--$4$ in the bottom contact). (a) Normalized right-lead critical current versus flux for the
  phase-biased (projection, solid) and floating ($I_B=0$, right phase controlled, dotted) bottom terminal; the inset resolves the first minimum
  with a $0.002\,\Phi_0$ step on $[1.00,1.06]$ and $0.01\,\Phi_0$ outside that interval. (b) Annotated matrices give first-minimum depth in percent (left) and flux
  position $\Phi_{\min}/\Phi_0$ (right) for each integer bottom-channel count.
  Rows distinguish phase-biased, floating with right phase controlled, and
  floating with right current biased protocols; both matrices use the same
  row order. Color increases with each displayed quantity. These are minima
  on the computed flux mesh; the underlying searches remain numerical lower
  bounds on protocol-specific critical currents. (c) Ratio of the floating (right phase controlled, stable states only) to the
  projected critical current versus flux; it is at most one by construction, falls to 0.39 inside the
  minimum and stays below one over most of the window, because the unconstrained maximum generally carries a nonzero
  bottom current while the floating one is restricted to stable states with $I_B=0$.}
  \label{fig:thirdlead}
\end{figure*}

Adding a bottom lead of contact width $b<L$ modifies the Fraunhofer interference of the right-lead critical current~\cite{ozguler2020aps},
whose magnetointerferometry for multiterminal junctions of various aspect ratios was computed by M\'elin \emph{et al.}
\cite{melin2024magneto}.
The value of $I_R^c(\Phi)$ depends on what is done with the third terminal. The author's thesis
\cite{ozguler2020thesis} maximized $I_R$ over both phases. This corresponds to a phase-biased bottom terminal held at
the optimal phase, a projection of the reachable set. A floating bottom
terminal imposes $I_B=0$ instead, and $I_R^c$ is then the maximum of $I_R$ over stable states with $I_B=0$. With the
right phase controlled, these are the branches of the $I_B=0$ curve on which $I_B$ increases with $\theta_B$ (positive
curvature along the bottom phase). With the right terminal current biased as well, only the part of those branches
that is also stable against $\theta_R$ fluctuations remains, and it ends at a fold of the Hessian. We compute all three, refine the flux near the first minimum with a
step of $0.01\,\Phi_0$, then $0.002\,\Phi_0$ on $[1.00,1.06]$, and locate each
maximum with searches that resolve the current jumps of nearly transparent channels [Fig.~\ref{fig:thirdlead}]. In these
clean devices the current--phase relations are sawtooth-like. $I_R$ rises to a maximum at the edge of a jump, where the
lowest Andreev level nearly reaches zero (below $10^{-3}\Delta$ at 116 of the 176
three-terminal projection maxima in the first scan), and a search seeded from the $41\times41$ grid can miss such an edge. For the projection we therefore scan $I_R$ densely in $\theta_R$ along every grid row, refine the best
rows in both phases onto the jump edges and also optimize from every floating optimum; this raises the grid-seeded
maximum at 75 of the 176 three-terminal flux points, by up to
19.9 percent, and at 14 of the 44 two-terminal points, by up to
4.5 percent. Every maximum is checked for continuity,
requiring some point within $10^{-6}$~rad to carry the same current to within half a percent. This rejects spikes resolved by the test, including possible
current inflation near the eigensolver's round-off floor; it does not certify global maximization. For a
floating bottom terminal with the right phase controlled the admissible states form the rising branches of the $I_B=0$
curve, where $I_B$ changes sign from negative to positive with $\theta_B$ (a minimum of the energy along $\theta_B$,
including the cusps at jumps). We follow the eight highest rising branches over $\pm1.5$ grid steps, refine the best point
and compare with a local optimization along the branch; every search returns an admissible state, so the floating
maximum is the largest of them. If the right terminal is current biased as well, a state must be a local minimum of the
tilted energy $U=E-(\hbar/2e)I_R\theta_R$ in both phases. Finite-difference Hessians are ill conditioned at the jumps, so we
test stability on $U$ itself, which stays continuous there. $U$ must rise in all $16$ directions on a ring of radius
$2\times10^{-3}$ around the state. The maxima over such states from the branch scan and from a local optimization differ
by up to 24.69 percent; we retain the largest admissible value. Every current-biased stable state is admissible under phase control and every
state under the projection, so the three critical currents are ordered by construction. The searches
provide numerical lower bounds on the protocol maxima; their finite coverage and the finite-radius stability test
remain limitations. An independent singular-value calculation reproduces all 220 device--flux points of that
scan. The further $0.002\,\Phi_0$ flux scan covers 155 device--flux points, using 61 right-phase samples,
161 bottom-phase samples, the 24 highest sampled rising roots and the candidates of the coarser scan as refinement starts.
For current-biased states it additionally requires a positive Hessian recomputed at the final state and positive tilted-energy increments
in 32 directions at radius $10^{-3}$ rad. Matrix-trace current and one-sided checks are repeated for the final
states. At the 35 flux--device points shared with the coarser grid, its candidates are rechecked with the same local tests and the largest admissible value from either search is retained; outside the fine window the values of the
coarser scan are kept. A fixed root bracket can miss a narrow rising branch, so all 124 three-terminal fine-flux points receive an additional refinement from the floating seeds of the coarser scan using bracket radii $10^{-4}$, $10^{-3}$, $10^{-2}$, $0.05$ and $0.25$ rad. For the widest contact, four points at $\Phi/\Phi_0=1.012$--$1.018$ also use nine right-phase starts displaced by up to $0.04$ rad. Only candidates that pass the same independent protocol tests may raise a maximum. Neither the minimum over this finite flux mesh nor the maximization at each flux is a global certificate, and fine ordering of nearly equal depths is not claimed.
Comparing $b=0$ to $b>0$ at
fixed $W,L,\mu$ isolates the third-lead contribution. On the sampled flux mesh, refined to a step of $0.002\,\Phi_0$, the first minimum of the two-terminal control sits at $\Phi/\Phi_0=1.018$ with normalized depth $0.020$. For the three-terminal devices the minimum lies at $N_B=1$: $1.032$ (depth $0.028$), $1.026$ (depth $0.019$) and $1.026$ (depth $0.019$); $N_B=2$: $1.042$ (depth $0.045$), $1.024$ (depth $0.024$) and $1.022$ (depth $0.024$); $N_B=3$: $1.040$ (depth $0.048$), $1.042$ (depth $0.047$) and $1.042$ (depth $0.047$); $N_B=4$: $1.044$ (depth $0.063$), $1.014$ (depth $0.047$) and $1.014$ (depth $0.046$) for the bottom terminal phase biased at its optimum, floating with the right phase controlled, and floating with the right terminal current biased, respectively. The normalized minimum depth grows monotonically with $N_B$, reaching $0.063$ in the first protocol, and does not grow monotonically with $N_B$, reaching $0.047$ with a floating bottom terminal and the right phase controlled; with the right terminal current biased it does not grow monotonically with $N_B$, reaching $0.047$. The minimum moves by at most $0.026\,\Phi_0$, and the protocols differ by up to $0.030\,\Phi_0$ in position and $0.021$ in depth, so how the third lead changes the minimum depends on how it is biased.

\section{Inversion-symmetry breaking under flux}
\label{sec:inversion}
\begin{figure*}[!t]
  \includegraphics[width=\textwidth]{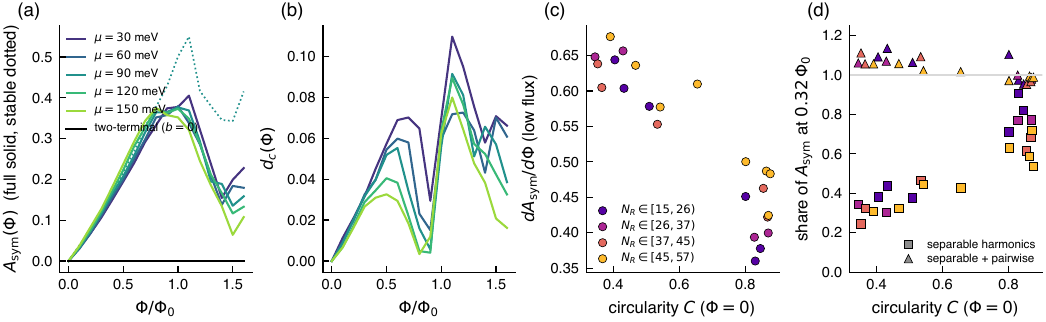}
  \caption{Flux-driven inversion-symmetry breaking and its decomposition. (a) Whole-contour asymmetry
  $A_{\rm sym}(\Phi)$ for the square-like fixed-width contours ($W=300$~nm, $L=300$~nm; colored), from the full image
  (solid) and the stable image (dotted), and the two-terminal control (black). (b) Centroid offset $d_c(\Phi)$ of the
  same contours. (c) Low-flux slope of $A_{\rm sym}$ versus zero-flux circularity for the $(\mu,W,L)$ grid, colored by
  right-lead channel number. (d) Share of $A_{\rm sym}$ at $\Phi=0.32\,\Phi_0$ carried by the separable harmonics alone (squares) and by the
  separable plus pairwise harmonics (triangles), versus zero-flux circularity.}
  \label{fig:inversion}
\end{figure*}

At zero flux time-reversal symmetry, $E_{\rm gs}(\bm\theta)=E_{\rm gs}(-\bm\theta)$, makes the current map odd and the
critical-current contour centrosymmetric under $(I_R,I_B)\to(-I_R,-I_B)$. Over the device set, $A_{\rm sym}$ at zero flux never exceeds
$10^{-9}$, and the two-terminal ($b=0$) control stays below $10^{-9}$ at all fluxes
[Fig.~\ref{fig:inversion}(a)]. A perpendicular flux breaks the current-inversion symmetry of the contour of every tested device at every nonzero
flux sampled. $A_{\rm sym}(\Phi)$ and $d_c(\Phi)$ rise from the first nonzero flux, with $A_{\rm sym}$ reaching a
maximum of 0.47 (median over the grid) within $\Phi/\Phi_0\le1.6$ [Figs.~\ref{fig:inversion}(a,b)]. The stable-image
contours break in the same way as the full-image ones (dotted curves). Up to $0.32\,\Phi_0$ the two asymmetries differ
by at most 0.006, below the 90th-percentile grid residual of $A_{\rm sym}$ for 100 percent
of the points, and over all fluxes with a stable image they rank the device--flux points alike (Spearman correlation
+0.96) with a median difference of 0.001 and a maximum of 0.25 at the highest
fluxes of the square device.

$A_{\rm sym}$ is even in $\Phi$ (time reversal maps the image at $-\Phi$ to the inverted image at $\Phi$) and
vanishes at $\Phi=0$, and the symmetric-difference measure grows linearly in $|\Phi|$ when the boundary shifts
linearly. Over 38 devices (15 of the fixed-width family at $0.1$--$0.3\,\Phi_0$, 23 of
the $(\mu,W,L)$ grid at $0.16$--$0.32\,\Phi_0$) the ratio $[A_{\rm sym}(\Phi_2)/A_{\rm sym}(\Phi_1)]/(\Phi_2/\Phi_1)$
between the two smallest nonzero fluxes is 1.09 (range 0.93--1.44), with slopes of
0.33--0.72 per flux quantum. The asymmetry saturates at 0.35--0.52
between 0.8 and 1.1$\,\Phi_0$ and decreases beyond, as the Fraunhofer factors of the
individual paths pass their first minima.

What breaks the symmetry is not a displacement of independent Fraunhofer factors. An exactly separable energy,
$E=-J_R(\Phi)\cos[\theta_R+\delta_R(\Phi)]-J_B(\Phi)\cos[\theta_B+\delta_B(\Phi)]$, has a centered rectangle as its image
for every field, however the amplitudes and phase shifts depend on it, so ``independent Fraunhofer factors displacing
the corners of a rectangle'' cannot produce any asymmetry. Two ingredients can. The first is nonreciprocity of the
marginal current--phase relations, $\max I_R\neq-\min I_R$ along a cut of the torus, the Josephson diode effect of each
terminal's own path. The flux breaks time reversal, and the bottom lead breaks the transverse mirror that would reverse the flux while preserving the left--right path. Correa and Nowak showed that phase biasing the
third terminal alone produces such a nonreciprocity without a field \cite{correa2024universal}, and field-free
nonreciprocity has been measured in graphene three-terminal junctions with broken mirror symmetry
\cite{zhang2024fieldfree}; here the third terminal is not phase biased and the flux supplies the time-reversal
breaking. Seen as a network of nonsinusoidal two-terminal couplings threaded by a flux, the three-terminal contour is
the contour-level counterpart of the supercurrent-interferometer diode of Souto, Leijnse and Schrade
\cite{souto2022diode} and of the flux-tunable diode measured in a four-terminal InAs/Al junction, where local fluxes
break time reversal and spatial inversion together \cite{coraiola2024diode}. The second is the coupling between the two
currents under flux. We separate the two by rebuilding the currents from successively larger sets of Fourier harmonics,
which gives three current maps, each imaged and given its own alpha-shape contour. The first is the separable part of the
currents, the mean of $I_R$ over $\theta_B$ and of $I_B$ over $\theta_R$, which is the derivative of the separable
harmonics of the energy. The second adds the pairwise part, the mean of the remainder along the lines
$\theta_R-\theta_B={\rm const}$. The third is the full map. The separable part alone produces a nonzero $A_{\rm sym}$ because its marginal current--phase
relations are nonreciprocal under flux (diode efficiency up to 0.051 at $0.32\,\Phi_0$), the Josephson diode
effect of each terminal's own path. At $\Phi=0.32\,\Phi_0$ it carries a median 59 percent of the
full asymmetry (range 25--91 percent) and at $0.8\,\Phi_0$ a median
73 percent. Adding the pairwise harmonics reproduces the full asymmetry to within 0.046
and the box fill to within 0.034, differences slightly above the 90th-percentile grid residuals
(0.033 and 0.028), so the multiterminal irreducible harmonics contribute a small but not always negligible part
[Fig.~\ref{fig:inversion}(d)]. Because $A_{\rm sym}$ is a nonlinear functional of the contour, these shares characterize the specified surrogate contours.
For example, the wide--short device at $0.32\,\Phi_0$ has sampled $A_{\rm sym}=0.0501$,
$0.2269$ and $0.2043$
for the separable, pairwise and full maps, so adding a sector can overshoot the full value.
The three finite-grid maps therefore give successive approximations to a nonlinear observable. It is the transport-only counterpart
of the field-driven diode response measured in three-terminal HgTe junctions \cite{thieme2026hgte}.

\emph{Scope of this comparison.} These surrogate comparisons describe the unshifted phase-grid maps, with the sampling
limitations stated in Appendix~\ref{sec:diagnostics}; they do not provide the current-accuracy or stable-domain
overlap certificate for the pairwise approximation. The refined clean-current controls and selected region
comparisons in Appendix~\ref{sec:pairwise-accuracy} provide the quantitative evidence.

We then test whether circularity predicts the response of the whole current region to flux in our model family.
On the
$(\mu,W,L)$ grid, where the aspect ratio sets the circularity and $\sqrt{\mu}\,W$ sets the channel number so that $C$
and channel number are nearly independent (rank correlation +0.14), the low-flux slope of $A_{\rm sym}$ against
$C$, with the right-lead channel number conditioned out by rank regression, has partial Spearman correlation
\mbox{$-$}0.88 (\mbox{$-$}0.87 for the stable-image contours), and the whole-flux integral of $d_c$
has \mbox{$-$}0.95 [Figs.~\ref{fig:inversion}(c)]; the raw correlations are \mbox{$-$}0.83 and \mbox{$-$}0.95.
Within this family the more separable, rectangle-like contours break more, and the correlated, ellipse-like contours
less. Stronger measured pairwise coupling therefore does not imply greater flux-induced asymmetry. We state this as
an association within a designed family of 23 clean rectangular junctions in which the bottom channel
number, the contact geometry and the path lengths vary together with the aspect ratio. The two
metrics are computed from the same boundary and therefore provide complementary confirmations.

\section{Beyond the clean, static contour}
\label{sec:beyond}
\begin{figure*}[!t]
\includegraphics[width=\textwidth]{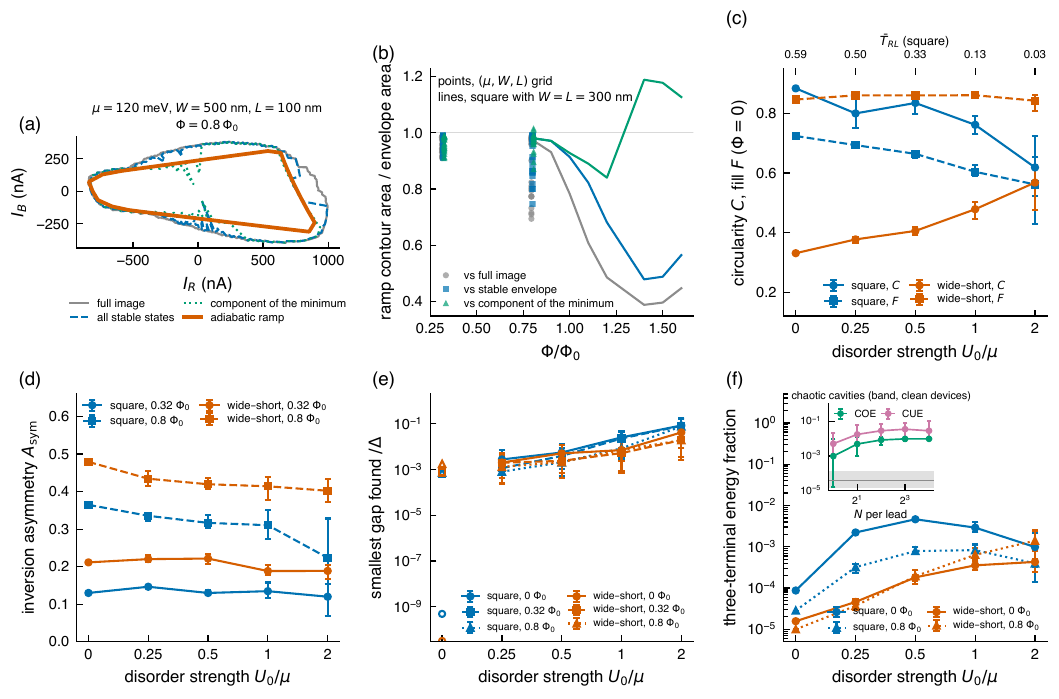}
\caption{Current ramps, disorder and chaotic cavities. (a) Switching contour reached by an adiabatic current ramp
from zero bias (thick) for the device whose ramp contour encloses the smallest fraction of the full-image area at
$0.8\,\Phi_0$, with the radial envelopes of the
full image, of all stable states and of the stable component containing the zero-bias minimum. (b) Area of the ramp
contour relative to the three envelopes, for the $(\mu,W,L)$ grid at $0.32$ and $0.8\,\Phi_0$ (points, one per device)
and for the square fixed-width device up to $1.6\,\Phi_0$ (lines). (c) Circularity and fill of the square and wide--short
devices under on-site disorder of box width $U_0$ (medians and 10--90\% ranges over 20 realizations; the clean values
at $U_0=0$); the upper axis gives the mean normal transmission per channel of the left--right path of the square
device. (d) Inversion asymmetry of the same devices at $0.32$ and $0.8\,\Phi_0$. (e) Torus minigap (refined minimum
over the torus). (f) Irreducible three-terminal fraction of the energy versus disorder strength; the inset shows the
same fraction for chaotic-cavity junctions (circular orthogonal and unitary ensembles with $N$ channels per lead;
medians and 10--90\% ranges over 50 draws) with the 10--90\% band of the clean survey devices.}
\label{fig:beyond}
\end{figure*}

The contours above are equilibrium objects of clean junctions. Three computations test what carries over to a
measurement. An adiabatic current ramp selects the stable branch that a bias sweep reaches, a disorder ensemble
uses the same device sizes, and chaotic cavities benchmark the harmonic decomposition.

\subsection{Adiabatic current ramp}
\label{sec:ramp}

We use an overdamped circuit with equal conductances from R to grounded L and from B to L, and no R--B shunt.
The ramp uses instantaneous equilibrium occupations at $T=0$; it does not include quasiparticle
occupation dynamics, conserved-parity evolution, or Landau--Zener transitions. Here adiabatic denotes slow bias continuation in the circuit model.
Its conductance matrix in $(\theta_R,\theta_B)$ coordinates is proportional to the identity, so the phases
follow the gradient flow of
$U(\bm\theta)=E_{\rm gs}(\bm\theta)-(\hbar/2e)\bm I_{\rm bias}\cdot\bm\theta$.
We increase $\bm I_{\rm bias}=\lambda(\cos\varphi,\sin\varphi)$ from a numerically located zero-bias minimum,
using tangent prediction and damped Newton correction. Every root on the path has a positive Hessian,
evaluated at that root, and a current residual below $10^{-6}$ of the device current scale. Ordinary continuation steps are limited
to $0.1$ rad in phase displacement; a larger step is rejected and the bias increment halved. This step limit matters. Without it, Newton
correction can reach a different stable minimum and miss an intervening relaxation. If the continuation criterion
is not met at the minimum bias step, a fixed-bias gradient-flow calculation starts;
it yields a numerical transition interval. The largest interval width is bounded by $0.1$ percent
of the current scale in this protocol.

We integrate $d\bm\theta/dt=(\bm I_{\rm bias}-\bm I(\bm\theta))/I_{\rm scale}$ with adaptive RK45
(relative tolerance $10^{-7}$, absolute tolerance $10^{-9}$, maximum time step 2), recording trapping or a
$2\pi$ displacement of either phase. A trapped state is corrected to a stable root before continuation resumes.
Four consecutive slips without trapping define \emph{observed running}; a trajectory reaching neither event within
5{,}000 time units per segment remains unresolved. This finite-time criterion does not prove indefinite running.
The scan contains 3{,}960 rays over 55 device--flux points, with 3{,}958 observed-running
endpoints and 2 other endpoints. Re-trapping occurs on 847 rays. Only the
53 devices with all 72 switching estimates resolved enter the polygon-area comparison; no missing
ray is interpolated. Each estimate is the midpoint between the last stable bias and the observed-running test bias;
these intervals quantify the bias resolution.

At $0.32\,\Phi_0$, the 23 completed grid devices cover 91.0--98.7 percent of the sampled
full-image envelope area; at $0.8\,\Phi_0$, the 22 devices cover 69.4--98.9 percent
(median 86.3 percent), 74.7--99.3 percent of the stable envelope, and
87.1--101.6 percent of the envelope of the component containing the zero-bias minimum
[Fig.~\ref{fig:beyond}(a,b)]. The square and near-square devices ($W/L\leq1.5$) span 95.9--98.9 percent
of the full-image area, and the wide, short devices ($W/L\geq3$) span 69.4--82.4 percent.
These are protocol-dependent comparisons with sampled static envelopes. Neither area averaging nor apparent
agreement establishes continuity of individual switching points. Neighboring-ray deviations from the mean of the
two adjacent rays reach 13.0 percent at $0.32\,\Phi_0$ and 18.3 percent at
$0.8\,\Phi_0$. The static envelopes also use angular binning and a running maximum, so individual-ray ratios
can be affected by both calculations.

Four targeted rays (three previously unresolved rays and one with a detected branch jump) were rerun with a $0.05$-rad phase-step limit, halved bias steps, tenfold tighter root and integration tolerances, a halved Hessian difference step, DOP853 integration with maximum time step $0.5$, and eight slips. All four reproduce the number of recorded re-trappings and have overlapping switching intervals; the largest midpoint change is 0.025 percent of the current scale. At their final stable endpoints, multiscale Hessian and one-sided current checks provide local numerical evidence. These targeted tests do not certify every continuation branch or indefinite running.
An independent matrix-trace evaluator checks the stable endpoints of all 3{,}960 rays, including their
current residuals, Hessian signs and one-sided current limits; sampled tilted energies decrease along the computed
overdamped trajectories within numerical tolerance.
For the two devices selected by the largest historical neighbouring-ray deviations, nested 72-, 144- and 288-direction polygons change by at most 0.89 and 0.21 percent relative to the 288-direction area. Both meet the prespecified 2 percent and 1 percent angular criteria. These nested grids test angular quadrature only; the same solver settings are used at each direction.
The same independent endpoint and sampled-flow checks also pass for all 576 rays of these angular scans.
For the fixed-width square device, the completed points between $0.9$ and $1.6\,\Phi_0$ span
38.8--97.5 percent of the full-image envelope, 47.8--98.1 percent of the stable envelope, and
84.1--118.9 percent of the component envelope. A static image alone therefore does not specify a
switching protocol. Thermal and quantum activation, unequal shunt conductances, and an exhaustive certification
of all continuation branches lie outside this calculation.

The ramp endpoints above are deterministic limits for the specified initial state and circuit. Switching-current distributions additionally depend on fluctuations. At finite temperature, thermally activated escape can occur before the metastable minimum loses stability. In an activated regime its exponential dependence on the barrier divided by $k_BT$ makes a small barrier change potentially important, even if a phase-averaged energy or current error is small. Thus pairwise and full energies can give similar static contours yet different escape rates or switching distributions. Quantifying the resulting stochastic rates and distributions requires a bath model, damping, ramp rate and initial-state preparation beyond the deterministic calculation. Thermal-noise effects on magnetization reversal in an anomalous ferromagnetic junction~\cite{r2csf142110384} and on its switching-current distributions~\cite{r2apl123152602}, and noisy breather dynamics in spatially extended sine--Gordon systems~\cite{r2csf199116641}, illustrate distinct fluctuation-sensitive Josephson settings. Transferring these fluctuation treatments to the short planar model requires adapting their magnetic or extended-junction degrees of freedom.

\subsection{Disorder ensemble}
\label{sec:disorder}

Both illustrative devices were recomputed with on-site Anderson disorder, uniformly distributed in
$[-U_0/2,U_0/2]$ on every site of the normal region, for $U_0/\mu\in\{0.25,0.5,1,2\}$ and 20
realizations each at $\Phi/\Phi_0\in\{0,0.32,0.8\}$ (480 contours). The mean normal transmission per
channel of the left--right path, from the same scattering matrix at the Fermi energy, falls from 0.59
(clean) to 0.50, 0.33, 0.13 and 0.03 for the square device and from
0.91 to 0.88, 0.81, 0.61 and 0.31 for the wide--short one,
so the disorder series runs from a slightly perturbed ballistic junction to a strongly scattering one. This is disorder
on a fixed $a=5$~nm lattice. For independent on-site potentials, the integrated two-dimensional covariance
scales as $U_0^2a^2/12$; holding $U_0/\mu$ fixed while changing $a$ would not keep physical disorder fixed.
No continuum-disorder extrapolation is inferred. Four results
follow [Fig.~\ref{fig:beyond}(c--f)]. (i) The median geometric ordering survives all tested disorder strengths. The
zero-flux circularity
of the square device stays above that of the wide--short one at every $U_0$ (0.80 versus
0.38 at $U_0=0.25\mu$, 0.62 versus 0.57 at $2\mu$; clean 0.88 versus
0.33), and the fill of the wide--short device stays near the product-set value (0.86 at
$0.25\mu$ and 0.84 at $2\mu$, against 0.69 and 0.56 for the square). Weak disorder lowers
the circularity of the square device from 0.88 to 0.80 and raises that of the wide--short
device from 0.33 to 0.38; strong disorder brings the two sampled shape distributions
closer. Two geometries do not separate transmission changes from geometric effects or establish a causal mechanism. (ii) The inversion asymmetry is robust. At $0.32\,\Phi_0$ the
square device gives medians of 0.146, 0.130, 0.135 and 0.120
(clean 0.130; 10--90\% range 0.07--0.15 at $2\mu$) and the wide--short device
0.220, 0.222, 0.188 and 0.189 (clean
0.211); at $0.8\,\Phi_0$ the medians fall from 0.34 to 0.22
(square) and from 0.43 to 0.40 (wide--short) between $0.25\mu$ and $2\mu$ (clean
0.36 and 0.48), so the more separable device breaks more at every disorder
strength, as in the clean family. (iii) The harmonic content also changes. The pairwise fraction of the square device changes from 24
percent (clean) to 29, 36, 45 and 45 percent, and the
irreducible three-terminal fraction changes from 0.01 percent to 0.22,
0.46, 0.29 and 0.10 percent at zero flux. The three-terminal share is nonmonotone in disorder strength for this device,
peaking in median at $U_0=0.5\mu$ among the sampled strengths. Across all realizations and fluxes it remains below
0.52 percent. At $U_0=2\mu$, the square-device medians are
0.09 percent at $0.32\,\Phi_0$ and 0.04 percent at $0.8\,\Phi_0$;
the wide--short median at $0.8\,\Phi_0$ is 0.14 percent. All sampled disorder shares remain
below the $N=16$ median of either chaotic ensemble in Appendix~\ref{sec:chaos}. Disorder can raise the
irreducible energy share relative to a clean device, but these data do not show a transition to the
percent-level shares of those cavities. An energy-power share alone does not bound a current or contour error.

(iv)
Disorder raises the optimized gap estimates. The locally refined minimum, below $5\times10^{-6}\,\Delta$ for the
clean device at every finite flux, has medians of 0.8--2.8$\times10^{-3}\,\Delta$ at
$U_0=0.25\mu$ and 0.02--0.08$\,\Delta$ at $2\mu$, and at $U_0\ge0.5\mu$ a fraction
70--$100$ percent of the realizations (median 92 percent per group) gives an
estimate above $10^{-3}\Delta$. These local searches supply upper bounds on the true torus minimum.
For the diagnostic realization below, adaptive coverage of the entire two-phase torus gives a positive gap interval $[0.00233,0.00259]\,\Delta$, smaller than the local-search estimate. The singular-value perturbation inequality gives $g(\bm\theta)\ge g(\bm\theta_c)-2\sin(h/2)\|s\|_2$, with $g=\sigma_{\min}(A)$, on every cell of half-width $h$ centered at $\bm\theta_c$. We include a numerical allowance and use ordinary floating-point arithmetic. A ground-state Chern number is well defined in principle for a globally gapped device, but has not
been established for this realization. The coarse $40\times40$ bound-state calculation gives
$+12$, whereas independent adaptive subdivision gives a noninteger phase sum $1.84$.
This adaptive implementation replaces selected plaquettes with their children without matching subdivided
edges to neighboring coarse cells. Its sum therefore lacks the shared-link cancellation that makes the
periodic Fukui--Hatsugai--Suzuki lattice invariant integer \cite{fukui2005chern}.
Indeed, the same routine gives $0.00786$ for a globally smooth trivial rank-one test projector, while
uniform $4\times4$ and $32\times32$ meshes give zero to roundoff (Appendix~\ref{app:strengthening}).
The noninteger adaptive sum is consequently not evidence of physical projector discontinuity.
Near-gap-edge levels occur at the diagnostic cells, but two-sided projector limits were not established.
Neither the coarse integer nor its adaptive replacement is a validated invariant of this device.
A continuum-inclusive formulation \cite{repin2019topological}, or a full Bogoliubov--de Gennes treatment,
would permit a separate investigation of ground-state topology; the closed-lattice example in Appendix~\ref{sec:topology}
is a different model.

\subsection{Chaotic-cavity benchmark}
\label{sec:chaos}

To calibrate the harmonic fractions we generated three-terminal spin-degenerate chaotic cavities, ballistic quantum dots whose classical electron
dynamics is chaotic, with scattering matrices from the circular orthogonal and unitary ensembles (COE and CUE)~\cite{beenakker1997rmt},
$N\in\{1,2,4,8,16\}$ channels per lead and 50 draws each, and decomposed the same ground-state
energy $E_{\rm gs}(\theta_R,\theta_B)=-\frac12\sum_k|E_k|$, summed over all eigenphases of the secular kernel so that
levels at the gap edge, which the orthogonal ensemble places on the line $\theta_R=\theta_B$, are counted consistently,
on the $41$-point grid. The irreducible three-terminal fraction is small with one channel per lead (0.1 and
0.5 percent in the orthogonal and unitary ensembles) and grows with the channel number, reaching
1.0 percent (10--90\% range 0.9--1.0) in the orthogonal and
2.8 percent (1.2--10.2) in the unitary ensemble at $N=16$; the pairwise fraction is then
28--33 percent [Fig.~\ref{fig:beyond}(f)]. The clean survey devices, with $15$ to $56$
channels, have a median three-terminal fraction of 0.004 percent (508 device--flux points),
243 times below the time-reversal-symmetric chaotic value at $N=16$ and 710 times below the
unitary one, and pairwise fractions in the same range as the cavities. This geometry--flux pool is not a matched
comparison at fixed channel count. On the sampled phase grids the contour shape does not separate the two
regimes. At $N=16$ the orthogonal-ensemble contour has fill 0.69 (0.67--0.72) and circularity
0.86 (0.82--0.88), a tilted ellipse close to that of the clean square device. Uniform continuum
accuracy of these contour statistics is not established. At comparable
channel number and flux, a measured three-terminal harmonic content of a percent or more, obtained from the measured
current--phase relations of both currents \cite{prosko2024flux}, therefore departs from the pairwise-additive description that the
clean rectangular devices follow. The many-channel chaotic ensembles reach this scale, whereas all sampled
disordered devices stay below it. A single value is not a classifier of the microscopic scattering mechanism.

\section{Direct pairwise couplings versus an internal island}
\label{sec:island-circuit}
A multiterminal irreducible harmonic need not require an irreducible
microscopic transfer process. For example, a classical superconducting island
with phase $\theta_0$, connected by ordinary junctions to the external
terminals, has
\begin{equation}
 E=-\sum_i J_i\cos(\theta_i-\theta_0),\qquad
 \min_{\theta_0}E=-\left|\sum_i J_i e^{i\theta_i}\right|.
 \label{eq:classical-island}
\end{equation}
The eliminated energy is generally nonpairwise in the external phases. It
has the same phase dependence as the isolated resonance at $x=0$ in
Eq.~\eqref{eq:resonance-energy}, despite a different microscopic construction.
With fixed $J_i$ a classical capacitive gate does not change this equilibrium
energy; predicting a gate response requires additional physics or an explicitly
gate-dependent element model.

A conventional charging island supplies one such degree of freedom. Its
Hamiltonian is $H=4E_C(\hat n-n_g)^2-\sum_i E_{J,i}
\cos(\hat\theta_0-\theta_i)$. Near the degeneracy of charge states $n=0,1$,
the two-state reduction gives, up to a phase-independent constant,
\begin{equation}
 \begin{gathered}
 F_{\rm isl}=-\sqrt{d^2+\left|\sum_i t_i e^{i\theta_i}\right|^2},\\
 t_i=E_{J,i}/2,\qquad d=4E_C(1/2-n_g).
 \end{gathered}
 \label{eq:quantum-island}
\end{equation}
This established circuit reduction \cite{melo2022multiplet} provides an alternative realization of the square-root phase dependence.
For fixed $E_{J,i}$, its charge-degeneracy energy is half the classical
minimum; the fitted energy scale below accounts for that factor.

We benchmark both island descriptions against the two finite-gap planar
devices of Sec.~\ref{sec:planar-resonance}. In dimensionless form the charging
model is $F_{\rm isl}/\Delta=-A\sqrt{(\lambda x)^2+
|\sum_i p_i e^{i\theta_i}|^2}$, with $A,\lambda>0$, $p_i>0$ and
$\sum_i p_i=1$. An additional, prespecified variant includes parallel direct
junctions, $F_{\rm bg}/\Delta=-\sum_{i<j}b_{ij}\cos(\theta_i-\theta_j)$,
$b_{ij}\ge0$. The classical model fits $A,p_i$ at $x=0$; each charging model
fits its parameters jointly at $x=0,1$. Parameters are then fixed at
$x=-4,-1,4$, without a gate-by-gate refit. There are respectively $n$, $n+1$
and $n+1+n(n-1)/2$ free parameters. No offset of the charge degeneracy is
fitted, so the circuit predictions are even in $x$; the microscopic device
need not share that symmetry.
The mapping is $E_{J,i}=2\Delta A p_i$ and $d=\Delta A\lambda x$;
$\lambda$ is a calibrated gate lever arm. This differs from the unfitted spectral intervention above.

This comparison is retrospective, since the devices and their gate
results were already known. Calibration uses 1,024 existing phase samples per
training gate. Validation recomputes the microscopic currents on an
independent phase set, checks frequency refinement, and doubles its size to
test the stability of the reported RMS errors. The direct-pair reference is
the converged arbitrary-harmonic pair projection at each target gate,
evaluated on these same validation phases. It is an optimistic approximation
baseline with access to target-gate information.
The benchmark for a circuit is a current error below 5 percent of the full current RMS at
each withheld gate.

\begin{figure*}[t]
 \centering
 \includegraphics[width=\textwidth]{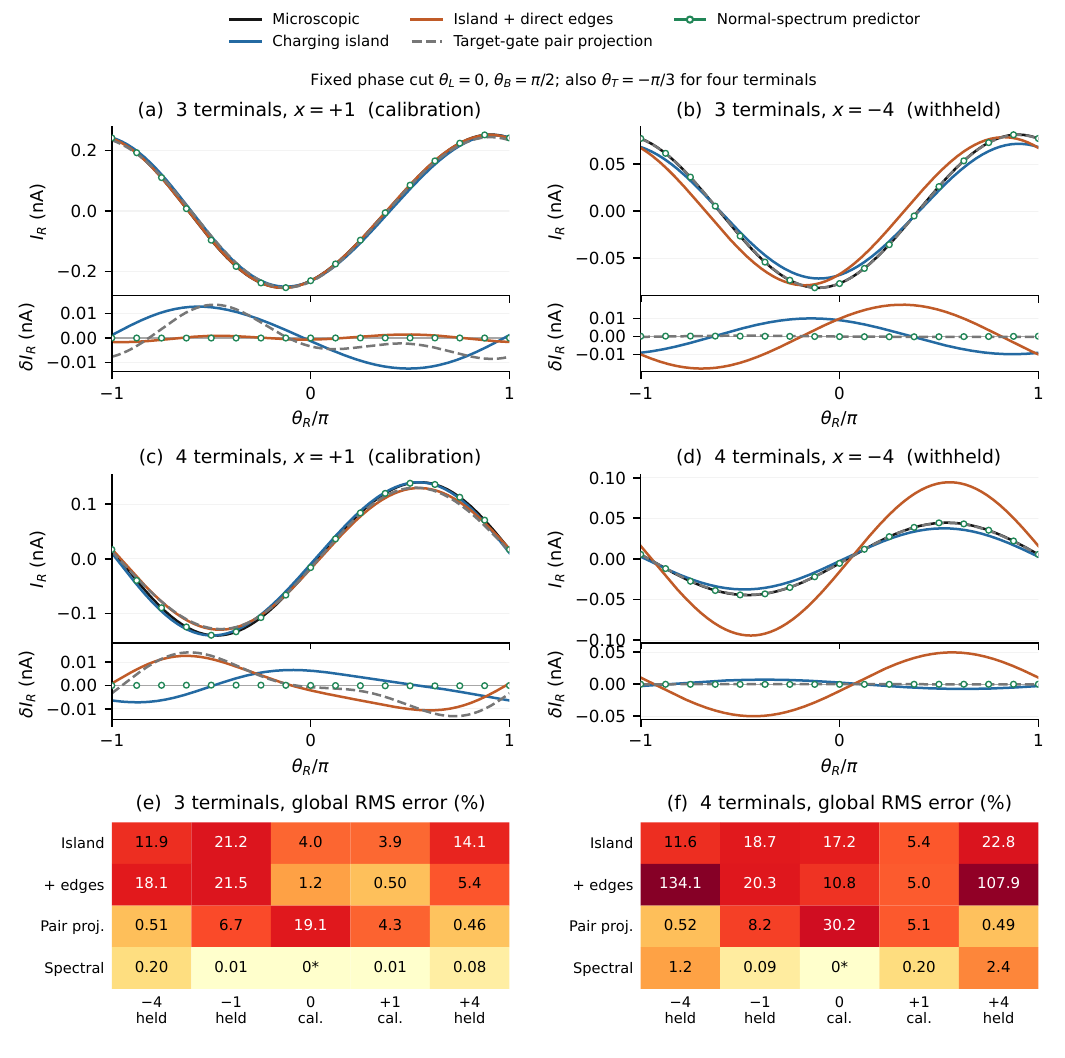}
 \caption{Circuit predictions compared with microscopic current waveforms.
 (a--d) Microscopic $I_R$ and fixed model predictions for the two clean
 finite-gap planar devices, with $\Delta=0.18$ meV. Sweep $\theta_R$ at
 $\theta_L=0$, $\theta_B=\pi/2$ and, for four terminals,
 $\theta_T=-\pi/3$. Each lower axis shows model minus microscopic current.
 The left cuts use the calibration gate $x=1$; the right cuts use the
 withheld gate $x=-4$, chosen retrospectively to illustrate the largest
 discrepancies. Each curve contains 513 calculated phases; green symbols show
 a subset of the spectral predictions for visibility.
 (e,f) All-terminal, full-phase RMS errors (percent) at the five prespecified
 gates; color increases logarithmically with error. ``Cal.'' and ``held''
 refer to circuit fitting only. The pair projection uses target-gate data;
 the spectral predictor uses normal-state information, with a zero at $x=0$ by
 construction (asterisk). The circuit benchmark is 5 percent at every withheld gate. Phase cuts and global errors measure different
 aspects of agreement.}
 \label{fig:island-circuit}
\end{figure*}

None of the three fitted circuit families meets this benchmark at all withheld
gates in either device (Fig.~\ref{fig:island-circuit}). The charging island's
largest withheld errors are 21.2 and
22.8 percent for $n=3,4$; adding direct edges gives
21.5 and 134.1 percent.
The fixed classical island is worse, reaching 591
and 655 percent. Calibration accuracy and prediction at withheld gates can therefore differ substantially. In the three-terminal device, adding edges
reduces the largest calibration-gate error from 4.0 to
1.2 percent, while its largest withheld error remains
above 20 percent. All eight optimizer starts terminate successfully for each fit. Global optimality
and parameter identifiability require further analysis.
In particular, two three-terminal background edges reach the near-zero bound.
The separate all-mode normal-spectrum predictor keeps errors below 2.5 percent
on the same validation phases, without fitting Josephson currents; its agreement at
$x=0$ holds by construction.

A post-scoring diagnostic identifies a restriction that optimization cannot
remove. Every fitted circuit above predicts the same current field at $x$
and $-x$. For any such field $\mathbf G$, the triangle inequality gives
\begin{equation}
 \max_{s=\pm1}\frac{\|\mathbf G-\mathbf I(sx)\|}{\|\mathbf I(sx)\|}
 \ge \frac{\|\mathbf I(x)-\mathbf I(-x)\|}
 {\|\mathbf I(x)\|+\|\mathbf I(-x)\|}.
 \label{eq:even-gate-obstruction}
\end{equation}
Here each norm includes all terminals and the same validation phases.
At the two withheld endpoints $x=\pm4$, the lower bound is
6.7 and 11.3 percent for $n=3,4$. Thus even an
optimally refitted field with this gate symmetry cannot bring the error below 5 percent
at both endpoints. This is an obstruction to the chosen centered gate law. A shifted charge degeneracy or gate-dependent
couplings would escape the argument and require new validation.
On 64 validation phases per gate, independent charge-basis calculations for the fitted parameters change currents
by less than 0.03 percent relative to the two-state expression at a fixed
$E_C=10^3\Delta A\max(1,4\lambda)$ per circuit. The finite-charge truncation
therefore does not explain the resolved errors in this specified large-$E_C$
regime; changing that regime would define another circuit model.

The benchmark compares direct external-pair additivity with specified circuit models
and calibration laws. Equilibrium current agreement concerns the tested observable;
identifying a physical island or relating spectra, dynamics and topology requires
additional evidence. Models with other internal nodes, gate-dependent couplings or
explicit multiphase energy terms require their own validation.
The charge-only circuit assumes suppressed quasiparticles; $\Delta$ here is the planar current normalization.
Our quantitative distinction is between direct external-pair additivity and
specified descriptions retaining internal degrees of freedom.

The phase cuts in Fig.~\ref{fig:island-circuit}(a--d) show the discrepancies in more
detail than an RMS score. At $x=-4$, the charging island
underestimates the peak-to-peak $I_R$ by 12 and 16 percent for $n=3,4$.
Adding direct edges brings the three-terminal amplitude within 4 percent,
but leaves a 22-percent relative error in this component along the cut, so
amplitude agreement alone misses the phase dependence. In the four-terminal
cut, the same model predicts 2.12 times the microscopic peak-to-peak amplitude.
These are illustrative one-component cuts; panels (e,f) give the all-terminal phase-space errors. No circuit parameters were
changed for these illustrations.

\section{Separate topology diagnostics and lattice illustration}
\label{sec:topology}
\begin{figure*}[!t]
  \includegraphics[width=\textwidth]{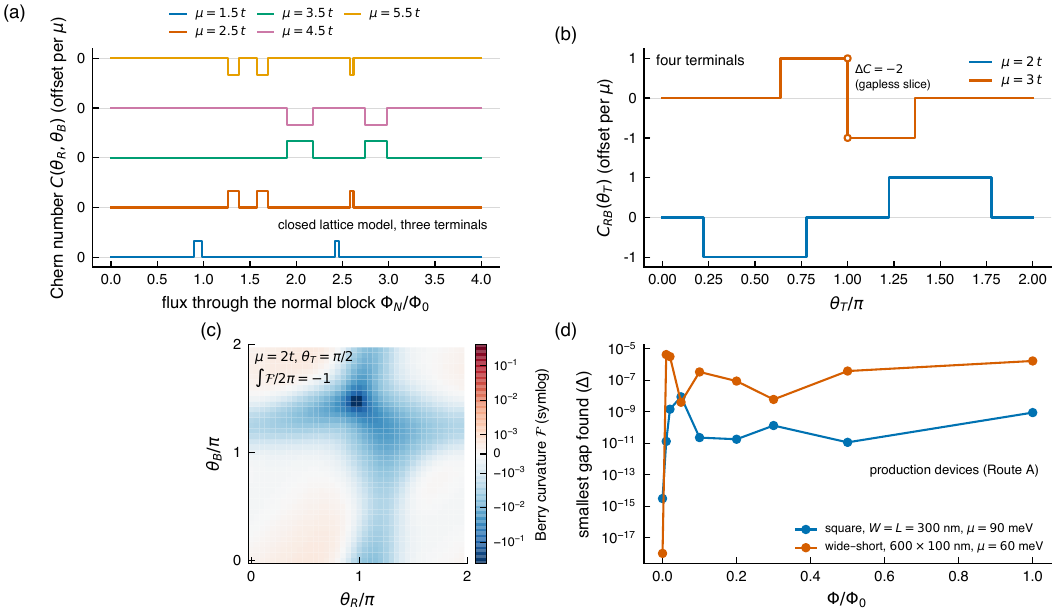}
  \caption{Band topology in the closed lattice model ($\Delta=0.4t$) and the gap of the illustrative devices. (a) Chern
  number of the occupied BdG manifold over $(\theta_R,\theta_B)$ for three terminals versus the flux through the normal
  block, for five chemical potentials (offset vertically); narrow $C=\pm1$ windows open above $\Phi_N\approx0.9\,\Phi_0$.
  (b) Chern number of the $(\theta_R,\theta_B)$ slice for four terminals at zero flux versus the third phase at $\mu=2t$
  and $3t$ (offset), changing by the net Weyl charge crossed; open circles mark the one-sided values at the gapless $\theta_T=\pi$ slice for $\mu=3t$, where $C$ is undefined. (c) Berry curvature of the slice at $\mu=2t$, $\theta_T=\pi/2$,
  integrating to $C_{RB}=-1$; it is concentrated within a few grid cells of the projection of the enclosed Weyl point
  (peak $\mbox{$-$}0.41$ per cell against $|\mathcal F|<10^{-2}$ on $93\%$ of the slice), so the color scale is logarithmic in
  $|\mathcal F|$ beyond $10^{-3}$. (d) Smallest Andreev gap found over the phase torus of the two illustrative devices
  (Route A) versus flux, using SVD and multiple local refinements. No estimate exceeds $5\times10^{-6}\,\Delta$ at the tested fluxes, and both zero-flux values are below $10^{-12}\Delta$. Values below $10^{-18}\Delta$ are placed at the plotting floor and do not resolve a nonzero gap. These are upper estimates of the true minima.}
  \label{fig:topology}
\end{figure*}
\begin{table}[!t]
\caption{Refined Weyl points of the four-terminal lattice model. Local gap minimization gives residuals below
$10^{-12}\,t$; charges are independently reproduced on cubes with 12 and 24 plaquettes per edge.
Charges sum to zero at each $\mu$, and time-reversed partners carry equal charge.}
\label{tab:weyl}
\begin{tabular}{@{}c c c c c@{}}
\hline\hline
$\mu/t$ & $\theta_R/\pi$ & $\theta_B/\pi$ & $\theta_T/\pi$ & charge \\
\hline
2 & 1.000 & 1.222 & 0.222 & $-1$ \\
2 & 1.000 & 1.778 & 0.778 & $+1$ \\
2 & 1.000 & 0.222 & 1.222 & $+1$ \\
2 & 1.000 & 0.778 & 1.778 & $-1$ \\
3 & 1.637 & 1.000 & 0.637 & $+1$ \\
3 & 0.363 & 1.363 & 1.000 & $-1$ \\
3 & 1.637 & 0.637 & 1.000 & $-1$ \\
3 & 0.363 & 1.000 & 1.363 & $+1$ \\
\hline\hline
\end{tabular}
\end{table}

A natural question is whether the contour morphology or the symmetry breaking is tied to a topological transition of
the Andreev band structure. We addressed it in two ways. First, for the two illustrative devices, we followed the
minimum of the Andreev gap over the $(\theta_R,\theta_B)$ torus as a function of flux, refining the minimum by local
optimization of the smallest singular value of $A$ from every sampled local minimum and previously located seed, which resolves small gaps to $10^{-15}$ where the square root of
the eigenvalues of $A^\dagger A$ does not. At zero flux it is $<10^{-12}\,\Delta$, zero to numerical precision, and at every
finite flux tested the smallest gap found lies between $10^{-11}\,\Delta$ and $5\times10^{-6}\,\Delta$
[Fig.~\ref{fig:topology}(d)]. These are small gaps found by an optimizer, and no gapped regime was resolved. The clean junction between leads of the same width carries transmissions close to one, whose
Andreev levels come close to zero energy at $\theta_R=\pi$ or $\theta_B=\pi$ even without a field (for a single channel
$E_{\min}=\Delta\sqrt{1-T}$, which vanishes only at $T=1$); the spin-degenerate junction
is in class C at a generic phase point and its zero-energy nodes have codimension three, but a perfectly transmitting channel is a non-generic point that the clean geometry realizes at zero flux and
approaches at finite flux. (The phase dependence of the proximity-induced gap
of diffusive multiterminal junctions has been mapped directly by tunneling spectroscopy \cite{wisne2024mapping}.) No topological threshold is established for these devices. The sampled small gaps do not
certify a closing, and the adaptive bound-state phase sum in Appendix~\ref{sec:disorder} is not an integer-preserving
lattice construction. Gap-edge touchings alone do not establish discontinuity of the selected projector.
A reliable ground-state Chern number requires a well-defined occupied manifold and independently
validated integration; neither follows from the present gap diagnostic.

Second, we illustrate the topology in a closed lattice model, a $3\times3$-site normal block with two-site
superconducting pads and $\Delta=0.4t$, by computing the Chern number of the occupied
Bogoliubov--de Gennes manifold over the phase torus with the Fukui--Hatsugai--Suzuki lattice method
\cite{fukui2005chern}, validated on the Qi--Wu--Zhang Chern insulator \cite{qi2006topological}. For three terminals
the Chern number vanishes at zero flux, as time reversal requires, and remains zero at the resolved sampled fluxes below
$\Phi_N=0.92\,\Phi_0$ through the normal block at every chemical potential studied; above that, narrow windows of
$C=\pm1$ open, bounded by gap closings; independent phase-grid checks are reported below
[Fig.~\ref{fig:topology}(a)], in agreement with Meyer and Houzet \cite{meyer2017nontrivial}. For four terminals the
three phases span a three-torus with Weyl points, zero-energy touchings of the class-C reduced block whose codimension
is three \cite{yokoyama2015singularities}, already at zero flux. The Chern number of a two-phase slice jumps by $\pm1$
as the third phase crosses an isolated Weyl point [Fig.~\ref{fig:topology}(b)], with four Weyl points of charges $(-1,+1,+1,-1)$
at $\mu=2t$ and four at $\mu=3t$ (Table~\ref{tab:weyl}); time reversal maps a Weyl point at $\bm\theta$ to one at
$-\bm\theta$ with the same charge, so charge neutrality requires at least four, which is what the table shows. At $\mu=3t$, two charge-$-1$ nodes have the same third phase
$\theta_T=\pi$; the slice is gapless there and its Chern number is undefined. The adjacent gapped slices differ by
$-2$, the net enclosed charge of the two nodes. The
measurable signature is a quantized transconductance. Under a small, mutually incommensurate bias the adiabatic
response $\langle\partial_{\theta_R}H\rangle=\partial_{\theta_R}E_{\rm gs}-\hbar\,\Omega_{RB}\,\dot\theta_B$, with
$\Omega_{RB}=-2\,\mathrm{Im}\langle\partial_{\theta_R}\Psi|\partial_{\theta_B}\Psi\rangle$ the Berry curvature of the
many-body ground state and $\dot\theta_B=2eV_B/\hbar$, pumps one Cooper pair per cycle per unit Chern number, so that
$\partial\langle I_R\rangle/\partial V_B=-(4e^2/h)\,C_{RB}$ in the convention of Eq.~\eqref{eq:supercurrent}, quantized
in units of $4e^2/h$ \cite{riwar2016multi,eriksson2017topological,meyer2017nontrivial}. The Berry curvature of the
slice at $\theta_T=\pi/2$ concentrates near the projection of the enclosed Weyl point and integrates to $C_{RB}=-1$
[Fig.~\ref{fig:topology}(c)]. A real-time simulation of the same lattice model at $\mu=t$ and $\Delta=1.6t$ (chosen to
widen the Andreev gaps), with $\theta_B$ swept at $\omega=0.003\,t/\hbar$ at $\theta_T=0.73\pi$, the equilibrium current
subtracted and the pumped charge averaged over twelve values of $\theta_R$, gives $0.991$ Cooper pairs per cycle for
$C_{RB}=-1$. The adiabatic prediction on the same twelve phases is $0.998$; halving the time step from
$0.05$ to $0.025\,\hbar/t$ changes the finite-speed mean by $5.8\times10^{-6}$ Cooper pairs. A single trajectory at fixed
$\theta_R$ is not quantized. Finite sweep speed and phase quadrature limit this comparison; nonadiabatic transitions
become important at faster sweeps or narrower gaps, so the numerical result is restricted to the stated parameters.

Fermion-parity switches are distinct from these Weyl touchings. At a generic phase point, a spin--orbit junction
is in class D and a spin-degenerate junction in class C \cite{van2014single}. Class-D odd-parity domains have
codimension-one boundaries, observed as curves in three-terminal Al/InAs spectroscopy \cite{coraiola2024spin}.
The class-C Weyl nodes considered here have codimension three and do not change many-body parity. No parity
boundary is therefore present in our spin-degenerate model. Contour smoothness still depends on folds and stability
of the current map, so a gapped spectrum alone does not establish it.

The closed lattice model has a $3\times3$-site normal block with two-site superconducting pads on
each contacted side, hopping $t=1$, $\Delta=0.4t$, on-site energy $4t-\mu$, and a Landau-gauge Peierls phase with flux
$f$ per plaquette in units of $h/e$; the flux through the four plaquettes of the normal block is $\Phi_N=8f\,\Phi_0$.
For three terminals we use $\mu\in\{1.5,2.5,3.5,4.5,5.5\}\,t$, $101$ flux values $f\in[0,0.5]$ and a $20\times20$
phase grid for the Chern number, independently rechecked at 40 points per axis and at 80 where the Chern value or
local mesh diagnostics did not match the prespecified comparison. For four terminals we use 48 values of
$\theta_T$, independently rechecked on 44-point slice grids and on 88-point grids where required. Known gapless
slices are omitted, and unresolved mesh checks are withheld.
Of 505 three-terminal flux samples and 96 four-terminal slices, 0 and 1, respectively, are withheld as gapless or unresolved. The remaining local mesh checks pass, with 0 Chern values changed from the base grids. Flagged cases are refined through 160 and, where needed, 320 points per axis; these are numerical mesh checks. The eight listed nodes are refined by local gap minimization and checked by direct flux through
six cube faces (half-width $0.08$ rad, 12 and 24 plaquettes per edge), with residual flux below $10^{-8}$,
maximum plaquette phase below $\pi/3$ and minimum overlap singular value above $0.1$.
The Berry-curvature map uses a $40\times40$ grid. The torus minigap of the two illustrative devices
(Route A) is searched on an $81\times81$ grid, with SVD-based Nelder--Mead refinement from every grid-local minimum and the previously located seeds, at all 18 displayed device--flux points. These searches do not provide a positive global lower bound.

\section{Disorder studies}
\label{app:disorder-ensemble}

\subsection{Disorder-enhanced current corrections}

\begin{figure*}[t]
 \includegraphics[width=\textwidth]{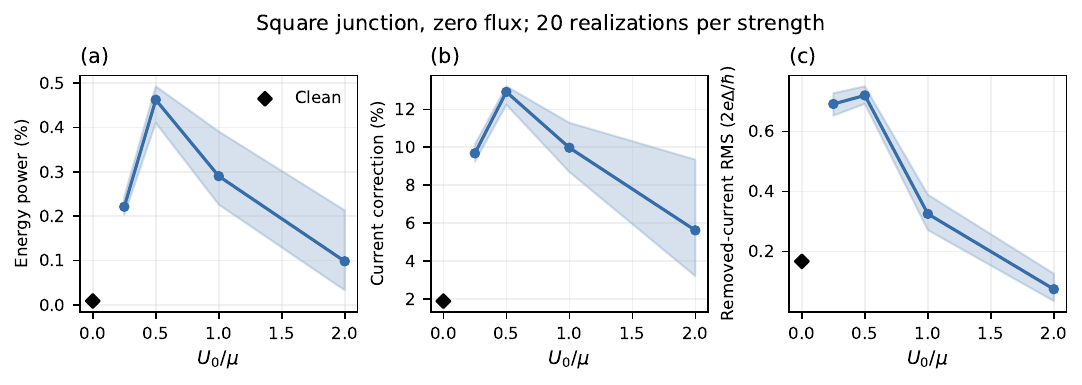}
 \caption{Energy dominance and current corrections in the square junction at zero flux.
 (a) Irreducible energy-power share; (b) RMS current error after pairwise projection,
 Eq.~\eqref{eq:disorder-current-error}; (c) the absolute RMS removed current in units of $2e\Delta/\hbar$.
 Blue points show medians over twenty disorder realizations at each strength; shading is the 10--90\%
realization range.
 Black diamonds are the clean reference. These current-map diagnostics do not measure contour displacement.}
 \label{fig:disorder-currents}
\end{figure*}

For the two-device disorder ensemble specified in Appendix~\ref{sec:disorder}, we project the phase-resolved currents onto the pairwise sectors and measure $\epsilon_I$ of Eq.~\eqref{eq:disorder-current-error}.
At zero flux in the square device, $\epsilon_I$ rises from 1.88 percent in the clean reference to
a median 12.93 percent at $U_0/\mu=0.5$, then falls at the larger tested strengths
[Fig.~\ref{fig:disorder-currents}]. The median absolute numerator increases 4.3-fold relative to
the clean reference, so the larger fraction is not explained solely by suppression of the total current.
The largest sampled ensemble correction is 13.58 percent RMS. Differentiation weights each energy harmonic
by its phase index, so an energy-power fraction is neither a current-amplitude fraction nor a contour-error bound.
The two geometries and sampled strengths do not determine a universal crossover or the mechanism of the
nonmonotonicity. These RMS corrections alone do not establish changes in stable or switching contours;
the selected region comparisons in Table~\ref{tab:pairwise-regions} are a separate test.

The current analysis uses $41$- and $81$-point grids per phase for all realizations, with $161$-point
checks on selected cases and targeted refinements. The final RMS change must be no larger than
$\max(0.002,0.05\epsilon_I)$. Clean zero-field grids intersect gapless lines; independent shifted
$81$, $161$ and $321$ grids avoid those lines. Direct marginal quadrature at 96 fixed off-grid phases in twelve
controls uses up to 513 points and a normalized prediction-change tolerance of 0.002. These are finite refinement tests.

\subsection{Which harmonics carry the disorder correction?}
\label{sec:harmonic-mechanism}

We resolve the current correction directly into phase harmonics.
Write $E/\Delta=\sum_{m,n}e_{mn}e^{i(m\theta_R+n\theta_B)}$. In units $I_0=2e\Delta/\hbar$,
Parseval's identity gives the squared norm of the irreducible current as
\begin{equation}
 P_{\rm irr}=\sum_{mn(m+n)\ne0}(m^2+n^2)|e_{mn}|^2.
 \label{eq:weighted-irreducible-power}
\end{equation}
The norm here sums the R and B components, as in the preceding RMS comparisons.
The lowest three-terminal charge vectors are permutations of $(2,-1,-1)$ and their negatives.
With L grounded they give the three conjugate mode pairs
$\mathcal Q=\{\pm(1,1),\pm(2,-1),\pm(1,-2)\}$, the equilibrium quartet harmonics
\cite{ohnmacht2024quartet}. Their presence denotes a contribution from multiterminal irreducible harmonics;
it is not, by itself, a unique identification of microscopic paths (Sec.~\ref{sec:intro}).

\begin{figure*}[t]
 \includegraphics[width=\textwidth]{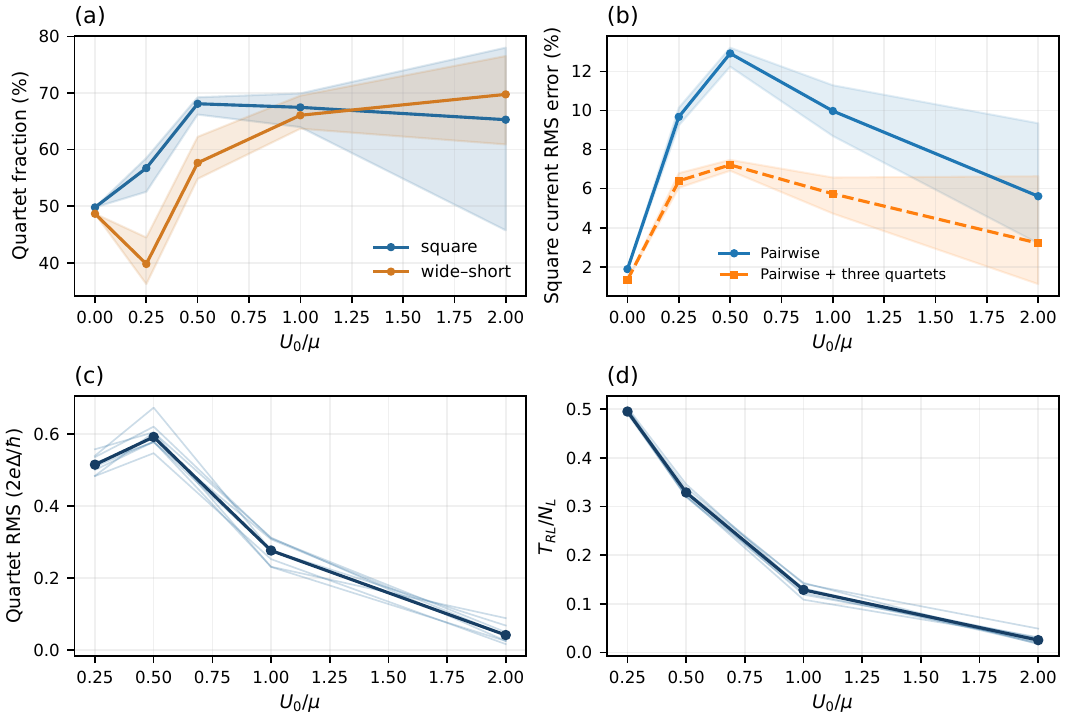}
 \caption{Mode-resolved disorder response at zero flux. (a) Fraction of omitted R/B current power in the
 three quartet mode pairs $\mathcal Q$, as medians and 10--90-percent realization ranges for 20 independent
 patterns per strength and geometry; the zero-disorder points are single clean controls.
 (b) Square-device median current RMS error and 10--90-percent realization ranges before and after restoring those modes.
 (c) Quartet-current RMS for eight additional matched spatial patterns, each scaled through all four
 disorder strengths; thin lines show individual patterns and the heavy line their median.
 (d) Normal transmission for the same patterns. Panels (c,d) use a distinct ensemble from (a,b).
 The mode correction is a phase-resolved current diagnostic.}
 \label{fig:disorder-harmonics}
\end{figure*}

Across the 20 square-device realizations at $U_0/\mu=0.5$, these six Fourier modes carry a median
68.1 percent of the omitted current power (bootstrap 95-percent interval for the median
67.5--68.4 percent), compared with 49.8 percent
in the clean control. The last grid refinement changes that ensemble median by
0.008 percentage points and any individual fraction by at most
0.469 percentage points; the bootstrap interval describes realization sampling
and does not include this numerical variation. Restoring only these modes to the pairwise current lowers the median current RMS
error from 12.93 to 7.22 percent [Fig.~\ref{fig:disorder-harmonics}(a,b)].
Thus the increased correction is substantially carried by the lowest multiterminal irreducible harmonics, although
higher harmonics remain necessary for quantitative current accuracy. This current-space projection does
not establish the accuracy of its Hessian, stable region, or switching boundary.
We analyze every zero-field realization of both geometries.
Energy-derived and directly evaluated current coefficients are checked independently;
25 cases required additional phase-grid refinement for these observables
(Appendix~\ref{app:strengthening}).

A second calculation holds the spatial random potential fixed while scaling its amplitude, using eight
further disorder patterns in the square device and the same four disorder strengths. The quartet-current RMS
increases between $0.25\mu$ and $0.5\mu$ in 8 of eight patterns and is smaller at $2\mu$
than at $0.5\mu$ in 8 of eight. Its medians are 0.515,
0.592, 0.276, and 0.041 in units $I_0$.
Over the same disorder series the median normal transmission $T_{RL}/N_L$ falls through 0.495,
0.329, 0.129, and 0.026, decreasing at every step in
8 of eight patterns [Fig.~\ref{fig:disorder-harmonics}(c,d)].
The nonmonotonic harmonic amplitude therefore survives a controlled change of disorder amplitude within
individual patterns, and cannot be explained solely by normalization to a smaller total current.
Normal transmission suppression alone also does not give a monotone predictor of the quartet amplitude.
These tests identify the leading phase-dependent contributions and their response to disorder;
they do not isolate a unique microscopic interference pathway or establish universal disorder scaling.

\subsection{Matched disorder ensembles}
\label{app:matched-ensembles}

To distinguish sample-dependent changes from typical response, we fix
256 independent onsite-disorder maps per device before evaluating the ensemble.
The devices are $120\times120$ and $180\times90$ nm rectangles with
$n=3,4$, $a=5$ nm, $\mu=20$ meV and 40 nm contacts, giving two channels per
terminal. Ideal leads remain clean. The onsite potential is uniform in
$[-U_0/2,U_0/2]$; each dimensionless spatial map is held fixed as
$U_0/\mu$ takes $0.25,0.5,1,2$. The same random map on common sites also
pairs the terminal-count controls. Device classes and strengths are not
pooled as independent samples. The four maps illustrated in Sec.~\ref{sec:multiterminal} are excluded from this ensemble.

These 4096 disordered calculations and four clean references use the
zero-temperature, energy-independent-$S(0)$ model. All scheduled cases
must satisfy the numerical checks; no disorder sample is selected by its
current amplitude or removed because of its result. Phase-grid
refinement, shifted grids, the differentiated-energy route and independent
Sobol validation test the pairwise and charge-two currents at a normalized
current discrepancy of 0.002; energy-residual fraction changes are tested
at 0.0002. The charge-two sector contains exactly the multiterminal irreducible harmonics of transferred-pair order two.

\begin{table*}[t]
\caption{Fixed matched disorder ensembles, with 256 independent realizations per device
evaluated at every strength. $A_2$ is irreducible charge-two current RMS; brackets in its column give the empirical 10--90\% sample range. Enhancement percentages compare with the corresponding single clean reference; their brackets are simultaneous 95\% Bonferroni intervals for the 16 enhancement probabilities, expanded by observed numerical sensitivities. These sensitivities are diagnostics. Current-error columns are ensemble medians.}
\label{tab:disorder-ensemble}
\begin{ruledtabular}
\begin{tabular}{ccclcc}
Device & $U_0/\mu$ & $A_2/A_2^{\rm clean}$ & Enhanced (\%) & Pair error (\%) & Corrected (\%) \\
Square, $n=3$ & 0.25 & 0.74 [0.38, 1.36] & 29.3 [21.3, 38.3] & 14.4 & 5.4\\
Square, $n=3$ & 0.5 & 0.47 [0.19, 1.07] & 13.7 [8.1, 21.1] & 10.4 & 3.5\\
Square, $n=3$ & 1 & 0.35 [0.12, 0.94] & 8.2 [4.0, 14.5] & 8.0 & 2.5\\
Square, $n=3$ & 2 & 0.42 [0.11, 1.11] & 14.8 [9.0, 22.4] & 9.5 & 3.6\\
\hline
Square, $n=4$ & 0.25 & 1.23 [0.44, 2.09] & 57.4 [48.0, 66.5] & 19.7 & 9.9\\
Square, $n=4$ & 0.5 & 1.11 [0.23, 1.84] & 54.7 [45.3, 63.9] & 17.7 & 8.5\\
Square, $n=4$ & 1 & 0.80 [0.24, 1.47] & 37.5 [28.8, 46.9] & 13.7 & 5.9\\
Square, $n=4$ & 2 & 0.76 [0.34, 1.52] & 32.8 [24.4, 42.0] & 14.7 & 6.3\\
\hline
Rectangle, $n=3$ & 0.25 & 0.95 [0.59, 1.25] & 44.5 [35.4, 53.9] & 18.6 & 11.8\\
Rectangle, $n=3$ & 0.5 & 0.77 [0.34, 1.24] & 26.2 [18.5, 35.0] & 15.6 & 9.1\\
Rectangle, $n=3$ & 1 & 0.56 [0.22, 1.06] & 12.5 [7.2, 19.7] & 13.2 & 7.5\\
Rectangle, $n=3$ & 2 & 0.51 [0.19, 1.02] & 12.5 [7.2, 19.7] & 13.6 & 6.1\\
\hline
Rectangle, $n=4$ & 0.25 & 1.00 [0.93, 1.08] & 51.2 [41.4, 60.5] & 20.8 & 12.7\\
Rectangle, $n=4$ & 0.5 & 1.01 [0.86, 1.18] & 52.3 [43.0, 61.6] & 20.9 & 12.6\\
Rectangle, $n=4$ & 1 & 0.98 [0.70, 1.38] & 48.0 [38.8, 57.4] & 21.2 & 11.9\\
Rectangle, $n=4$ & 2 & 0.94 [0.49, 1.41] & 44.1 [35.0, 53.6] & 20.3 & 10.5\\
\hline
\end{tabular}
\end{ruledtabular}
\end{table*}

\begin{figure*}[t]
 \centering
 \includegraphics[width=\textwidth]{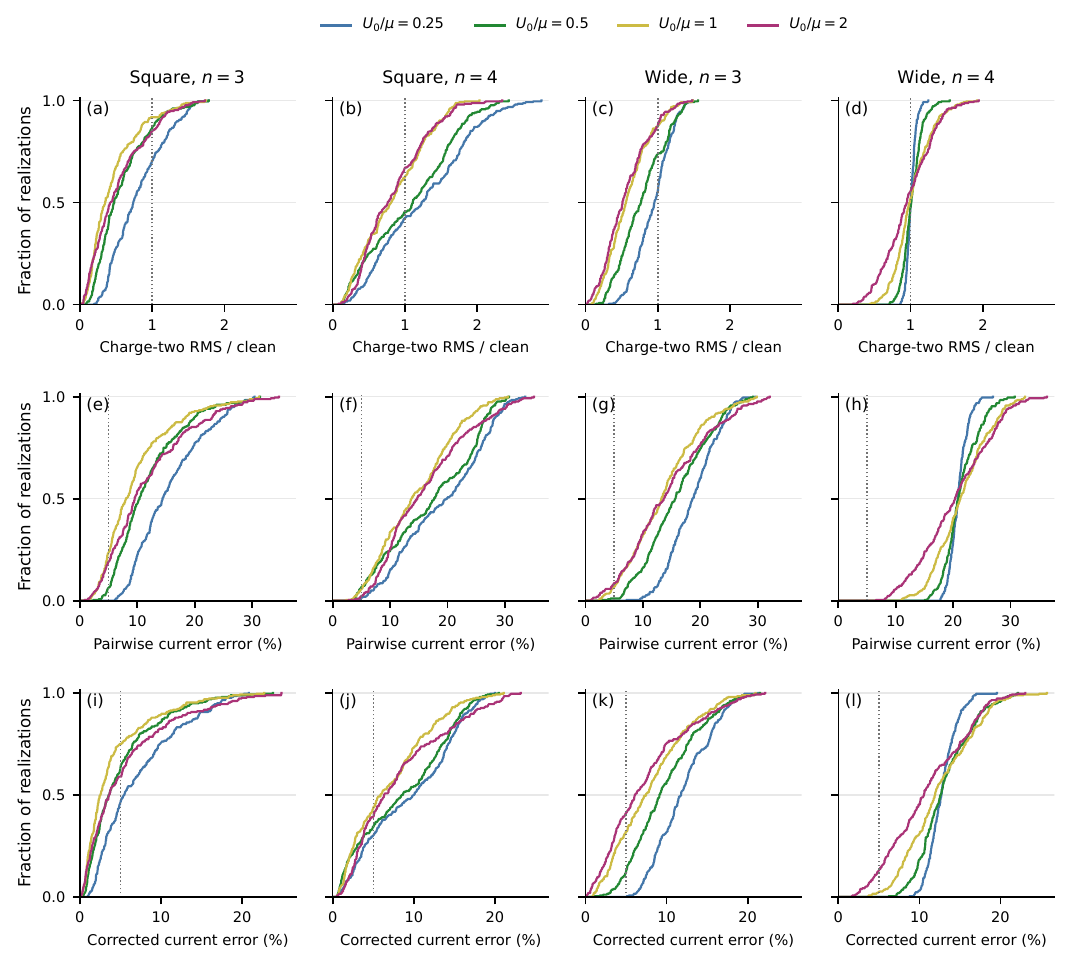}
 \caption{Matched disorder distributions, with 256 independent realizations
 per device and the same map followed across strengths. Columns show the square
 three-terminal, square four-terminal, wide three-terminal and wide four-terminal
 devices. Colors give the disorder strength. Each curve is the empirical fraction
 of realizations at or below the horizontal-axis value, with one step per sample
 and no density smoothing. (a)--(d) Irreducible charge-two current RMS relative to
 the corresponding clean reference. The vertical line at one separates suppression
 and enhancement. (e)--(h) All-terminal phase-current RMS error of the arbitrary-pair
 model. (i)--(l) Error after restoring all charge-two multiterminal irreducible harmonics. Dotted lines
 in the lower two rows mark the prespecified 5 percent numerical benchmark, not an experimental acceptance threshold. Table~\ref{tab:threshold} gives
 the threshold comparisons and Table~\ref{tab:disorder-ensemble} gives separate
 sampling uncertainty. The compared observables are phase-current RMS errors; critical and switching currents require their own diagnostics.}
 \label{fig:disorder-ensemble}
\end{figure*}

Empirical medians and 10--90\% ranges describe sample variability; exact
binomial/order-statistic intervals separately describe sampling uncertainty.
For enhancement relative to clean, Clopper--Pearson intervals use a
Bonferroni correction over the sixteen device/strength probabilities.
Observed numerical sensitivities of both the sample and clean reference
bracket ambiguous classifications, which remain in the denominator.
The resulting interval expansions are conditional sensitivity analyses. Other intervals are pointwise.
Matched differences compare the same seed across strengths and the same
full-current data before and after the charge-two correction.

Across the sixteen device/strength groups (Fig.~\ref{fig:disorder-ensemble} and
Table~\ref{tab:disorder-ensemble}), median pairwise current errors
range from 8.0 to 21.2 percent, compared with
2.5 to 12.7 percent after the charge-two correction.
Median energy-power remainders range from 0.2 to 1.3 percent;
on an RMS basis, the median energy errors range from 4.5 to 11.3 percent.
The fraction above 5 percent pairwise error ranges from 76.6 to
100.0 percent; the fraction below 5 percent after correction ranges
from 0.0 to 75.0 percent. These are empirical fractions for the stated ensembles.

Some prespecified 128-to-256 prefix-stability diagnostics fail, leaving ensemble-size sensitivity
in the affected summaries. We report their fixed-sample uncertainty intervals and full empirical distributions.

\begin{figure*}[t]
 \includegraphics[width=\textwidth]{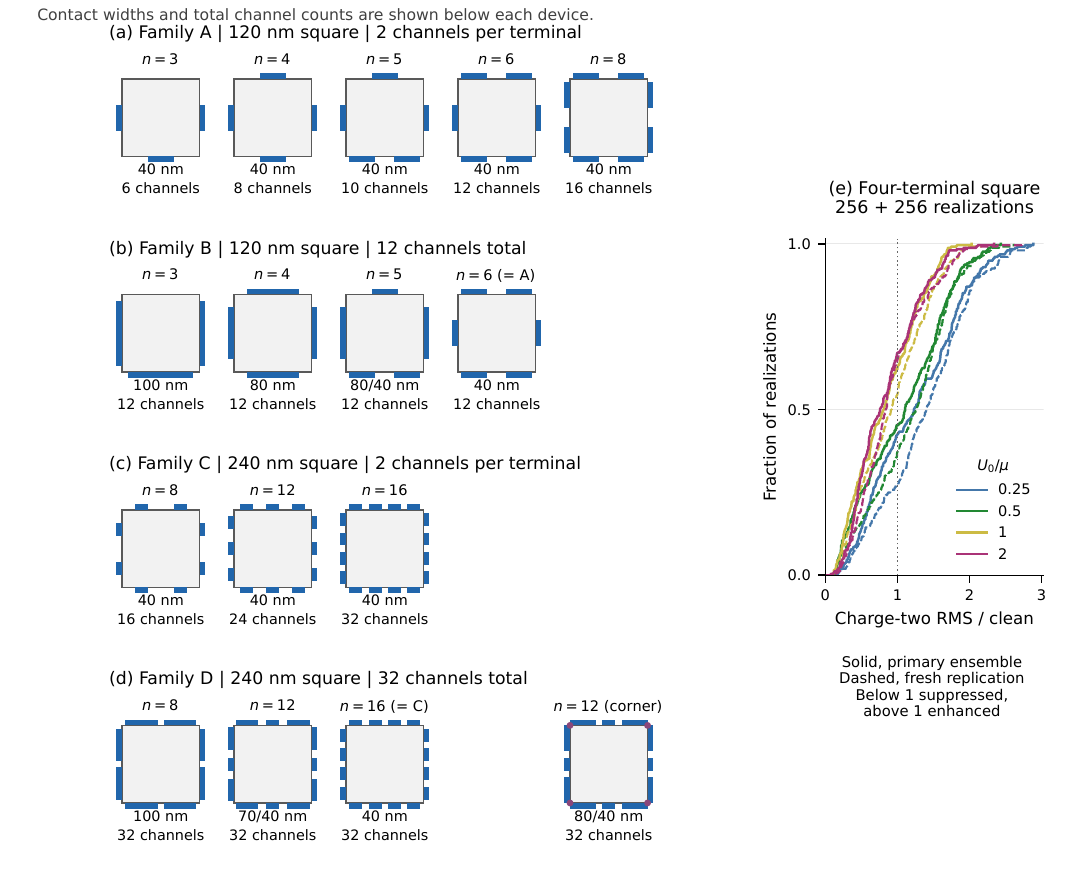}
 \caption{Contact geometries for all four families and separate disorder controls.
 (a,b) Families A and B on a 120 nm square. A keeps two channels per terminal,
 while B keeps twelve channels in total. (c,d) Families C and D on a 240 nm
 square. C keeps two channels per terminal, while D keeps 32 channels in total.
 Blue segments show superconducting contacts; widths and total channel counts
 appear below each device. When two widths are listed, the larger applies to
 the broader contacts shown. At $n=6$, B coincides with A; at $n=16$, D coincides
 with C. D includes both the main twelve-terminal control with separated
 interfaces (70/40 nm widths) and the additional corner-sharing control
 (80/40 nm widths); purple dots mark its four shared corner sites.
 Contacts are drawn to scale within each square, but the 120 and 240 nm
 devices use different physical scales. (e) Empirical cumulative distributions
 of irreducible charge-two current RMS divided by the clean value for the
 four-terminal A device. Each curve includes 256 matched disorder realizations
 at the indicated strength; solid curves show the primary ensemble and dashed
 curves an independent 256-realization replication. The same random potential
 map is scaled across strengths within each ensemble. Values below or above
 one indicate suppression or enhancement relative to clean. These curves show sample distributions; statistical intervals and the
 complete three-/four-terminal comparison are in
 Appendix~\ref{app:disorder-ensemble}. This disorder ensemble is not extended to the eight-, twelve- or sixteen-terminal devices.}
 \label{fig:multiterminal-controls}
\end{figure*}

The two primary median diagnostics that are not met occur in the square
four-terminal junction at $U_0/\mu=0.25$ and $2$. Before evaluating any of its maps, we fixed a separate
256-realization replication of that device at all four strengths
(1024 additional disordered calculations). Its seed set is disjoint from
the primary ensemble. Replication estimates (Table~\ref{tab:disorder-replication}) use the replication data separately;
the 512 samples remain in their original ensembles for fixed-sample inference.
Figure~\ref{fig:multiterminal-controls}(e) displays both ensembles separately.
The replication enhancement intervals use a family of four comparisons.

\begin{table*}[t]
\caption{Independent square four-terminal replication, 256 independent matched
realizations at every strength. The amplitude column gives its median
clean-normalized charge-two RMS and pointwise 95\% median confidence interval.
Enhancement intervals are simultaneous 95\% Bonferroni intervals across these
four replication probabilities. Both interval types are expanded by observed
numerical sensitivities, with the same qualification as
Table~\ref{tab:disorder-ensemble}. Error columns are medians.}
\label{tab:disorder-replication}
\begin{ruledtabular}
\begin{tabular}{cclcc}
$U_0/\mu$ & $A_2/A_2^{\rm clean}$ & Enhanced (\%) & Pair error (\%) & Corrected (\%) \\
0.25 & 1.40 [1.27, 1.50] & 72.7 [65.1, 79.4] & 21.8 & 11.4 \\
0.5 & 1.26 [1.13, 1.35] & 62.9 [55.0, 70.3] & 19.6 & 9.4 \\
1 & 0.91 [0.81, 1.02] & 45.3 [37.5, 53.3] & 15.3 & 6.9 \\
2 & 0.84 [0.77, 0.89] & 34.8 [27.5, 42.6] & 16.1 & 7.2 \\
\end{tabular}
\end{ruledtabular}
\end{table*}

The primary-versus-replication median diagnostics are not met at $U_0/\mu=0.25$, $0.5$
and $1$, nor is the enhancement-fraction diagnostic at $0.25$, where the empirical fraction changes from $147/256$ (57.4 percent) to $186/256$ (72.7 percent).
The replication enhancement intervals exclude one half at $0.25$ and $0.5$,
whereas the primary simultaneous intervals do not. The separate estimates and uncertainties display this variation between ensembles.
All replication 128-to-256 prefix diagnostics pass.

A comparability audit of all 2048 square four-terminal records finds the
same numerical source versions, geometry, contact and model parameters,
refinement rules, recorded software environment, and clean-reference
normalization in the two ensembles. The uniform onsite-disorder law and
seed construction also agree. The disjoint seed sets yield 512 distinct
recorded disorder maps, each held fixed across the four strengths;
an audit reproduces the maps and currents at eight boundary-seed
and strength combinations. The tested configurations, normalization, map construction and replay agree.
These implementation checks are distinct from statistical-independence assumptions and shared-code uncertainty.
The 15.2-percentage-point enhancement-fraction difference at $U_0/\mu=0.25$
remains unexplained. Since this device was selected for replication after
examining the primary stability results, the comparison is a conditional reproducibility diagnostic.
The two fixed ensembles remain separate for inference, with no additional outcome-dependent resampling.
Their numerical checks and sampling uncertainties address different sources of error.

These distributions characterize the specified finite-lattice, frozen-$S$ disorder ensembles.
Finite-gap response, continuum convergence and fabrication-specific disorder require separate validation;
the higher-terminal controls characterize selected devices rather than statistical trends at $n>4$.

Table~\ref{tab:disorder-diagnostics} lists the prespecified diagnostics with their limits and outcomes. The limits
were fixed before the ensembles were computed; the prefix comparisons are descriptive and were never used as stopping
rules. Between the primary and replication ensembles of the square four-terminal device, the pairwise-failure
fractions and the 10th and 90th percentiles agree within the limits at all four strengths; the medians at
$U_0/\mu=0.25$, $0.5$ and $1$ and the enhancement fraction at $0.25$ do not. Each group's 256 maps are held fixed
across the four strengths, so the 4096 calculations comprise 1024 device--map combinations; the three- and
four-terminal devices of each geometry share the maps on common sites.

The energy-independent model evaluates the normal scattering matrix at $E=0$. We assess its regime device
by device with two gauge-invariant indicators computed from the same lattice and disorder map. The first is the
dwell-time matrix $D_{ab}=\langle\psi_a|\psi_b\rangle$ of scattering states inside the normal region at $E=0$
\cite{choi2013dwell}. The states carry unit incident flux in the $\hbar=1$ normalization, so $\Delta$ times an
eigenvalue gives $\Delta\tau/\hbar$. Its trace reproduces the summed traversal times of a clean strip to machine precision. We quote
$\Delta\tau/\hbar$ for its largest eigenvalue and for the mean eigenvalue. The second is the largest change of a
terminal transmission probability between $E=0$ and $E=\pm\Delta$, per channel of the smaller terminal.
These indicators diagnose in-region dwell and transmission dispersion without bounding the full energy-dependent
current correction. In the 4096
ensemble devices, $\Delta\tau_{\max}/\hbar$ has median 0.77 with 10th and 90th percentiles 0.34 and 1.74,
$\Delta\tau_{\rm mean}/\hbar$ has median 0.25 with percentiles 0.15 and 0.44, and the transmission change
has median 0.083 with 90th percentile 0.20. The designed devices of Sec.~\ref{sec:multiterminal} have
dwell medians of 0.50 and 0.15. The clean three-terminal contour devices are closer to the
short-junction regime, with medians of 0.34 and 0.10 over 34 geometries, as are the 160 disordered
three-terminal devices of this appendix, with 0.35 and 0.07; the planar controls of
Appendix~\ref{app:planar-controls} give 0.42 and 0.13. The required separation $\Delta\tau/\hbar\ll1$ is absent
for much of the survey, whose statistics therefore characterize the energy-independent scattering model. Across the sixteen
groups the Spearman correlation between $\Delta\tau_{\max}/\hbar$ and the pairwise current error ranges from
\mbox{$-$}0.28 to 0.74 (\mbox{$-$}0.06 for the pooled ensemble), so long dwell times alone do not account for the larger errors.

Finite-gap calculations with dispersive BCS leads test the consequence directly for twelve ensemble devices selected
in two stages using their energy-independent errors (Table~\ref{tab:finitegap-ensemble}). In each geometry they are the
realizations whose energy-independent errors lie closest to the 10th and 90th percentiles of their group for $n=3,4$
at $U_0/\mu=1$, and closest to the median for $n=4$ at $U_0/\mu=0.25$ and $2$. The finite-gap pairwise errors differ
from the energy-independent ones by \mbox{$-$}0.3 to 4.3 percentage points, while the RMS currents fall by
20--39 percent. The largest change occurs in the weakly disordered square device with the longest dwell time. Doubling the frequency
quadrature changes these currents by less than $10^{-11}$ and retaining all anomalous channels by less than
$10^{-4}$, in relative terms.

Table~\ref{tab:paired-disorder} uses the matching of the maps directly. On the same maps, the four-terminal device has the
larger pairwise current error at every strength in both geometries, with median paired differences of
1.7--8.2 percentage points whose 95-percent bootstrap intervals exclude zero; it is larger for 65--85
percent of the maps. The paired strength steps are nonmonotonic. From $U_0/\mu=0.25$ to $1$ the error falls at both
steps in three classes and changes by less than $0.3$ percentage points in the wide four-terminal class; from $1$ to
$2$ it rises in both square classes and falls in the wide four-terminal class. A two-sample Kolmogorov--Smirnov
comparison of the primary and replication current errors of the square four-terminal device gives statistics $D$ of 0.09--0.14.
Only the comparison at $U_0/\mu=0.25$ has $p<0.05$ ($p=0.013$, before any correction for the four
comparisons; $p$ of 0.07--0.25 at the other strengths), and the replication medians are higher by
1.4--2.0 percentage points at every strength. Because each map enters all four strengths, these comparisons
are correlated. Exploratory permutation tests that exchange whole maps between the two sets give $p$ of 0.06--0.18 for the
individual strengths, 0.03 for their sum and 0.006 for four shifts sharing a sign and each having magnitude
at least the smallest observed shift, 1.39 percentage points. The pairwise current error correlates with total normal transmission, with
Spearman correlations of 0.55--0.89 across strengths; the replication maps transmit more (Mann--Whitney $p$ of
0.004--0.07). An exploratory least-squares regression including transmission reduces the estimated mean difference
between sets from 0.7--1.5 to \mbox{$-$}0.29 to 0.06 percentage points. This association is consistent with an imbalance in
transmission between the sampled map sets. Dwell-time comparisons do not detect a difference at the five-percent
level ($p\ge0.05$). The origin of the replication mismatch remains uncertain.

Near Andreev zero modes the level currents are ill conditioned, with condition number $\Delta/(2E_{\min})$. For
24 ensemble devices, the sixteen group-median realizations and eight selected by the earlier derivative-test
gap statistic, the
Hellmann--Feynman currents at the 64 lowest-gap and 64 smallest-current points of 4096 phases agree with central
differences of the energy within $1.3\times10^{-7}$~nA, at gaps down to 0.012$\Delta$ and condition numbers up to
43. The selection statistic uses the first twelve test phases with gaps above $0.05\Delta$.
The smallest sampled currents are 0.18~nA. Exact zero modes and current zeros remain outside this check.

\begin{table}[t]
\caption{Finite-gap check of selected ensemble devices. Pairwise current errors in the energy-independent model
and with dispersive BCS leads, the fall of the RMS current, and the largest dwell time.}
\label{tab:finitegap-ensemble}
\begin{ruledtabular}
\begin{tabular}{lcccccc}
Geometry & $n$ & $U_0/\mu$ & \shortstack{$S(0)$\\(\%)} & \shortstack{Finite gap\\(\%)} & \shortstack{Current\\fall (\%)} & $\Delta\tau_{\max}/\hbar$\\
\hline
Square & 4 & 1 & 6.2 & 7.1 & 26 & 0.59\\
Square & 4 & 1 & 23.7 & 24.9 & 31 & 1.32\\
Wide & 4 & 1 & 15.6 & 17.1 & 21 & 0.57\\
Wide & 4 & 1 & 28.0 & 30.3 & 23 & 0.58\\
Square & 3 & 1 & 3.6 & 4.0 & 26 & 0.40\\
Square & 3 & 1 & 18.4 & 18.1 & 39 & 1.53\\
Wide & 3 & 1 & 5.7 & 6.5 & 29 & 0.40\\
Wide & 3 & 1 & 21.8 & 23.0 & 25 & 0.85\\
Square & 4 & 0.25 & 19.7 & 24.0 & 36 & 1.90\\
Wide & 4 & 0.25 & 20.8 & 23.5 & 20 & 0.34\\
Square & 4 & 2 & 14.7 & 16.4 & 25 & 0.55\\
Wide & 4 & 2 & 20.3 & 23.2 & 26 & 0.54\\
\end{tabular}
\end{ruledtabular}
\end{table}

\begin{table}[t]
\caption{Paired changes of the pairwise current error in the matched ensemble, in percentage points. Each entry is
the median over the 256 maps of the paired difference, later minus earlier strength or four minus three terminals on
the same map, with a 95-percent bootstrap interval from 10000 resamples of maps.}
\label{tab:paired-disorder}
\begin{ruledtabular}
\begin{tabular}{lcc}
Comparison, $U_0/\mu$ & Square & Wide\\
\hline
$n=3$, $0.25\to0.5$ & $\mbox{$-$}3.5$ [\mbox{$-$}3.8, \mbox{$-$}3.2] & $\mbox{$-$}2.8$ [\mbox{$-$}3.3, \mbox{$-$}2.4]\\
$n=3$, $0.5\to1$ & $\mbox{$-$}1.9$ [\mbox{$-$}2.6, \mbox{$-$}1.3] & $\mbox{$-$}2.5$ [\mbox{$-$}3.1, \mbox{$-$}1.9]\\
$n=3$, $1\to2$ & ${+}1.7$ [{+}0.5, {+}2.5] & ${+}1.0$ [\mbox{$-$}0.8, {+}2.2]\\
$n=4$, $0.25\to0.5$ & $\mbox{$-$}2.0$ [\mbox{$-$}2.5, \mbox{$-$}1.5] & ${+}0.3$ [0.0, {+}0.4]\\
$n=4$, $0.5\to1$ & $\mbox{$-$}2.2$ [\mbox{$-$}3.1, \mbox{$-$}1.8] & $0.0$ [\mbox{$-$}0.4, {+}0.3]\\
$n=4$, $1\to2$ & ${+}1.4$ [{+}0.6, {+}2.5] & $\mbox{$-$}1.1$ [\mbox{$-$}1.9, \mbox{$-$}0.1]\\
\hline
$n=4$ minus 3, $0.25$ & ${+}2.8$ [{+}1.8, {+}3.9] & ${+}1.7$ [{+}0.9, {+}3.1]\\
$n=4$ minus 3, $0.5$ & ${+}4.8$ [{+}2.6, {+}5.9] & ${+}5.4$ [{+}4.1, {+}6.2]\\
$n=4$ minus 3, $1$ & ${+}4.1$ [{+}3.3, {+}5.0] & ${+}8.2$ [{+}7.0, {+}9.0]\\
$n=4$ minus 3, $2$ & ${+}3.7$ [{+}3.0, {+}5.2] & ${+}5.8$ [{+}4.5, {+}6.8]\\
\end{tabular}
\end{ruledtabular}
\end{table}

\begin{table*}[t]
\caption{Prespecified disorder-ensemble diagnostics, their limits and outcomes. Prefix stability compares
the first 128 with all 256 realizations of each group. Fractions are the enhancement and pairwise-failure fractions;
quantiles are the median $m$ and the 10th and 90th percentiles $q$ of $A_2/A_2^{\rm clean}$, evaluated on the full
ensemble. Changes and limits are absolute.}
\label{tab:disorder-diagnostics}
\begin{ruledtabular}
\begin{tabular}{p{0.2\textwidth}p{0.15\textwidth}p{0.17\textwidth}p{0.38\textwidth}}
Diagnostic & Statistics & Limit & Checks passed; failures\\
\hline
Primary prefix stability (16 groups) & fractions & $0.10$ & 32 of 32\\
 & quantiles & $0.10\max(1,m)$; $0.20\max(1,q)$ & 46 of 48; square $n=4$, median at $U_0/\mu=0.25$, $0.132>0.123$; median at $U_0/\mu=2$, $0.137>0.100$\\
Replication prefix stability (square $n=4$) & fractions, quantiles & as above & 20 of 20\\
Primary versus replication (square $n=4$) & fractions & $0.10$ & 7 of 8; enhancement at $0.25$, $0.152>0.10$\\
 & quantiles & $0.10\max(1,m)$; $0.20\max(1,q)$ & 9 of 12; median at $0.25$, $0.173>0.140$; median at $0.5$, $0.156>0.126$; median at $1$, $0.108>0.100$\\
\end{tabular}
\end{ruledtabular}
\end{table*}

Table~\ref{tab:threshold} gives these fractions at thresholds of 2, 5 and 10 percent.

\begin{table}[t]
\caption{Threshold dependence of the matched-ensemble fractions for the sixteen groups of 256 realizations.
``Pairwise above'' is the percentage of realizations whose pairwise current error exceeds the threshold, ``medians
above'' the number of groups whose median error exceeds it, and ``corrected below'' the percentage whose error after
the charge-two correction falls below it. Ranges run over the groups.}
\label{tab:threshold}
\begin{ruledtabular}
\begin{tabular}{lccc}
Threshold & \shortstack{Pairwise\\above (\%)} & \shortstack{Medians\\above} & \shortstack{Corrected\\below (\%)}\\
\hline
2\% & 97.7--100.0 & 16 of 16 & 0.0--40.6\\
5\% & 76.6--100.0 & 16 of 16 & 0.0--75.0\\
10\% & 34.4--100.0 & 14 of 16 & 3.1--89.5\\
\end{tabular}
\end{ruledtabular}
\end{table}

\section{Terminal-count comparisons}
\label{app:terminal-tables}
\begin{figure*}[t]
 \includegraphics[width=\textwidth]{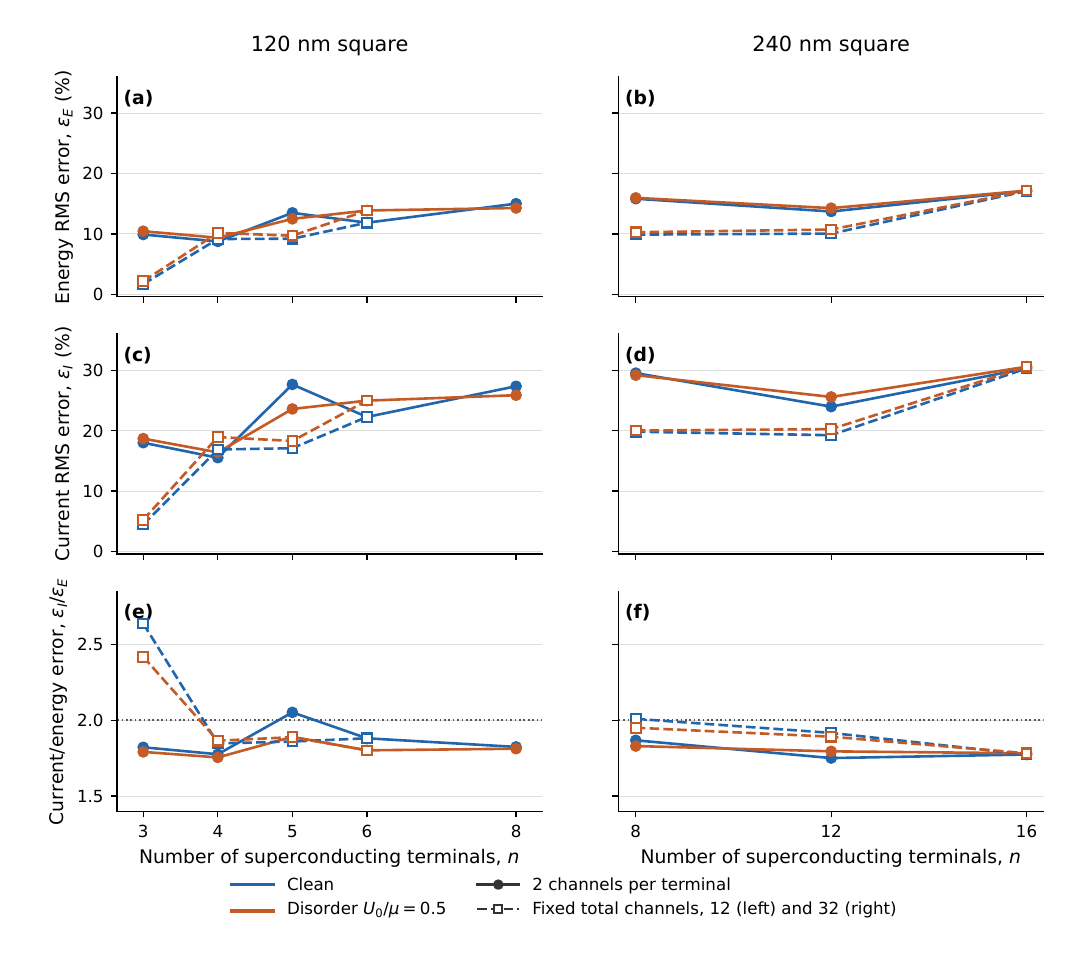}
 \caption{Energy dominance and current accuracy up to sixteen terminals.
 Columns keep the square size fixed at 120 nm (A/B) and 240 nm (C/D).
 (a,b) Energy RMS error $\epsilon_E$$=\sqrt{f_E}$. (c,d) Current RMS error $\epsilon_I$ using all
 terminal currents. (e,f) Their ratio $R=\epsilon_I/\epsilon_E$; the dotted $R=2$ line is a visual reference. Panels (a--d) use the same percentage scale; energy-power
 fractions are reported in the tables. Blue marks clean devices and orange one matched disorder pattern at
 $U_0/\mu=0.5$. Filled circles and solid lines keep two channels per terminal;
 open squares and dashed lines keep twelve total channels on the left and
 32 on the right. The coincident endpoints at $n=6$ (A/B) and $n=16$ (C/D)
 represent the same devices. The twelve-terminal D markers use the
 corner-separated contacts. Bars show numerical integration sensitivities; most are smaller than
 the symbols. The sensitivities in (a--d) are retained from the original comparison. Ratio bars in (e,f) use the largest changes formed from paired energy/current outputs under the available final quadrature and held-out refinements; they are not statistical confidence intervals.
 Lines connect results for the specified contact geometries.}
 \label{fig:multiterminal}
\end{figure*}

This appendix gives the numerical details of the terminal-count study of Sec.~\ref{sec:multiterminal}
(Fig.~\ref{fig:multiterminal} and Tables~\ref{tab:multiterminal} and \ref{tab:large-n}).

For the $n=3$--$6$ matrix, the finite pair approximation retains orders $k\le48$, increased to $96$
for the broad-contact clean three-terminal control. Coefficients are integrated
on two independently scrambled Sobol sequences, with $2^{14},2^{16},2^{18}$
phase points per sequence and recorded refinements to $2^{20}$ where needed.
An independently sampled, analytically integrated low-order control variate
reduces quadrature leakage; it does not discard physical harmonics.
All current and energy errors are evaluated on a separate $2^{15}$-point phase
sample. Synthetic pair and multiterminal energies, current conservation, gauge symmetry,
time reversal, finite-difference energy derivatives and an independent secular
energy calculation test the implementation. Direct conditional phase integration
provides a separate projection check for selected three-, four- and six-terminal
cases, including the largest current-error case.
Across these cases, the largest current prediction change on final
quadrature refinement is 0.50 percent of full-current RMS, the
independent-scramble difference is 0.66 percent and doubling the
Fourier cutoff changes the prediction by at most 0.31 percent.
The largest direct-projection discrepancy is 0.13 percent.
Refining the independent evaluation sample to $2^{17}$ points changes the selected RMS
errors by at most 0.04 percentage points. These are empirical sensitivity checks; across all designed
devices the largest recorded sample-size, scramble or cutoff sensitivity is 0.7 percentage points.
The designed devices carry total RMS currents of 5.5--134~nA, with median 39~nA, and nonpairwise RMS
currents of 0.04--26.6~nA, with median 9.7~nA. Finite-gap calculations with dispersive BCS leads for the two
eight-terminal family-A devices, using the same pair projection on $2^{16}$ Sobol phases, give pairwise errors of
28.1 and 26.7 percent, compared with 27.4 and 25.9 percent in the energy-independent model. Sample-size, scramble and cutoff
changes reach 0.8 percent of the current, comparable to that difference, and the RMS currents are
14--15 percent lower. A finite pair basis
gives an upper estimate of the irreducible residual; the reported cutoff and
quadrature tests measure sensitivity to the finite basis and integration.

For $n=8,12,16$, direct pair enumeration and trigonometric recurrence avoid
an exponential enumeration of unrelated charge vectors. Pair orders start at
$k\le16$ and are doubled when needed, reaching 32; two independent
Sobol sequences are refined up to $2^{22}$ points each. The same
acceptance thresholds apply. Every case is checked on a second, independent sequence, and a check outside tolerance triggers
refinement of both sequences. Selected clean and disordered C devices at each of these $n$ also receive direct
conditional integration at eight independent target phases, using scrambled Sobol
sequences or a shifted periodic lattice with exact low-order Fourier cancellation.
This check integrates
the microscopic residual after subtracting an independently sampled low-order
pair control, then adds that control's conditional integral exactly; it does
not use the final Fourier coefficients. For $n=8,12,16$, the largest prediction, independent-scramble and cutoff changes are 0.50,
0.53 and 0.27 percent of full-current RMS.
The direct-conditional discrepancy is at most 0.61 percent, and
the independently sampled current errors differ by at most 0.05
percentage points. Where the divide-and-conquer SVD does not converge, the QR-based LAPACK routine is used instead,
with reconstruction and independent secular-energy checks; no phase points are discarded.

The bars in Fig.~\ref{fig:multiterminal}(a--d) show the larger scalar-error change on final quadrature or on
refinement of the independent evaluation sample, where checked.
For panels (e,f), the ratio is recomputed from the paired energy and current errors at each recorded refinement, and its largest absolute change supplies the sensitivity bar. This preserves the dependence between the two errors, rather than propagating their scalar sensitivities independently. All 26 distinct plotted devices have a paired final projection-refinement comparison; 15 also have retained paired held-out refinements. The largest ratio sensitivity is $0.0107$, for the clean twelve-terminal family-D device. These are empirical numerical sensitivities, not statistical confidence intervals.

We separately check Eq.~\eqref{eq:weight-ratio} in the finite surrogate containing the retained pair and charge-two coefficients for 14 devices with $n=3$--$6$, using two archived coefficient estimators for each device: integration by parts from the current and direct energy integration. Direct integration of the surrogate energy and all-terminal current norms on a phase lattice certified free of Fourier aliasing agrees with the separately computed coefficient-weighted identity to relative discrepancy below $10^{-14}$ in all 28 checks. This validates the identity and its implementation for the retained pair-plus-charge-two surrogate. A quantitative decomposition of the full-error ratios in panels (e,f) would additionally require the complete higher-order omitted spectrum, which is absent from these coefficient sets.

Tables~\ref{tab:multiterminal} and \ref{tab:large-n} list the matched phase-current comparisons behind
Fig.~\ref{fig:multiterminal}, for three to six and for eight to sixteen terminals.
\begin{table*}[t]
\caption{Matched zero-field phase-current comparisons for $n=3$--$6$. Family A has two channels per terminal; family B has twelve total channels. B at $n=6$ is the same device as A. $f_E$ is the residual energy-power fraction; all other error columns are RMS percentages using all terminal currents. Disorder columns use one fixed spatial pattern at $U_0/\mu=0.5$. The last column retains all charge-two multiterminal irreducible harmonics as well as the pair terms. These are phase-current errors.}
\label{tab:multiterminal}
\begin{ruledtabular}
\begin{tabular}{ccccccr}
Family & $n$ & Channels & clean $100f_E$ & clean $100\epsilon_I$ & disordered $100\epsilon_I$ & corrected $100\epsilon_I$ \\
A & 3 & 2,2,2 & 0.978 & 18.01 & 18.69 & 7.38 \\
A & 4 & 2,2,2,2 & 0.765 & 15.53 & 16.41 & 6.40 \\
A & 5 & 2,2,2,2,2 & 1.817 & 27.66 & 23.60 & 13.36 \\
A & 6 & 2,2,2,2,2,2 & 1.403 & 22.28 & 24.98 & 13.85 \\
B & 3 & 4,4,4 & 0.028 & 4.45 & 5.24 & 4.18 \\
B & 4 & 3,3,3,3 & 0.837 & 16.89 & 18.95 & 10.40 \\
B & 5 & 3,3,2,2,2 & 0.846 & 17.10 & 18.30 & 10.72 \\
B & 6 & 2,2,2,2,2,2 & 1.403 & 22.28 & 24.98 & 13.85 \\
\end{tabular}
\end{ruledtabular}
\end{table*}

\begin{table*}[t]
\caption{Extension to eight, twelve and sixteen terminals. A has a 120 nm square; C and D have a 240 nm square. A and C have two channels per terminal; D has 32 total channels and coincides with C at $n=16$. Its main $n=12$ control uses 70 nm outer contacts with separated interfaces; the additional corner row uses 80 nm contacts sharing four corner interface sites. Current errors include all terminals; numerical error columns are percentages. Columns report clean and matched $U_0/\mu=0.5$ disorder cases, with no charge-two correction.}
\label{tab:large-n}
\begin{ruledtabular}
\begin{tabular}{ccrcccc}
Family & $n$ & Total channels & clean $100f_E$ & disordered $100f_E$ & clean $100\epsilon_I$ & disordered $100\epsilon_I$ \\
A & 8 & 16 & 2.250 & 2.039 & 27.358 & 25.878 \\
C & 8 & 16 & 2.501 & 2.547 & 29.534 & 29.196 \\
C & 12 & 24 & 1.878 & 2.033 & 23.984 & 25.588 \\
C & 16 & 32 & 2.922 & 2.947 & 30.322 & 30.592 \\
D & 8 & 32 & 0.975 & 1.054 & 19.841 & 20.020 \\
D & 12 & 32 & 1.009 & 1.148 & 19.253 & 20.267 \\
D (corner) & 12 & 32 & 1.119 & 1.297 & 20.418 & 21.749 \\
\end{tabular}
\end{ruledtabular}
\end{table*}

\section{Planar controls and held-out predictor tests}
\label{app:planar-controls}

The mechanism test of Sec.~\ref{sec:planar-resonance} used three criteria fixed before any current was evaluated;
Table~\ref{tab:mechanism-criteria} lists them with the achieved values.
\begin{table*}[t]
\caption{Prespecified criteria of the planar mechanism test and achieved values. Waveform is the largest
relative RMS error of the predicted full current over the five gates, in percent. Pairwise error is the largest
difference between the predicted and computed pairwise current errors, and gate contrast is the largest pairwise
error at $|x|\le1$ minus the largest at $|x|=4$, both in percentage points. The uniform-gate and local-gate columns
use the frozen-$S$ model for the eight devices of Fig.~\ref{fig:planar-resonance}(a,b); the local gate acts on a
central patch of the normal region (Appendix~\ref{app:local-gate}) and is scored with its first-order predictor. The
finite-gap columns use dispersive BCS leads for the clean and the disordered $140\times100$-nm$^2$ devices with
$n=3,4$ under the uniform gate, and for the clean devices under the local gate. Every waveform value above threshold
belongs to the disordered four-terminal device.}
\label{tab:mechanism-criteria}
\begin{ruledtabular}
\begin{tabular}{lclllll}
 & & \multicolumn{2}{c}{Frozen $S$} & \multicolumn{3}{c}{Finite gap}\\
Criterion & Threshold & Uniform & Local & Uniform, clean & Uniform, disordered & Local, clean\\
\hline
Waveform & $\le10$ & 0.07--1.97, 18.99 & 0.60--3.62, 23.41 & 0.20 and 2.41 & 0.22 and 16.19 & 1.93 and 2.28\\
Pairwise error & $\le3$ & $\le0.012$ & $\le0.076$ & $\le0.012$ & $\le0.021$ & $\le0.037$\\
Gate contrast & $\ge5$ & 15.2--40.2 & 15.2--40.2 & 18.6 and 29.8 & 26.6 and 20.5 & 18.6 and 29.8\\
All three met & & 7 of 8 & 7 of 8 & 2 of 2 & 1 of 2 & 2 of 2\\
\end{tabular}
\end{ruledtabular}
\end{table*}

\begin{table*}[t]
\caption{Status and reuse of the resonance tests. The eight reserved configurations span two geometries and share
one disorder map per geometry across terminal-count variants. They are not eight independent disorder draws.
Protocol labels refer to saved internal plans; the later tests reuse devices. The manual timing annotations of the
finite-gap local-gate follow-up remain partly unresolved. The numerical coverage of the added tests is reported separately.}
\label{tab:evidence-status}
\begin{ruledtabular}
\begin{tabular}{p{0.28\textwidth}p{0.26\textwidth}p{0.17\textwidth}p{0.21\textwidth}}
Analysis & Protocol status & Device reuse & Role\\
\hline
Five-gate frozen-$S$ test & Original prespecified selection and criteria & Eight reserved configurations & Primary mode intervention\\
45-gate frozen-$S$ extension & Retrospective & Same eight & Resolve sampled peaks\\
Clean uniform-gate finite-gap test & Planned model upgrade & Two-device subset & Test lead-energy dispersion\\
Clean dense finite-gap scan & Retrospective & Same clean pair & Resolve sampled peaks\\
Disordered uniform-gate finite-gap test & Follow-up with fixed criteria & Two-device subset & Disorder extension\\
Frozen-$S$ local-gate test & Follow-up with saved patch and criteria & Same eight & Test nonuniform gating\\
Finite-gap local-gate test & Planned follow-up; timing qualification above & Same clean pair & Combined gate/model test\\
Spectator-mode expansion & Retrospective & Same clean pair & Mechanism diagnosis\\
\end{tabular}
\end{ruledtabular}
\end{table*}

\subsection{Control matrix}
We hold contact modes fixed in square $120\times120$ and wide
$180\times90~\mathrm{nm}^2$ geometries at $a=5~\mathrm{nm}$.
Centered $40$-nm contacts have two propagating channels per terminal,
with $n=3,4$ and ideal-lead $\mu=20~\mathrm{meV}$. Each geometry uses a
clean case and two fixed disorder realizations with $U_0=10~\mathrm{meV}$.
Five boundary barriers $V_b=0,40,80,160,320~\mathrm{meV}$ at zero gate
and four additional normal-region gates $V_g=-4,-2,2,4~\mathrm{meV}$ at
zero barrier give 108 scattering systems. Positive gate
raises the normal onsite energy; ideal leads are unchanged. Disorder is
uniform in $[-U_0/2,U_0/2]$. Barriers act on the boundary row beneath each
contact; $V_b$ is not a transmission probability.

For each fixed normal $S(0)$, the free energy is
\begin{equation}
 F/\Delta=-t_T\sum_\nu\log[2\cosh(\sigma_\nu/2t_T)],
 \qquad t_T=k_BT/\Delta,
 \label{eq:planar-thermal}
\end{equation}
where $\sigma_\nu$ are the singular values in the existing spin/channel
convention. Its zero-temperature limit is $-\sum_\nu\sigma_\nu/2$.
We use $t_T=0,0.03,0.1,0.25$, fixed $\Delta$, and zero flux. This tests
thermal occupations without self-consistent gap suppression or heating.
All 432 computed errors in this matrix pass the stated numerical checks. Pairwise
terms retain at least 96.49 percent of energy-variation
power, while current error reaches 32.5 percent.

\begin{figure*}[t]
 \centering
 \includegraphics[width=\textwidth]{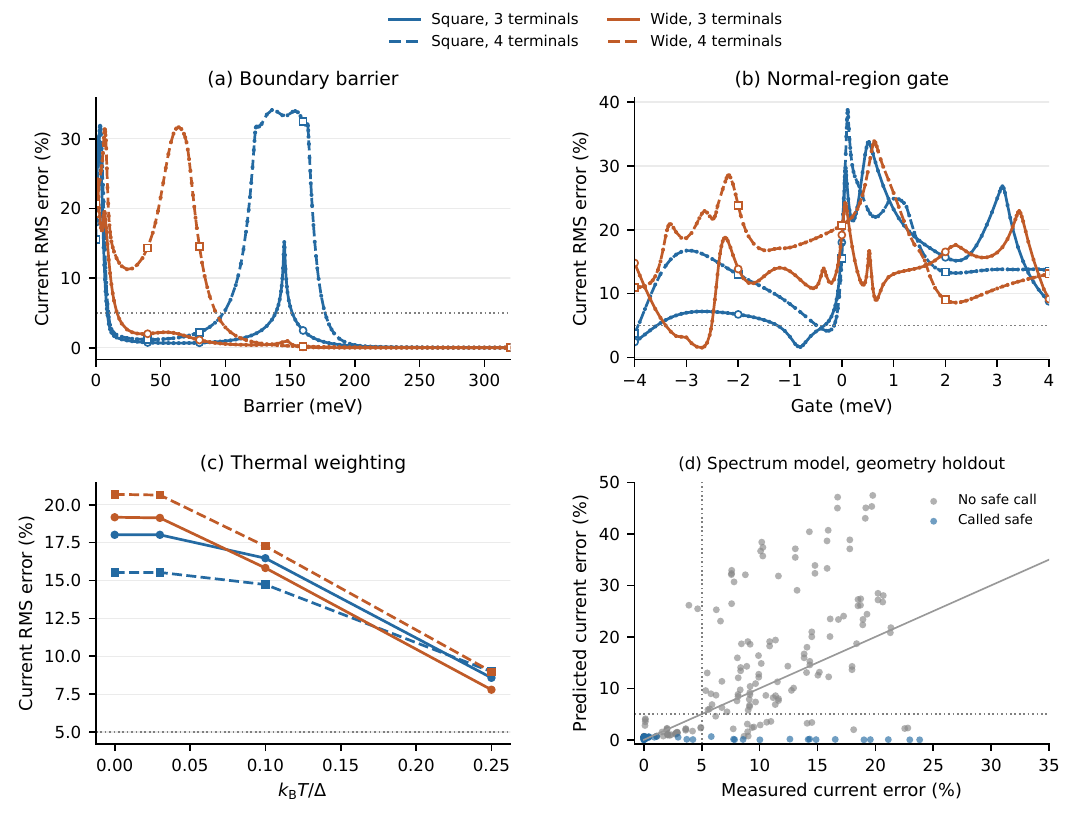}
 \caption{Controlled planar tests in the energy-independent-$S$ model.
 Panels (a--c) show every clean geometry and terminal count at the indicated
 one-at-a-time control, with the other controls at zero. Panels (a,b) use
 dense zero-temperature scans with adaptive midpoint calculations; small
 dots mark calculated values and open symbols mark the five
 control values of the matrix. Lines join calculations, without smoothing. Panel (c) shows the four temperatures of the matrix. Dotted horizontal lines mark
 5 percent current error.
 Panel (d) shows all 216 geometry-holdout errors, one per system and temperature, for
 the fixed sparse-spectrum predictor; its raw prediction is on the vertical axis.
 Blue points are called safe only after adding the training-calibrated
 margin. Fifteen such calls exceed the 5 percent benchmark. The diagonal is
 equality. These unsafe calls concern this predictor and split.}
 \label{fig:predictive-planar}
\end{figure*}

Barrier dependence need not be monotonic. The clean square four-terminal
zero-temperature errors at the five increasing $V_b$ values of the matrix are
15.53, 1.17, 2.22, 32.47, 0.00017 percent [Fig.~\ref{fig:predictive-planar}(a)].
The increase at $160~\mathrm{meV}$ is compatible with resonant scattering,
but the error sequence alone does not establish that mechanism. Gate and
temperature controls further change the approximation error within the same
model [panels (b,c)]. These observations rule out using the nominal barrier
alone as a monotonic accuracy criterion.

Retrospective clean zero-temperature scans resolve these control dependences
with initial spacings of $4~\mathrm{meV}$ in barrier and
$0.1~\mathrm{meV}$ in gate, followed by up to four midpoint-refinement
rounds. The additional calculations reveal narrow gate-dependent features between the control values of the matrix. For example, the square four-terminal
gate scan reaches a sampled current error of 38.8 percent at
$V_g=0.1125~\mathrm{meV}$. The additional
errors are excluded from predictor training and held-out scoring. These finite scans resolve sampled features;
they do not certify extrema or exclude narrower resonances.

There is a material restriction on physical interpretation. Recomputing
normal transmission matrices at the prespecified energies $\pm0.09$ and
$\pm0.18~\mathrm{meV}$ gives changes up to 54.5 percent
relative to the zero-energy Frobenius norm in clean zero-barrier controls.
Thus the test does not validate the short-junction approximation at
$\Delta=0.18~\mathrm{meV}$ across this matrix. The reported errors describe
the energy-independent-$S$ model. Quantitative finite-gap device predictions
require an energy-dependent scattering or BdG calculation, or an independently
verified smaller-gap limit. Even small transmission changes alone would not
certify all scattering phases or dwell times.

\subsection{Tests on held-out devices}
\begin{table}[t]
\caption{Held-out prediction tests. The mean absolute error (MAE) is in percentage points of relative
current error. Safe calls include the empirical training margin. False-safe
counts exceed 5 percent even after subtracting the numerical error diagnostic.
Temperature labels of the same device are correlated.}
\label{tab:prediction}
\begin{ruledtabular}
\begin{tabular}{llrrr}
Holdout & Inputs & MAE & Safe calls & False-safe\\
Disorder & Constant & 8.25 & 0/72 & 0 \\
Disorder & Controls & 5.74 & 8/72 & 0 \\
Disorder & Normal & 4.36 & 13/72 & 0 \\
Disorder & Spectrum & 5.00 & 11/72 & 0 \\
Geometry & Constant & 7.61 & 0/216 & 0 \\
Geometry & Controls & 5.02 & 24/216 & 4 \\
Geometry & Normal & 6.08 & 37/216 & 0 \\
Geometry & Spectrum & 6.87 & 56/216 & 15 \\
\end{tabular}
\end{ruledtabular}
\end{table}
Four predictors were fixed before the computed errors were inspected. They use a
constant baseline, device controls, normal transport features, and the latter augmented by sparse spectral
features. The computed error of each system at each temperature is the label to be predicted. Standardized ridge regression with fixed unit
penalty fits $\log(\epsilon_I+10^{-6})$, with an unpenalized intercept and no
feature search. Control inputs are terminal count, $U_0/\mu$, $V_b/\mu$,
$V_g/\mu$ and $t_T$. Normal inputs replace the two potentials by mean escape
probability, the second-to-first moment ratio of offdiagonal transmission-block
eigenvalues, and exit-terminal branching entropy. They retain $n,U_0/\mu,t_T$.
The spectrum model adds the median and minimum sampled gap and the fraction
below $0.1\Delta$ from 256 independently selected phases. These sampled
minima are not whole-torus gap certificates; no current or Fourier coefficient from the phase
grids enters a predictor.

Training uses 36 square systems (clean and disorder seed 0), giving 144
temperature labels. The 18 square systems with seed 1 give 72 disorder-holdout
labels; all 54 wide systems give 216 geometry-holdout labels. Every system's
temperatures stay together. A training-only margin is the largest positive
log-error residual under leave-one-system-out refitting, including the
numerical error diagnostic. A prediction is called safe only when its
margin-adjusted error is below 5 percent. This empirical envelope has no
probabilistic coverage guarantee. Predictions were recorded before the
held-out labels were scored, and altering those labels leaves the fitted models unchanged.

The prespecified geometry test requires mean absolute error at most two
percentage points, at least 25 percent safe-call coverage and no false-safe
case. We also require these conditions on the disorder holdout before claiming
overall transfer. None meets even the geometry test
(Table~\ref{tab:prediction}). The normal model avoids false-safe calls but
has inadequate accuracy and coverage; adding the sampled spectrum does not
repair transfer and produces false-safe calls. This result is specific to the fixed feature sets, linear log-error model and
selected splits. It does
not prove that normal scattering or spectra contain no predictive information.
The controlled dot bound remains a model-specific sufficient criterion;
we have not obtained a reliable general planar predictor.

\FloatBarrier
\section{Additional convergence and mode controls}
\label{app:strengthening}
\subsection{Frequency and channel refinement}
\label{sec:finitegap-refinement}
An additional follow-up tests all five previously reported gates for the two disordered uniform-gate
and two clean local-gate finite-gap devices. Full and frozen-background predictor Hamiltonians are checked
separately, giving forty logical cases. The protocol and thresholds were fixed before these new calculations.
The center models coincide by construction; the comparison follows the same four devices.

Each case uses a separately scrambled Sobol set of 1,024 phase points, distinct from production, and additional
conditioning probes. The latter include neighborhoods of zero/$\pi$ corners and fixed offsets from four
locally optimized low-singular-value points of the all-channel dressed Schur matrix at
$i\omega=i10^{-7}$~meV. All generated offsets are retained. These probes target poorly conditioned sampled neighborhoods. Excitation-gap certification,
exact zero-temperature crossing derivatives and uniform phase-space coverage require separate analyses.

Frequency orders 64, 128 and 256 are tested with both four and all nine anomalous channels per lead;
all normal channels remain included. For each of the two lower orders relative to 256, the Sobol waveform
RMS and the maximum probe-vector difference must each be at most $0.1$ percent of the reference current RMS.
The four-versus-nine-channel Sobol waveform discrepancy must be strictly below $0.1$ percent.
Both the corresponding full-target denominator and each model's own denominator must pass.
The probe maximum for the channel comparison is recorded separately.
All forty cases meet these unchanged criteria at the original 64-node/four-channel production settings
(Table~\ref{tab:finitegap-refinement}). Consequently the historical phase-grid projections and their criteria
are retained. Their largest separately recorded phase/projection sensitivity is below $0.034$ percentage points.

An independent contact-Nambu determinant agrees with the all-channel Schur algebra on eight Sobol and four
corner-neighborhood points per case. A supplementary comparison also tests the four lowest-score saved
noncentral conditioning probes in every case. The largest supplemental discrepancies are
$1.3\times10^{-17}$ in $I/(2e\Delta/\hbar)$ and $1.6\times10^{-13}$ in $F/\Delta$, below the fixed
$10^{-9}$ threshold. These are independent algebraic forms sharing the same self-energies and quadrature.
Recomputing sixteen historical production points per case reproduces the saved currents and energies
within $2.0\times10^{-13}$ in those units.

On the matched independent Sobol sample, changing 64/four to 256/nine changes the full-waveform predictor
error by less than $0.002$ percentage points. The disordered four-terminal failure remains approximately
$16.2$ percent at $x=-4$. These matched comparisons quantify waveform sensitivity to frequency quadrature and channel truncation;
the previously converged best-pairwise projections supply the scalar errors. The refinements support
the stated numerical settings for these four finite Hamiltonian/gate families. Spatial discretization,
disorder populations and material calibration are separate validation steps.

\subsection{Spatial sensitivity of one selected resonance}
\label{sec:resonance-lattice}
We tested the clean $140\times100$-nm three-terminal resonance of
Sec.~\ref{sec:planar-resonance} at finer lattice spacing, keeping
$\mu=20$~meV, nominal $40$-nm contacts and a $200$-meV barrier.
The original barrier occupies one boundary row, so retaining that prescription
(A) changes its physical thickness with $a$. A second prescription (B) samples
a fixed $5$-nm inward strip on the nested $a=5$ and $2.5$~nm meshes; its
$a=5$ Hamiltonian is identical to A. Neither prescription fixes the effective
hard-wall boundary exactly. At $a=3.5$~nm, rounding also changes the transverse
device span to $98$~nm and the contact site span to $35$~nm.

The original selected eigenmode is tracked by bidirectional interpolated
squared overlap, with no replacement based on current outcomes. The fixed-strip
comparison B has forward and reverse overlaps $0.924$ and $0.923$; both exceed
the prespecified $0.8$ threshold and the competing-mode separation exceeds
$0.5$. The A sequence has unresolved mode-identity uncertainty at $3.5$~nm,
which propagates to its $2.5$-nm comparison. Each five-gate scan is centered on
the tracked Hermitian mode and uses its own linewidth in $x=2\epsilon/\gamma$.
This follow-up tests the originally selected mode's persistence on each mesh.

The propagating-channel count falls from two to one per lead under both
refinements. This follows independently from the ideal-strip thresholds:
the second channel opens at $19.40$, $27.78$ and $24.51$~meV for
$a=5$, $3.5$ and $2.5$~nm, respectively, straddling the fixed $\mu$.
The tracked mode is odd under left--right reflection, while the remaining
bottom-lead fundamental is even. Its bottom-terminal width therefore vanishes
by symmetry. The normal weights change from approximately
$(0.197,0.197,0.606)$ to $(0.5,0.5,0)$, failing the original requirement
of three weights $p_i\ge0.1$. In B the linewidth decreases from $11.94$ to
$0.296~\mu$eV, while the Hermitian-mode-center frozen-$S$ current RMS changes from
$24.18$ to $36.35$~nA; these are short-junction model currents.

\begin{figure*}[t]
 \centering
 \includegraphics[width=\textwidth]{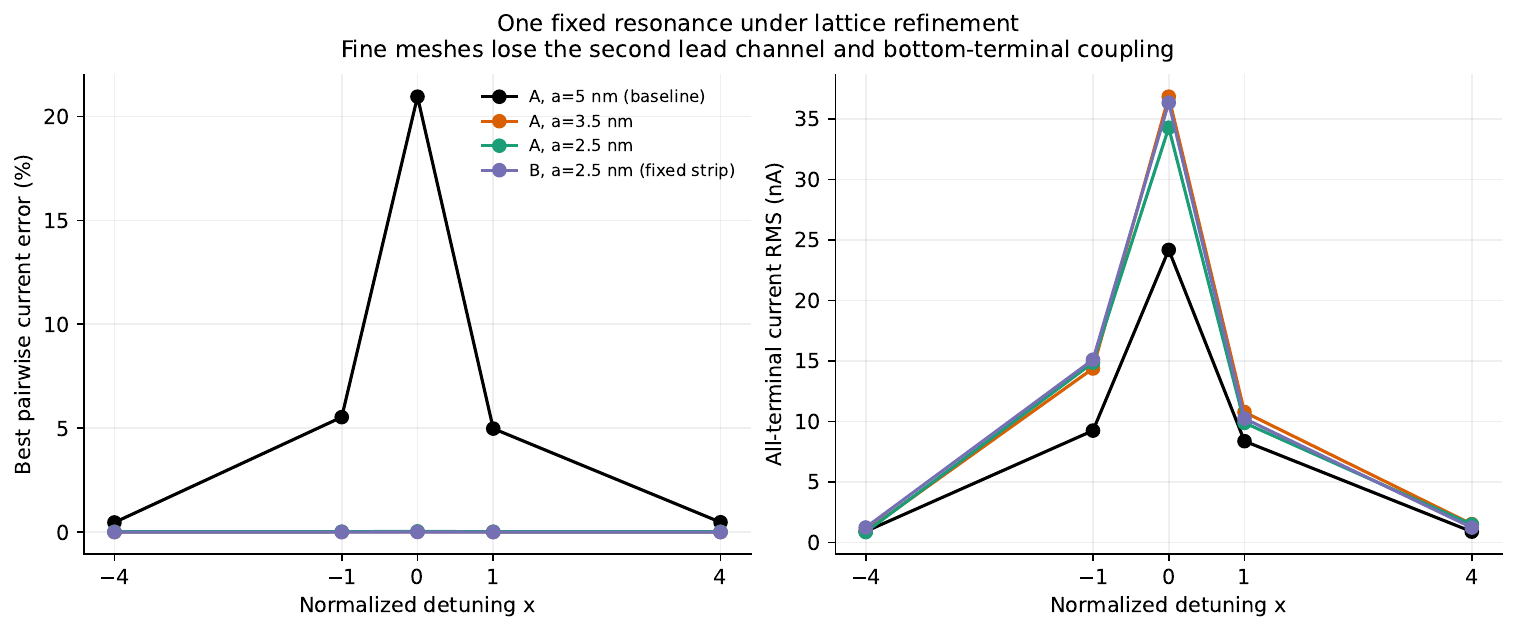}
 \caption{Spatial sensitivity of the fixed clean three-terminal resonance.
 Left: full frozen-$S$ pairwise-current error; right: all-terminal current RMS
 at $T=0$, using $\Delta=0.18$~meV for current conversion. A retains the
 original one-row barrier; B samples a fixed $5$-nm inward barrier strip.
 The five gates use each tracked mode's own center and linewidth; connecting
 segments add no calculations. The second lead channel closes on the finer
 meshes and the tracked mode loses its bottom-terminal coupling.
 Mode tracking is uncertain for the two refined A cases; B meets the fixed
 overlap criteria. Observed numerical sensitivities reach $0.12$ percentage
 points, so the tiny refined residuals are not assigned precise values here.}
 \label{fig:resonance-lattice}
\end{figure*}

The baseline Hermitian-mode center has $\epsilon_I=20.96$ percent. All three refined estimates
are below $0.04$ percent, with observed phase/projection changes up to
$0.12$ percentage points and an unchanged $0.5$-percent numerical tolerance.
These checks resolve the loss of the large nonpairwise response; the precision
of each tiny residual remains unresolved. An independent shifted $129^2$ current fast-Fourier-transform (FFT) projection confirms that loss. All refined scans fail the unchanged
five-percentage-point resonance-contrast criterion
[Fig.~\ref{fig:resonance-lattice}]. The historical $a=5$ full currents were
also reproduced at all five gates on all $16{,}384$ saved phases.
In this near-threshold device, spatial refinement changes the contact spectrum
and removes the selected mode's three-terminal character. The original resonance
therefore does not persist as a spatially stable three-terminal realization.

A spatially stable shared mode requires channel-resolved overlap at three or more contacts. The physical contact widths and confinement must be represented consistently, the relevant propagating channels must stay open, and their transverse symmetry must permit overlap with the tracked state. This near-threshold example fails that test. Alternative carrier densities or contact placements can be investigated by tracking the mode's terminal weights and separation from neighboring levels while refining the spatial model and contact profile.

\subsection{Contour, harmonic and topology checks}
\begin{table*}[t]
\caption{Achieved stable-region numerical changes, in percent for each separately normalized region comparison. Columns $h_f,h_p$ and $a_f,a_p$ are the final refinement changes of the full and pairwise images in $100d_H$ and $100d_A$. Columns $\delta_H,\delta_A$ are the changes of the full--pairwise discrepancy, in percentage points. All grids use the same pairwise zero-mode subtraction. These are finite-grid diagnostics.}
\label{tab:achieved-refinement}
\begin{ruledtabular}
\begin{tabular}{lcrrrrrr}
Device, $U_0/\mu$ & Last grids & $h_f$ & $h_p$ & $a_f$ & $a_p$ & $\delta_H$ & $\delta_A$\\
Square, $0$ & $161\to 321$ & 0.123 & 0.179 & 0.084 & 0.163 & 0.123 & 0.009\\
Wide--short, $0$ & $161\to 321$ & 0.035 & 0.158 & 0.042 & 0.135 & 0.148 & 0.028\\
Square, $0.5$ & $81\to 161$ & 0.166 & 0.219 & 0.125 & 0.148 & 0.003 & 0.021\\
Wide--short, $0.5$ & $81\to 161$ & 0.151 & 0.237 & 0.167 & 0.116 & 0.028 & 0.027\\
Square, $2$ & $81\to 161$ & 0.194 & 0.184 & 0.157 & 0.131 & 0.030 & 0.031\\
Wide--short, $2$ & $81\to 161$ & 0.196 & 0.203 & 0.133 & 0.174 & 0.038 & 0.046\\
\end{tabular}
\end{ruledtabular}
\end{table*}
\begin{table*}[t]
\caption{Full--pairwise model discrepancies across the three saved grids, with the constant pairwise current removed independently on every grid. The clean and disordered grid sequences differ as indicated. All entries are percentages.}
\label{tab:comparison-sequence}
\begin{ruledtabular}
\begin{tabular}{lcrrrrrr}
Device, $U_0/\mu$ & Grids & \multicolumn{3}{c}{$100d_H$} & \multicolumn{3}{c}{$100d_A$}\\
Square, $0$ & 81,161,321 & 2.102 & 2.156 & 2.033 & 2.532 & 2.574 & 2.565\\
Wide--short, $0$ & 81,161,321 & 1.272 & 1.339 & 1.191 & 2.111 & 2.135 & 2.107\\
Square, $0.5$ & 41,81,161 & 4.209 & 3.809 & 3.805 & 8.020 & 8.130 & 8.152\\
Wide--short, $0.5$ & 41,81,161 & 0.986 & 0.989 & 0.962 & 2.752 & 2.776 & 2.749\\
Square, $2$ & 41,81,161 & 4.294 & 3.893 & 3.863 & 8.727 & 8.897 & 8.866\\
Wide--short, $2$ & 41,81,161 & 2.710 & 2.645 & 2.607 & 5.891 & 5.932 & 5.978\\
\end{tabular}
\end{ruledtabular}
\end{table*}

Tables~\ref{tab:achieved-refinement} and \ref{tab:comparison-sequence} separate achieved numerical
changes from the full--pairwise model differences. The pairwise constant current is removed independently
at every grid. Each model retains its own stability mask. The alternate triangulation and fourth-order
Hessian checks are unchanged by this constant translation and remain within the stated tolerances.
The sequences need not converge monotonically, so no Richardson extrapolation or formal continuum error
bar is assigned. Values beyond the precision supported by these variations should not be interpreted.

For the mode analysis, normalized discrete Fourier transforms of the directly evaluated currents give
an exact finite-grid Parseval decomposition. Energy coefficients provide an independent check of
$\mathbf I_{mn}/I_0=i(m,n)e_{mn}$ for the six quartet modes; the largest relative discrepancy of their
combined coefficient norm in the accepted grids is 1.95 percent.
An exactly representable pairwise-plus-quartet energy validates the mode accounting to roundoff.
All 160 zero-field disorder realizations and two shifted clean controls are included. The initial
41/81 grids are supplemented by 161, 321, or 641 points per axis when the quartet fraction changes by
more than two percentage points, the low-mode derivative discrepancy exceeds two percent, or the current RMS refinement criterion is not met. The eight matched-pattern
controls use unequal irrational grid offsets, independently evaluated off-grid phase derivatives, and
charge-conservation checks. Bootstrap intervals resample the 20 independent realizations within a
fixed geometry and strength with 10,000 draws; they quantify sampling of that ensemble.

The adaptive-topology control uses the globally smooth periodic spinor
$q=(1,z)^T/\sqrt{1+|z|^2}$, with
\begin{align}
 z(x,y)={}&0.3+0.6\sin(x+0.31)+0.5i\sin(y+0.72)\nonumber\\
 &+0.2e^{i(x+y+0.4)}.
\end{align}
Its rank-one projector is topologically trivial. The adaptive routine used for the disordered device, with a four-point base grid, depth three and phase threshold 0.15, returns 0.00786; uniform four- and 32-point grids return
zero within $10^{-15}$. All candidate dyadic overlap magnitudes exceed 0.57. This counterexample
isolates the missing shared-edge cancellation without a rank change or singular link; it makes no claim
about the physical projector of that device.

\FloatBarrier
\vspace*{8pt}
\section{Current response to omitted normal modes}
\label{app:omitted-modes}
The clean uniform-gate finite-gap calculations of Sec.~\ref{sec:planar-resonance} and of this appendix use the following channel
settings. All nine transverse channels per lead enter the normal self-energy; four enter the anomalous self-energy
in the phase-grid runs. An independent 256-phase comparison with all nine anomalous channels gives an
\emph{observed} waveform discrepancy below 0.04 percent. Phase and frequency refinements are separate.

An isolated-mode model and a model that freezes spectator energies make
different approximations. The former discards a contribution to the contact
resolvent; the latter discards its gate variation. Both can be treated as a
perturbation of the equilibrium current. This follows the methodological
lesson of spectator-mode analyses in driven circuit QED
\cite{ozguler2024spectator}, where an omitted contribution is propagated
to the observable. Here the corresponding observable is the equilibrium current, with the response formula and numerical
criteria developed below.

In the real eigenbasis of $H_0$ defined in
Sec.~\ref{sec:planar-resonance}, let $v_a$ be the row of contact amplitudes
of normal mode $a$. Its contribution to the electron contact resolvent is
$v_a^T v_a/(i\omega-\epsilon_a-g)$, where $g$ is the uniform gate energy;
the hole contribution has $\epsilon_a+g\to-\epsilon_a-g$.
The two resolvents form a block matrix $\mathcal G$.
With the dispersive BCS lead self-energy $\mathcal S$, the channel determinant
is $M=1-\mathcal S\mathcal G$. Here the real normal self-energy already
absorbed in $H_0$ is subtracted from the electron block of $\mathcal S$
and added to its hole block. All transverse channels are retained below.

Let $\mathcal G_0$ describe the chosen reference approximation and
$\delta\mathcal G=\mathcal G-\mathcal G_0$. Define
\begin{align}
 M_0&=1-\mathcal S\mathcal G_0, &
 \delta M&=-\mathcal S\delta\mathcal G,\nonumber\\
 X&=M_0^{-1}\delta M, &
 X_i&=\partial_{\theta_i}X.
 \label{eq:omitted-response-matrix}
\end{align}
The matrices $X$ and $X_i$ are dimensionless. For invertible channel matrices
and differentiable phase points where differentiation under the convergent
frequency integral is justified, the spin-reduced determinant free energy gives
\begin{equation}
 \delta j_i=-\frac{1}{\pi\Delta}\int_0^\infty
 d\omega\,\operatorname{Re}\operatorname{Tr}
 \left[(1+X)^{-1}X_i\right].
 \label{eq:omitted-exact-current}
\end{equation}
Here $j_i=I_i/(2e\Delta/\hbar)$; this identity includes the continuum.
The derivative is
\begin{equation}
 X_i=M_0^{-1}\partial_i\delta M
      -M_0^{-1}(\partial_i M_0)X.
 \label{eq:omitted-vertex}
\end{equation}
The second term accounts for the reference junction's response and cannot
in general be discarded. Expanding $(1+X)^{-1}$ yields
\begin{align}
 \delta j_i^{[1]}&=-\frac{1}{\pi\Delta}\int_0^\infty
 d\omega\,\operatorname{Re}\operatorname{Tr}X_i,\nonumber\\
 \delta j_i^{[2]}&=-\frac{1}{\pi\Delta}\int_0^\infty
 d\omega\,\operatorname{Re}\operatorname{Tr}[(1-X)X_i].
 \label{eq:omitted-orders}
\end{align}
These are corrections to $j_i^{(0)}$.
Because $\delta\mathcal G$ is a sum over omitted modes, the first-order
correction resolves their additive contributions, including cancellations.

A sufficient expansion condition is $r=\|X\|_2<1$ throughout the frequency
and phase domain of interest. For $d=\dim M_0$, the remainder after order
$k=1,2$ obeys
\begin{equation}
 |\delta j_i-\delta j_i^{[k]}|
 \le \frac{d}{\pi\Delta}\int_0^\infty
 d\omega\,\frac{\|X_i\|_2 r^k}{1-r},
 \label{eq:omitted-bound}
\end{equation}
provided the right-hand side is integrable. This follows from the Neumann
remainder and $|\operatorname{Tr}AB|\le d\|A\|_2\|B\|_2$;
no commutativity or normality is assumed. A relative waveform bound also
requires a nonzero current scale. If $\|\mathbf I-\mathbf I_{\rm app}\|\le B$
and $\|\mathbf I_{\rm app}\|>B$, the relative error is at most
$B/(\|\mathbf I_{\rm app}\|-B)$.
The dressed resolvent and its phase derivative determine how omitted spectral weight
enters the current response. Applying the bound to a device requires control over
the full frequency and phase domain; sampled values of $r$ provide local tests.

We test both omissions retrospectively on the same two clean
$140\times100$-nm devices, with $n=3,4$, $\Delta=0.18$~meV and the five
gates $x=-4,-1,0,1,4$. All nine transverse channels per terminal enter both
normal and anomalous lead self-energies. Mode-deletion references retain the
1, 4 or 16 modes nearest zero at the central gate, selected before these
current tests. The other reference retains every mode but freezes spectator
energies. No current is fitted. Table~\ref{tab:omitted-mode-response} reports
the largest errors over the five gates. Second order improves every tested
reference, while first order can increase its error. The corrected one-mode
four-terminal model still misses the exploratory 1-percent target at both
$|x|=4$ endpoints. The $x=0$ identity of the
frozen-spectator model is constructional.

\begin{table}[t]
\caption{Largest relative full-current waveform errors (percent) over the
five gates. Corrections are added to each reference current. The comparison
uses all terminals and the same 128 phase samples.}
\label{tab:omitted-mode-response}
\begin{ruledtabular}
\begin{tabular}{lrrrr}
Reference & $n$ & Bare & Order 1 & Order 2 \\
Frozen spectators & 3 & 0.20 & 0.004 & $<0.001$ \\
Frozen spectators & 4 & 2.42 & 0.098 & $<0.001$ \\
One mode & 3 & 13.52 & 8.04 & 0.55 \\
One mode & 4 & 28.22 & 23.27 & 1.03 \\
Four modes & 3 & 10.43 & 13.22 & 0.53 \\
Four modes & 4 & 20.40 & 15.83 & 0.71 \\
Sixteen modes & 3 & 5.73 & 0.91 & 0.13 \\
Sixteen modes & 4 & 4.52 & 0.28 & 0.13 \\
\end{tabular}
\end{ruledtabular}
\end{table}

The final estimates use 128 independent scrambled Sobol phases and 256 frequency
nodes. Comparing 128 and 256 nodes on the common 64 phases changes any
tested current waveform by less than 0.00005
percent. Doubling the phase sample at fixed frequency resolution changes
the second-order error by at most 0.009
percentage points and preserves every 1-percent classification.
Independent full-lattice BCS, phase-derivative, gauge and current-conservation
checks pass. Scaling the omitted resolvent verifies the quadratic and cubic
remainders of the first- and second-order corrections in the tested
small-perturbation regime. These checks support the expansion on the stated phases. The method evaluates
current corrections from normal spectral information and phase-resolved integration;
its computational cost and use in device selection require separate assessment.

\FloatBarrier
\section{Energy of an isolated short-junction resonance}
\label{app:isolated-resonance}
This appendix derives Eq.~\eqref{eq:resonance-energy}. Neglecting the background reaction matrix, the selected mode
gives $K=-x^{-1}\hat v\hat v^{\mathsf T}$ at $E=0$ in Eq.~\eqref{eq:resonance-reaction}, where $\hat v=v/|v|$ is the
normalized vector of its contact amplitudes, $|v|^2=\gamma$ and $x=2\epsilon/\gamma$. Because $K$ has rank one,
\begin{equation}
 S=1+(e^{i\eta}-1)\,\hat v\hat v^{\mathsf T},\qquad e^{i\eta}=\frac{ix-1}{ix+1},
 \label{eq:rank-one-s}
\end{equation}
which is unitary and reduces to $1-2\hat v\hat v^{\mathsf T}$ at resonance. Summing $\hat v_a^2$ over the channels
of terminal $i$ gives its weight $p_i$. The normal conductance between terminals $i\ne j$ follows directly,
\begin{equation}
 G_{ij}=\frac{2e^2}{h}\sum_{a\in i,\,b\in j}|S_{ab}|^2=\frac{2e^2}{h}\,\frac{4p_ip_j}{1+x^2},
 \label{eq:resonance-conductance}
\end{equation}
which is the Breit--Wigner form used in the experimental protocol of Sec.~\ref{sec:discussion}.

Beenakker's secular equation $\det[1-\alpha(E)^2M]=0$ with $M=R^\dagger S^*RS$ (Sec.~\ref{sec:methods}) involves
only the space spanned by $\hat v$ and $w=R^\dagger\hat v$. Since
$R^\dagger S^*R=1+(e^{-i\eta}-1)ww^\dagger$, $M$ equals the identity on the orthogonal complement, which gives
$E=\Delta$ and no phase dependence. On the two-dimensional space $M$ is unitary with unit determinant, so its
eigenvalues are $e^{\pm i\varphi}$. With $|w^\dagger\hat v|=|\zeta|$, $\zeta=\sum_ip_ie^{i\theta_i}$, the trace of $M$
gives
\begin{equation}
 \cos\varphi=1-(1-\cos\eta)(1-|\zeta|^2)=1-\frac{2(1-|\zeta|^2)}{1+x^2}.
 \label{eq:resonance-phase}
\end{equation}
The secular equation then requires $\alpha^2=e^{\mp i\varphi}$, so the Andreev level is $E=\Delta|\cos(\varphi/2)|$ and
\begin{equation}
 \frac{E^2}{\Delta^2}=\frac{1+\cos\varphi}{2}=\frac{x^2+|\zeta|^2}{1+x^2}.
\end{equation}
With $E_{\rm gs}=-\sum_{E_p>0}E_p$ (Sec.~\ref{sec:conventions}), this is Eq.~\eqref{eq:resonance-energy} up to a
phase-independent constant. For 200 random rank-one scattering matrices, with two to six terminals, one to three
channels per terminal and random phases, Eq.~\eqref{eq:resonance-energy} agrees with the spectrum of
$A^\dagger A$ in Eq.~\eqref{eq:Aop} to $10^{-14}$, and Eq.~\eqref{eq:resonance-conductance} holds to machine
precision.

The infinite-gap level of Proposition~\ref{prop:dot} with $\Gamma_i=\gamma_i/2$ has, at zero temperature,
$F=-\Gamma_\Sigma\sqrt{(\epsilon_d/\Gamma_\Sigma)^2+|\zeta|^2}$ with $\epsilon_d/\Gamma_\Sigma=x$. The two energies
differ only by the phase-independent factors $\Delta/\sqrt{1+x^2}$ and $\Gamma_\Sigma$. The normalized pairwise
current error of an isolated resonance is therefore the same in the short-junction limit $\gamma\gg\Delta$ and in the
infinite-gap limit $\gamma\ll\Delta$, and Eq.~\eqref{eq:dot-sufficient} applies with $|\epsilon_d|/\Gamma_\Sigma$
replaced by $|x|$. The absolute current differs between the limits. The selected planar modes have
$\gamma\approx12~\mu$eV$\ll\Delta=0.18$~meV, where the short-junction model overestimates the current scale;
background modes, lead dispersion and the continuum are what the finite-gap calculations of
Sec.~\ref{sec:planar-resonance} add. For intermediate $\gamma/\Delta$ the isolated resonance has no equally simple
form.

\section{Local-gate test}
\label{app:local-gate}
The gate of Sec.~\ref{sec:planar-resonance} is uniform over the normal region, so it shifts every normal mode by the
same energy and leaves the mode shapes unchanged. To test a nonuniform gate, we apply the potential $g$ only to the
normal-region sites of a central patch spanning the middle half of the length and of the width, 27 percent
of the normal sites. At each gate all normal modes, their contact couplings and $S$ are recomputed in the frozen-$S$
model, and the pairwise error is evaluated as in Sec.~\ref{sec:planar-resonance}. The plan and the three criteria
were fixed before this calculation.

Let $u$ be the selected mode at the center gate and $f=\langle u|P|u\rangle$ its weight in the patch, where $P$
projects onto the patch sites; $f$ ranges over 0.20--0.26 in the eight devices. The gates are chosen from first-order
perturbation theory, $g=x\gamma/(2f)$, and are therefore about $1/f$ times larger than the uniform gates. The mode is
followed across the sweep by its overlap with the mode at the previous gate, which exceeds 0.998 at every step, and
it reaches the target detunings within 0.09 in $x$. Both predictors retain the center-gate background.
The first uses only center-gate normal-state data and shifts the selected mode by $fg$ with every other mode and
the lead couplings frozen, as in the uniform-gate predictor. The second also uses the tracked mode energy at each target gate.

\begin{table}[t]
\caption{Local-gate test in the frozen-$S$ model. $f$ is the weight of the selected mode in the gated patch.
Resonant is the largest pairwise current error at $|x|\le1$ and detuned the largest at $|x|=4$. The waveform
columns give the largest relative RMS error of the full current predicted with the first-order (1) and
tracked-energy (2) mode shifts. All errors are in percent.}
\label{tab:local-gate}
\begin{ruledtabular}
\begin{tabular}{lccccccc}
\shortstack{Size\\(nm$^2$)} & $n$ & \shortstack{$U_0$\\(meV)} & $f$ & Resonant & Detuned & \shortstack{Wave-\\form 1} & \shortstack{Wave-\\form 2}\\
\hline
$140\times100$ & 3 & 0 & 0.23 & 21.0 & 0.48 & 2.0 & 2.0\\
$140\times100$ & 3 & 10 & 0.25 & 32.6 & 0.56 & 3.5 & 3.4\\
$140\times100$ & 4 & 0 & 0.20 & 40.8 & 0.57 & 3.3 & 3.3\\
$140\times100$ & 4 & 10 & 0.21 & 26.3 & 0.46 & 23.4 & 23.5\\
$160\times120$ & 3 & 0 & 0.26 & 30.4 & 0.61 & 1.3 & 1.6\\
$160\times120$ & 3 & 10 & 0.25 & 15.6 & 0.44 & 3.6 & 5.2\\
$160\times120$ & 4 & 0 & 0.25 & 38.6 & 0.64 & 0.6 & 0.9\\
$160\times120$ & 4 & 10 & 0.24 & 22.3 & 0.45 & 0.9 & 1.2\\
\end{tabular}
\end{ruledtabular}
\end{table}

Table~\ref{tab:local-gate} lists the results. Over the sweep the local gate changes the terminal weights $p_i$ by at
most 0.03 and the mode width by at most 4 percent. Resonant errors of 15.6--40.8 percent fall to at most
0.65 percent at $|x|=4$. Both predictors meet all three criteria in 7 of the eight devices. The
exception is again the disordered $140\times100$-nm$^2$ four-terminal device, whose waveform errors of 23.41
and 23.54 percent exceed the 10-percent tolerance while its pairwise-error prediction and gate contrast
pass. The selected modes extend over the whole normal region, so a gate on a quarter of its area mainly moves their
energies. Gates that strongly reshape the mode, or move several modes through a resonance by different amounts,
are not tested here.

With finite-gap leads, the local-gate test of the two clean $140\times100$-nm$^2$ devices meets all three criteria,
with largest waveform errors of 1.93 and 2.28 percent, pairwise-error differences of at most 0.037 percentage points
and gate contrasts of 18.6 and 29.8 percentage points (Table~\ref{tab:mechanism-criteria}); its absolute full-current
errors are at most 0.003~nA. These calculations use 64 frequency nodes and four anomalous channels per lead,
inherited from the clean uniform-gate checks. The independent frequency and channel follow-up in
Appendix~\ref{sec:finitegap-refinement} now tests both models at every reported local gate and meets the unchanged
criteria at these settings. Along the uniform and the local sweeps, the relative background change is
$\|K_b(x)-K_b(0)\|_2/\|K_b(0)\|_2$, with $\|\cdot\|_2$ the operator norm and $K_b$ containing all modes except the
selected one. It is at most 3.6 and 5.8 percent in seven
devices and reaches 20 and 27 percent in the disordered four-terminal device. That device's waveform
failure is thus consistent with a background that does not stay frozen, as the predictor assumes. Across all tests
the predictor's absolute full-current errors are at most 0.20~nA in the frozen-$S$ model, 0.26~nA under
the local gate and 0.028~nA with finite-gap leads. The largest recorded phase-sampling sensitivity of the
resonance errors is 0.2 percentage points.

\FloatBarrier
\bibliography{refs}

\end{document}